\documentstyle[twoside]{article} 

\def\d{\partial}
\def\dh{\mathop{\vphantom{\odot}\hbox{$\partial$}}}
\def\dl{\dh^\leftrightarrow}
\def\sqr#1#2{{\vcenter{\vbox{\hrule height.#2pt\hbox{\vrule width.#2pt 
height#1pt \kern#1pt \vrule width.#2pt}\hrule height.#2pt}}}}
\def\w{\mathchoice\sqr45\sqr45\sqr{2.1}3\sqr{1.5}3\,}

\def\=d{\,{\buildrel\rm def\over =}\,}

\def\i3p{\p32\int d^3p}

\def\As{A\hbox to 1pt{\hss /}}
\def\np4{\int d^4p_1\cdots d^4p_{n-1}\, }

\def\nx4{\int d^4x_1\ldots d^4x_n\, }
\def\xnn{x_1,\ldots ,x_n}

\def\Dr{D^{\rm ret}}
\def\Da{D^{\rm av}}
\def\kon#1#2{\vbox{\halign{##&&##\cr
\lower4pt\hbox{$\scriptscriptstyle\vert$}\hrulefill &
\hrulefill\lower4pt\hbox{$\scriptscriptstyle\vert$}\cr $#1$&
$#2$\cr}}}
\def\lra{\longleftrightarrow}
\def\konv#1#2#3{\hbox{\vrule height12pt depth-1pt}
\vbox{\hrule height12pt width#1cm depth-11.6pt}
\hbox{\vrule height6.5pt depth-0.5pt}
\vbox{\hrule height11pt width#2cm depth-10.6pt\kern5pt
      \hrule height6.5pt width#2cm depth-6.1pt}
\hbox{\vrule height12pt depth-1pt}
\vbox{\hrule height6.5pt width#3cm depth-6.1pt}
\hbox{\vrule height6.5pt depth-0.5pt}}
\def\konu#1#2#3{\hbox{\vrule height12pt depth-1pt}
\vbox{\hrule height1pt width#1cm depth-0.6pt}
\hbox{\vrule height12pt depth-6.5pt}
\vbox{\hrule height6pt width#2cm depth-5.6pt\kern5pt
      \hrule height1pt width#2cm depth-0.6pt}
\hbox{\vrule height12pt depth-6.5pt}
\vbox{\hrule height1pt width#3cm depth-0.6pt}
\hbox{\vrule height12pt depth-1pt}}

\def\konw#1#2#3{\hbox{\vrule height12pt depth-1pt}
\vbox{\hrule height12pt width#1cm depth-11.6pt}
\hbox{\vrule height6.5pt depth-0.5pt}
\vbox{\hrule height12pt width#2cm depth-11.6pt \kern5pt
      \hrule height6.5pt width#2cm depth-6.1pt}
\hbox{\vrule height6.5pt depth-0.5pt}
\vbox{\hrule height12pt width#3cm depth-11.6pt}
\hbox{\vrule height12pt depth-1pt}}

\def\eh{{\scriptstyle{1\over 2}}}

\def\lap{\bigtriangleup\,}

\def\i{{\rm int}}

\def\r{{\rm ret}}
\def\a{{\rm av}}

\def\m3{{\mu_1\mu_2\mu_3}}

\def\p{{(+)}}

\def\tu{\tilde u}
\begin{document}

\thispagestyle{empty}
{\bf Z\"urich-University-Preprint ZU-TH-36/94}\\
{\bf hep-th/9411080}\\
{\bf (published in Annals of Physics 244 (1995) 340-425)}\\
\vskip0,5cm
\vbox to 2,5cm{ }
\centerline{\Large\bf Nonabelian Gauge Theories } \vskip 0.3cm
\centerline{\Large\bf The Causal Approach} \vskip2.0cm
\centerline{\large\bf Tobias Hurth$^*$} 
\vskip 0.5cm
\centerline{\large\it
Institut f\"ur Theoretische Physik der Universit\"at Z\"urich}
\centerline{\large\it Winterthurerstr. 190, CH-8057 Z\"urich, Switzerland}
\vskip 2,5cm
{\bf Abstract.} - We present the causal construction of perturbative Yang-Mills theories in four(3+1)-dimensional space-time.\\
We work with free quantum fields throughout. The inductive causal method by Epstein and Glaser leads directly to a finite 
perturbation series and does not rely on an intermediary regularization of the theory. The causal method naturally separates the physical infrared problem of massless theories from ultraviolet-sensitive features like normalizability by regarding the distributional character of the S-matrix. \\
We prove the normalizability of the Yang-Mills theory with fermionic matter fields and study the discrete symmetry transformations in the causal formalism. We introduce a definition of nonabelian gauge invariance which only involves the free asymptotic field operators and give mathematically rigorous and conceptually simple proofs of nonabelian gauge invariance and of the physical unitarity of the S-matrix in all orders of perturbation theory.
\vskip 0.5cm
{\bf PACS.} 11.10 - Field theory, 12.35C-General properties of quantum
chromodynamics.
\vskip 0.5cm
$^*)$ Emailaddress: hurth@physik.unizh.ch  or hurth@insti.physics.sunysb.edu
\vskip 0.3cm

\vfill\eject

\newpage
{\bf

\vskip1.5cm

{\Large\bf Contents}

\vskip1.5cm

1.  Introduction   \hskip0,5cm  3
\vskip0,3cm
2.  Nonabelian Gauge Invariance in the Causal Approach \hskip0,5cm 5 

\hspace{0,4cm} a) Definition of Nonabelian Gauge Invariance \hskip0,5cm 

\hspace{0,4cm} b) General Proof of Nonabelian Gauge Invariance Apart from Splitting \hskip0,5cm 

\vskip0,3cm
3.  Gauge Invariance in Second Order Perturbation Theory \hskip0,5cm 11

\hspace{0,4cm} a) Loop Terms \hskip0,5cm 

\hspace{0,4cm} b) Tree Terms \hskip0,5cm 

\vskip0,3cm
4. Normalizability of Nonabelian Gauge Theories \hskip0,5cm 19
\vskip0,3cm
5. Discrete Symmetries \hskip0,5cm 22

\hspace{0,7cm}(a) Abelian Case \hskip0,5cm 

\hspace{0,7cm}(b) Nonabelian Case \hskip0,5cm 
\vskip0,3cm

6. Proof of Nonabelian C-Number Gauge Invariance \hskip0,5cm 31

\hspace{0,7cm}(a) General Inductive Step \hskip0,5cm 

\hspace{0,7cm}(b) Distribution Splitting \hskip0,5cm 
\vskip0,3cm

7. Proof of the Unitarity in Nonabelian Gauge Theories \hskip0,5cm 45 

\hspace{0.7cm}(a) Preliminaries \hskip0,5cm  

\hspace{0.7cm}(b) Pseudo-Unitarity of S(g) \hskip0,5cm 

\hspace{0.7cm}(c) Definition of the Physical Subspace \hskip0,5cm 

\hspace{0.7cm}(d) Proof of Physical Unitarity \hskip0,5cm 

\vskip0,5cm

Appendix A \hspace{0,5cm} $SU(N)$-Group Theory \hskip0,5cm 53 
\vskip0,5cm
Appendix B \hspace{0,5cm} Derivation of the Cg-Identities \hskip0,5cm 56
\vskip0,5cm
Appendix C \hspace{0.5cm} The Causal Approach to QFT \hskip0,5cm 66
\vskip0,5cm 
References

}

\newpage

\rm

{\Large\bf 1. Introduction }
\vskip0,6cm

The main concern of elementary particle physics is to understand the basic dynamical structures of matter. Up to now three out of four of the fundamental interactions known today can be described as quantized gauge field theories. The local gauge principle thus now appears as the most important structure of the dynamics of matter. The latest precision measurements in the electro-weak sector have confirmed the predictions of the standard model with a high degree of accuracy.

This phenomenological success of today's particle physics is contrasted by the fact that our knowledge of local quantum field theory is quite limited. Up to now the main problem, namely the question whether the principles of relativistic local quantum field theory are consistent with a nontrivial exact S-matrix in four(3+1)-dimensional space-time, is not completely solved [1]. The only real example of the axioms of relativistic local quantum field theory in four(3+1)-dimensional space-time is the theory of free fields in which the S-matrix is just the identity. Nevertheless, there are considerable advances in constructive quantum field theory [2] (see in particular Balaban's work [3]).

Besides the lattice approach [4], perturbation theory is the only systematic method to make quantitative predictions for scattering processes. The naive application of the standard Feynman technique above the classical tree level leads to divergent integrals. In the standard renormalization program one separates the divergences by introducing an intermediate regularization and absorbs them by a redefinition of the physical parameters.

The main idea of the standard renormalization procedure seems to be simple, the technical details, however, are somewhat more complicated, in particular due to the problem of disentangling the overlapping divergences. (In fact the original proof of the renormalizability of QED, given by Ward and Dyson, breaks down at the 14th order of perturbation theory due to these overlapping divergences, as shown by Yang and Mills [5].)

Therefore, the traditional proof of the renormalizability of nonabelian gauge theories given by \\'t Hooft and Veltman [6] is quite involved due to ultraviolet and also infrared problems. Detailed graphical and combinatorial arguments are needed.

The proof given by Becchi, Rouet and Stora [7] (see also [8,9]) uses the quantum action principle and is more elegant. But one has to prove at first by a detailed analysis that the chosen renormalization and regularization scheme preserves the validity of this principle. This was done for the dimensional regularization by Breitenlohner and Maison [10]. There is the well-known troublesome $\gamma_5$-problem in this scheme. In the BPHZ-scheme a revisited analysis of the quantum action prinicple - using the forest-formula - was recently given [11].

We are going to present an alternative approach to perturbative nonabelian gauge theories using methods of Epstein and Glaser [13,12]. This approach is governed by the fundamental concept of causality. In an exemplary mode, we explicitly construct the $SU(N)$-Yang-Mills theory with fermionic matter fields in four (3+1) space-time dimensions by the causal method. We prove the most important properties as normalizability, gauge invariance and unitarity. We show that a straightforward construction of the (finite) perturbation theory without any intermediate modifications is possible, even in the nonabelian gauge theory.

The ultraviolet and infrared problems are fully under control: the former by careful splitting of causal distributions and the latter by regarding the distributional character of the S-matrix. We directly construct the S-matrix in the Fock space of free field operators.

In contrast to the usual approaches, the whole analysis is carried out in configuration space which is more suitable for general considerations.

By using the causal method, one simplifies the analysis of normalizability, gauge invariance and unitarity in the nonabelian gauge theory:

We introduce a definition of nonabelian gauge invariance as a simple commutator condition at every order of perturbation theory separately. This condition only involves the free asymptotic field operators. Therefore, we only need the concept of the linear (abelian) BRS transformations of the free field operators. It is nontrivial that such a condition expresses the full content of nonabelian gauge symmetry of the theory. Now conceptually simple inductive  proofs of this operator gauge invariance  and of the unitarity of the S-matrix in the physical subspace can be given.

In the causal formalism, the physical infrared problem is naturally separated by adiabatic switching of the S-matrix $S(g)$ with a test function $g$. Thus the analysis concerning gauge invariance and unitarity is well-defined, even in a massless theory. Of course it remains to study the physical infrared problem by taking the limit $g \longrightarrow 1$ in the right physical quantities (see [14], [15], [16]). This limit of course is related to the confinement problem.

Our proof of perturbative normalizability of the Yang-Mills theory with fermionic matter fields is simply based on rigorous power counting arguments and does not require any analysis of combinatorial or topological properties of Feynman graphs. 

The causal approach leads to a perturbative series ordered in powers of the coupling constant. So it is not directly suitable to treat the Higgs mechanism that is usually developed in a loop-ordered approach.

The causal method still stands in the context of perturbation theory at this point. Perturbative results are limited to statements about formal power series.
Relativistic quantum field theory is only completely understood if one can also
assign a well-defined meaning to the S-matrix in a nonperturbative sense. This is still an open question.

In recent years other approaches have been developed in order to improve the mathematical and conceptual status of perturbative quantum field theory. These approaches are quite different from the causal method and still use regularizations at intermediate stages. They include Polchinski's method of the continuous renormalization group [17] and the Gallavotti-Nicolo tree formalism [18]. Using these methods, several studies deal with the abelian gauge theory ([19], [20], [21]). Quite recently Bonini et al. have presented also an analysis of the Slavnov-Taylor identities of a pure $SU(2)$-Yang-Mills theory, at least at one-loop level [22]. Moreover, there are some studies about the asymptotic completeness in the abelian theory avoiding the usual infrared regularizations [23].

The paper is organized as follows: In Chapter 2 we give our definition of nonabelian gauge invariance and inductively prove this operator gauge invariance apart from distribution splitting. In Chapter 3 we explicitly study gauge invariance in the second order of perturbation theory. In Chapter 4 we prove the normalizability of the Yang-Mills theory with fermionic matter fields. This proof contains the abelian case as a by-product. In Chapter 5 we study the discrete symmetry transformations in the causal formalism. In Appendix B we express the full content of gauge invariance by relations between C-number distributions and prove them in Chapter 6. Finally, in Chapter 7 we prove the most important property of the nonabelian theory, namely the unitarity of the S-matrix in the physical subspace. In Appendix A we analyze the numerically invariant tensors which transform according to the r-fold tensor product of the adjoint representation of $SU(N)$.
In Appendix C we give a brief introduction to the Epstein-Glaser method, present a solution to the crucial problem of distribution splitting and state some 
results which are decisive especially for the causal construction of massless theories.

\newpage

{\Large\bf 2. Nonabelian Gauge Invariance in the Causal Approach}
\vskip0,8cm
We introduce the concept of nonabelian gauge invariance in the causal construction. We are able to establish the gauge invariance condition as a commutator relation on the Fock space of asymptotic free fields in every order of perturbation theory (Subchapter (a)). This operator condition is sufficient to prove the unitarity of the nonabelian gauge theory, as shown in Chapter 7. We give a general proof of this operator condition by induction on the order n, except for the distribution splitting (Subchapter (b)). The latter problem is treated in Chapter 6.
\vskip0.3cm
{\bf (a) Definition of Nonabelian Gauge Invariance}
\vskip0,3cm
In the causal approach, as described in Appendix C, the S-matrix is directly constructed in the Fock space of the free asymptotic fields in the form of a formal power series
$$S(g)=1+\sum_{n=1}^{\infty}{1\over n!}\int d^4x_1...
d^4x_n\,T_n(x_1,...,x_n)g(x_1)...g(x_n),\eqno(2.1)$$
where $g(x)$ is a tempered test function which switches the interaction. The central objects are the n-point distributions $T_n$ which may be viewed as mathematically well-defined time-ordered products. The defining equations of the theory in the causal formalism are ({\bf i}) the fundamental (anti-)commutation relations of the free field operators, ({\bf ii}) their dynamical equations and ({\bf iii}) the specific coupling of the theory $T_{n=1}$. The n-point distributions $T_n$ in (2.1) are then constructed inductively from the given first order $T_{n=1}$. We want to consider a nonabelian theory. In an exemplary mode, we choose the Yang-Mills theory  with fermionic matter fields in four space-time dimensions. The corresponding specific coupling in the Feynman gauge is 
$$T_1=igf_{abc}({1\over 2}:A_{\mu a}A_{\nu b}F^{\nu \mu}_c:
-:A_{\mu a}u_b\d^\mu\tilde u_c:)+$$
$$+i\frac{g}{2}:\bar \psi_\alpha (\lambda_a)_{\alpha \beta}\gamma^\mu \psi_\beta:A_{\mu a}.\eqno(2.2)$$
All field operators herein are well-defined free fields and these are the only quantities appearing in the whole theory. The double dots denote their normal
ordering. The second term in $T_1$  is needed for the implementation of the gauge invariance in the first order of perturbation theory (see below (2.11)). The specific coupling $T_{n=1}(x)$ of the theory does not contain a quadrilinear term proportional to $g^2$, the four-gluon-vertex. We will explicitly show (Chapter 3b) how this term is automatically generated in second order by our gauge invariance condition (see below (2.12)). This mechanism is essential in order to introduce such a condition of gauge invariance as commutator relation (2.12) in every order n of perturbation theory separately.\\
$A_{\mu a}(x)$ are the (free) gauge potentials satisfying the commutation relations (Feynman gauge)
$$[A_a^{(-)\mu}(x),A_b^{(+)\nu}(y)]=i\delta _{ab}
g^{\mu\nu}D_0^{(+)}(x-y),\eqno(2.3)$$
where $A^{(\pm)}$ are the emission and absorption parts of $A$ and $D^{(\pm)}_0$ the (mass zero) Pauli-Jordan distributions. $u_a(x)$ and $\tilde u_a(x)$ are the free massless fermionic ghost fields fulfilling the anti-commutation relations
$$\{u_a^{(\pm)}(x),\tilde u_b^{(\mp)}(y)\}= -i\delta _{ab}D_0^{(\mp)}(x-y).\eqno(2.4a)$$
The gauge fields are minimally coupled to spinor fields $\psi_\alpha,\bar{\psi}_\beta$. The latter satisfy the anti-commutation relation
$$ \{\psi_\alpha^{(-)}(x), \bar{\psi}_\beta^{(+)}(y)\}=\delta_{\alpha \beta} \frac{1}{i} S^{(+)} (x-y) \eqno(2.4b)$$
where $S^{(+)}=(i\gamma_\mu \d^\mu + m) = D_m^{+}$. $f_{abc}$ denotes the usual antisymmetric structure
constants of the gauge group, say $SU(N)$; $(-i/2) \lambda_a$ are the generators of the fundamental representation of the Lie algebra of the gauge group. The time-dependence of $A,u$ and $\tilde u, \psi$ and $\bar{\psi}$ is given by the wave equation
$$\w A_a^\mu(x)=0,\quad \w u_a(x)=0,\quad \w \tilde
u_a(x)=0, \eqno(2.5)$$
respectively by the free Dirac equation
$$i\gamma_\mu \d^\mu \psi_\alpha(x)=M_{\alpha \beta} \psi_\beta(x)  \eqno(2.6)$$
with a real and diagonal mass matrix $M_{\alpha \beta}$ which satisfies $\lambda_a M = M \lambda_a $. For a simple gauge group like $SU(N)$ $M_{\alpha \beta}$ is a multiple of the unit matrix (2.4b) - according to Schur's lemma. In the following we omit colour indices for the spinor fields in order to simplify the notation. We define
$$F_a^{\mu\nu}\=d \d^\mu A_a^\nu -\d^\nu A_a^\mu. \eqno(2.7)$$
$$j_a^\mu = : \bar{\psi} \gamma^\mu \lambda_a \psi:. \eqno(2.8)$$
According to (2.6), $j_a^\mu$ is a conserved current: $ \quad \d_\mu j_a^\mu = 0 $\\
Besides unitarity, nonabelian gauge invariance is the most important property to the S-matrix. In the causal formalism, the starting point of the analysis are the linear (abelian !) BRS transformations of the free asymptotic field operators.

The generator of the abelian operator transformations is the charge
$$Q\=d \int d^3x\,(\d_{\nu}A_a^{\nu}{\dl}_0u_a), \quad Q^2=0,
\eqno(2.9)$$
with the (anti-)commutation relations
$$[Q,A_\mu^a]_-=i\d_\mu u_a,\quad \{ Q,\tilde{u}_a \}_+ = -i\d_\nu A_a^\nu,\quad \{ Q,u_a \}_+ =0,$$
$$[Q,\psi]_-=0,\quad [Q,\bar{\psi}]_-=0,\quad [Q,F_{\mu \nu}^a]_-=0. \eqno(2.10)$$

Now nonabelian gauge invariance means that the commutator of the specific coupling (2.2) with the charge $Q$ is a divergence (in the sense of vector analysis):
$$[Q,T_{n=1}]= i \d_\nu [igf_{abc} (:A_\nu^a u_b F_c^{\nu \mu} :- \eh : u_a u_b \d^\nu \tilde{u}_c:)+igj_a^\nu u_a] \=d i\d_\nu T_{1/1}^\nu \eqno(2.11)$$
The second term in (2.2) (the gluon-ghost-coupling) is essential that
 $[Q,T_{n=1}]$ can be written as a divergence. Note the different compensation
of terms in the invariance equation (2.11) compared with the invariance of the
Yang-Mills Lagrangean under the full BRS-transformations of the interacting fields in the conventional formalism.

Having defined gauge invariance in the first order of perturbation theory by this relation, we can similarly express the condition of nonabelian operator gauge invariance in every order of perturbation theory separately by a simple commutator relation of the n-point distributions $T_n$ with the charge $Q$, the generator of the free operator gauge transformations (2.9):
$$[Q,T_n(x_1,...,x_n)]=i\sum_{l=1}^n\d_\mu^{x_l}T^
\mu_{n/l}(x_1,...,x_n),\eqno(2.12)$$
where $T_{n/l}^\nu (x_1, \ldots, x_n)$ are n-point distributions of an extended theory which contains, in addition to the usual Yang-Mills couplings $T_{n=1}(x)$ (2.2), the so-called $Q$-vertex $T_{1/1}^\nu (x)$ which already occurs in (2.11) as a divergence-representation of [$Q, T_1$].
The first order S-matrix of the extended theory is equal to
$$S_1(g_0,g_1)\=d \int d^4x \,[T_1(x)g_0(x)+
T_{1/1}^\nu (x)g_{1\nu}(x)].\eqno(2.13)$$
Since $T_{1/1}^\nu$ contains odd numbers of fermionic ghost operators, $g_1=(g_{1\nu})_{\nu=0,1,2,3}\in ({\cal S}({\bf R}^{\it 4}))^4$ must be an anti-commuting C-number field. The higher orders are determined by the usual inductive construction up to local normalization terms (see Appendix C).
The $T^\mu_{n/l}$ are the n-point
distributions of the extended theory with one $Q$-vertex at $x_l$, all other $n-1$ vertices are ordinary Yang-Mills vertices (2.2).\\
The simple operator condition (2.12) involving only well-defined asymptotic field operators expresses the full content of the nonabelian gauge structure of the quantized theory and can be proved by induction on the order n of perturbation theory following the causal construction of $T_n$ and $T_{n/l}^\nu$ (see Subchapter (b)).The nonabelian character is implemented in the theory by the specific couplings (2.2), especially by the structure constants $f_{abc}$.

Thus, the concept of abelian gauge transformations of the free field operators is sufficient in order to derive all consequences of nonabelian gauge invariance in perturbative quantum field theory, in particular the most important one, namely the unitarity of the S-matrix in the physical subspace (i.e. the decoupling of the unphysical degrees of freedom in the theory) - as shown in Chapter 7 of the present study.
\vskip0,1cm
{\bf Further Comments:}\\
$\bullet$ The representation of $[Q, T_{n=1}]$ (2.11) and also of $[Q, T_n]$ (2.12) as a divergence is in general not unique. The most general $Q$-vertex $\tilde T_{1/1}^\nu$ with the same mass dimension and ghost number (see below (2.20)) as $T_{1/1}^\nu$ in (2.11) is the following:
$$[Q,T_1]=i\d_\nu [T_{1/1}^\nu + \alpha B_{1/1}^\nu] \=d i \d_\nu \tilde{T}_{1/1}^\nu$$
$$\mbox{with} \qquad B_{1/1}^\nu=igf_{abc} \d_\mu (: u_a A_b^\mu A_c^\nu:), \quad \d_\nu B_{1/1}^\nu =0, \quad \alpha \in \mbox{C} \quad \mbox{free.} \eqno(2.14)$$
Below we choose $\alpha = 0$, if not otherwise mentioned. This choice has just practical reasons and has no physical consequences. Note that when $\alpha$ is fixed in (2.14) the representation of $[Q, T_n]$ as a divergence, $T_{n/l}^\nu$ (2.12), is also uniquely determined up to local normalization terms.

$\bullet$ So far, we have chosen a specific $T_1$ (2.2) (Faddeev-Popov) and the field operators are fixed in the Feynman gauge (2.3). The generalisations of our procedure to other covariant gauges are straightforward. Furthermore, one can easily derive the most general gauge invariant (in the sense of (2.11)) specific coupling $T_1$ (see [24]). 

$\bullet$ Analogously to the charge $Q$ (2.9) we can define an anti-charge $\bar Q$
$$\bar Q \=d \int d^3 x(\d_\nu A_a^\nu \stackrel{\leftrightarrow}{\d}_0 \tilde u_a) \quad \mbox{with} \quad \bar Q^2=0. \eqno(2.15)$$
It generates the following operator transformations:
$$[\bar Q, A_\mu^a]_-= i\d_\mu \tilde u_a,\quad \{\bar Q, u_a \}_+ =+i\d_\nu A_a^\nu, \quad \{\bar Q, \tilde u_a \}_+=0,$$
$$[\bar Q, \psi]_-=0, \quad [\bar Q, \bar \psi]_-=0, \quad [\bar Q, F_{\mu \nu}^a]_-=0 \eqno(2.16)$$
Anti-gauge invariance means, analogously to (2.11):
$$[\bar Q, T_{n=1}]_-= i\d_\nu[igf_{abc}(:\tilde u_a A_\kappa^b F_c^{\kappa \nu}:-:A_\nu^a \d_\kappa A_b^\kappa \tilde u_c: +: \tilde u_a \d_\nu u_b \tilde u_c:$$
$$-\eh \d_\nu(:\tilde u_a u_b \tilde u_c:) + igj_a^\nu \tilde u_a]\quad+\quad\beta\quad i\d_\nu [igf_{abc} \d_\mu(:A_a^\nu \tilde u_b A_c^\mu :)]$$
$$\=d i\d_\nu [\bar T_{1/1}^\nu + \beta \bar B_{1/1}^\nu]. \eqno(2.17)$$
Again we arrive at the general condition in every order n of perturbation theory 
$$[\bar Q,T_n(x_1,...,x_n)]=i\sum_{l=1}^n\d_\mu^{x_l}\bar T^\mu_{n/l}(x_1,...,x_n),\eqno(2.18)$$
In the present study we focus on the operator gauge invariance (2.12) corresponding to the charge $Q$ only. It is sufficient to prove the crucial property of physical unitarity (see Chapter 7).

$\bullet$ Some brief remarks on the algebraic structure of (2.12):

In addition to the charge $Q$ (2.9), we define the ghost charge
$$Q_c:=i \int d^3x :(\tilde u \stackrel{\leftrightarrow}{\d}_0 u): \eqno(2.19)$$
In the algebra, generated by the fundamental field operators, we introduce a gradation by the ghost number $G (\hat A)$ which is given on the homogenous elements by
$$[Q_c, \hat A]= - G(\hat A) \cdot \hat A. \eqno(2.20)$$
Now we can define an anti-derivation $d_Q$ in the graded algebra by 
$$d_Q \hat A:= Q \hat A-(e^{i\pi Q_c} \hat A e^{-i\pi Q_c}) Q\eqno(2.21)$$
The anti-derivation $d_Q$ is obviously homogenous of degree (-1) and satisfies $d_Q^2 = 0$. Note ${G} (T_n) = 0, {G} (T_{n,l}^\nu) = (-1)$. For further details and consequences see [24].

$\bullet$ For the purposes of illustration, we analyze the condition of operator gauge invariance (2.12) in the abelian case (QED): The corresponding charge $Q$ is similar to (2.9)
$$Q\=d\int d^3x\,(\d_\mu A^\mu {\dl}_0 u) \eqno(2.22)$$
where $u$ is a bosonic ghost field operator 
$$[u(x)^\mp, \quad u(x)^\pm]_-=(-i)D^\pm(x-y) \eqno(2.23)$$
Since
$$[Q,A_\mu]=i\d_\mu u  \eqno(2.24)$$
and $Q$ commutes with $\psi$ and $\overline \psi$, we obtain
$$[Q,T_1]=ie[Q,:\overline \psi \gamma ^\nu\psi:A_\nu]= -e:\overline \psi \gamma ^\nu\psi:\d_\nu u=$$
$$=\d_\nu (-e:\overline \psi \gamma ^\nu\psi:u)\=d i\d_\nu T^\nu_{1/1}. \eqno(2.25)$$
Due to the fact that there is only one ghost operator, the distributions with one $Q$-vertex have the form
$$T^\mu_{n/l}(x_1,...,x_n)=\tau^\mu_l(x_1,...,x_n)u(x_l), \eqno(2.26)$$
where $\tau^\mu_l$ has the external field operators $A,\psi$ and $\overline \psi$ but no $u$. For a fixed $x_l$ we consider from $T_n$ all terms with the external field operator $A_\mu(x_l)$
$$T_n(x_1,...,x_n)=:t^\mu_l(x_1,...,x_n)A_\mu (x_l):+... \eqno(2.27)$$
(The dots stand for the terms without $A_\mu(x_l)$.). It follows from (2.25) that the difference between the $Q$-vertex $x_l$ and an ordinary external vertex consists of the replacement of the external field operator $A_\mu (x_l)$ by $u(x_l)$, hence
$$\tau_l^\mu =t_l^\mu .\eqno(2.28)$$
Now we pick out from (2.12) all terms containing $u(x_l)$
$$it^\mu_l(x_1,...,x_n)\d_\mu u(x_l)=i\d_\mu^{x_l}[\tau^\mu_l(x_1,...,x_n) u(x_l)].$$
Writing this as
$$\d_\mu^l[t_l^\mu u(x_l)]-(\d_\mu^l t_l^\mu)u(x_l)= \d_\mu^l[\tau^\mu_l u(x_l)]$$ 
and using (2.28), we get
$$\d_\mu^l t^\mu_l(x_1,...,x_n)=0.\eqno(2.29)$$
This is the usual gauge invariance condition of refs. [12,25,26]. We see that in QED, each term $\sim [Q,A(x_l)]$ on the left hand side of (2.12) corresponds to the precisely one term $\d^{x_l}$ in the divergence on the right hand side with $Q$-vertex $\sim u(x_l)$. In Yang-Mills theories this is not true: Divergences with regard to inner or other external vertices appear in (2.12). This stems from the fact that $u_a$ in the charge (2.9) is an operator field which interacts with the gluons. It is the nonabelian character of the gauge group which makes gauge invariance much more complicated.

\vskip0,5cm
{\bf (b) General Proof of Nonabelian Gauge Invariance apart from Distribution Splitting}
\vskip0,5cm
In order to prove the gauge invariance condition in the causal approach
$$d_QT_n=[Q,T_n]=i\sum_{l=1}^n \d_\mu^l T_{\mu /l}^\mu \eqno(2.30)$$
we have to prove another relation simultaneously [27].

In the inductive construction of $T_n,\,T_{n/l}$ the n-point
distributions $\tilde T_{n'},\,\tilde T_{n'/l}$ of $S^{-1}$ for $n'<n$ appear (see Appendix C). Therefore our inductive proof of (2.30) only works if we prove in addition
$$d_Q\tilde T_n(x_1,\ldots,x_n)=[Q,\tilde T_n(x_1,...,x_n)]=i\sum_{l=1}^n\d_\mu^{x_l} \tilde T^\mu_{n/l}(x_1,...,x_n).\eqno(2.31)$$
Note that $\tilde T_n,\,\tilde T_{n/l}$ are directly given in terms of the $T_{n'},\,T_{n'/l}$ for $n'\leq n$. Due to $\tilde T_1=-T_1,\,\tilde T_{1/1}=-T_{1/1}$, the case $n=1$ is completely covered by (2.11). We turn to the step from $n-1$ to $n$.
$A'_n$ is constructed from the $T_k,\tilde T_k,\quad 1\leq k\leq n-1$, by means of
$$A'_n(x_1,...;x_n)=\sum_{X,Y} \tilde T_k (X) T_{n-k}(Y,x_n),\eqno (2.32)$$
where $X\=d\{x_{i_1},...,x_{i_k}\},\quad Y\=d\{x_{i_{k+1}},...,x_{i_{n-1}}\}$ and $X\cup Y =\{x_1,...,x_{n-1}\}$ is a partition with $1\leq k\equiv \mid X\mid \leq n-1$. Moreover, we set $x_{i_n}\=d x_n.$ Assuming that (2.31) and (2.30) hold in all lower orders $\leq n-1$, we obtain
$$d_Q A'_n(x_1,\ldots,x_n)=[Q,A'_n(x_1,...;x_n)]=i\sum_{X,Y}\{ \sum_{l=1}^k \d_{\nu}^{x_{i_l}} [\tilde T^{\nu}_{k/l}(X) T_{n-k}(Y,x_n)]+$$
$$+\sum_{l=k+1}^n\d_{\nu}^{x_{i_l}} [\tilde T_k (X) T_{n-k/l-k}^{\nu}(Y,x_n)]\}\=d i\sum_{X,Y}\sum_{l=1}^n\d_{\nu}^{x_{i_l}}[\tilde T_k (X) T_{n-k}
(Y,x_n)]_l^{\nu}.\eqno (2.33)$$
The lower index $l$ means that we have exactly one $Q$-vertex at the l-th argument in $(X,Y,x_n)=(x_{i_1},...,x_{i_n})$, which is $x_{i_l}$. Exchanging the two sums, (2.33) becomes
$$d_Q A'_n(x_1,\ldots,x_n)=i\sum_{l=1}^n\d_{\nu}^{x_l}\sum_{X,Y}[\tilde T_k (X)T_{n-k}
(Y,x_n)]_{l'}^{\nu},\eqno (2.34)$$
with $l'$ being the position where $x_l$ appears in  $(X,Y,x_n)=(x_{i_1},...,x_{i_n})$. Drawing on the analogous formula to (2.32) for $A'_{n/l}$, we see that
$$d_Q A'_n(x_1,\ldots,x_n)=i\sum_{l=1}^n\d_{\nu}^{x_l}A_{n/l}^{\prime\nu}(x_1,...;x_n).\eqno (2.35)$$
One proves (2.30) for $(R'_n,R'_{n/l})$ in the same way. Altogether we conclude for the $D=R'-A'$ distributions that 
$$d_Q D_n(x_1,\ldots,x_n)=[Q,D_n(x_1,...;x_n)]=i\sum_{l=1}^n\d_{\nu}^{x_l}D_{n/l} ^\nu(x_1,...;x_n). \eqno(2.36)$$
Next we want to split both sides of (2.36). On the l.h.s. this is done in the usual manner by normal ordering of $D_n$ using Wick's theorem, splitting the numerical distributions and then calculating the commutator with $Q$. On the r.h.s. we split $D^\nu_{n/l}$ in the usual manner and then take the distributional derivative of the retarded part $R^\nu_{n/l}$. Since $R_n=D_n,\,R_{n/l}=D_{n/l}$ on $\Gamma ^+ \setminus \{(x_n,...,x_n)\}$ and $R_n=0,\,R_{n/l}=0$ on $(\Gamma ^+)^c$, gauge invariance can be violated in this process only in the single point $(x_n,...,x_n)$, i.e. by local terms. It is exactly this point where the $R$-distributions are not completely determined, they have some normalization freedom here. One has to show that $$d_Q R_n(x_1,...,x_n)=[Q,R_n(x_1,...;x_n)]=i\sum_{l=1}^n\d_{\nu}^{x_l}R_{n/l}
^\nu(x_1,...;x_n) \eqno(2.37)$$
can be achieved, if the $R$-distributions are normalized in a suitable way, respecting all desired properties, in particular Lorentz covariance. Then, the $T'=R-R'$ distributions fulfill (1.8) and it is easy to see that this gauge invariance equation survives the symmetrizations
$$T_n(x_1,...,x_n)=\sum_{\pi}{1 \over n!} T'_n (x_{\pi 1},...,x_{\pi n})$$
$$T_{n/l}(x_1,...,x_n)=\sum_{\pi}{1 \over n!} T'_{n/\pi ^{-1}l}
(x_{\pi 1},...,x_{\pi n}).\eqno(2.38)$$
This finishes the inductive step for $T_n$. $\tilde T_n$ may be written as a sum of products of $T_k\,'$s, $1\leq k\leq n$ (C.9), for which (2.30) is proven. Using these facts, one can prove (2.31) with a calculation similar to (2.33). {\bf q.e.d.}\\
Note that this inductive proof is independent of the special choice of the divergence-representation $\tilde{T}_{1/1}^\nu$ (2.14).\\

Summing up, we have proven nonabelian gauge invariance (2.12,2.30) by induction on the order n of perturbation theory, except for the distribution splitting. It remains to be proven that nonabelian gauge invariance is preserved under this operation.

There are essentially two ways to show that (2.12) can be preserved after distribution splitting.

$\bullet$ The first one is to prove directly the operator gauge invariance (2.12) by purely algebraic methods. This will be discussed in a forthcoming paper [24].
One should keep in mind that the operator condition (2.12) is sufficient to show the unitarity in the physical subspace (see Chapter 7).

$\bullet$ In Chapter 6 of the present study we proceed in another way:
We express the operator gauge invariance by a set of identities between C-number distributions analogous to the Slavnov-Taylor identities. In this way we develop the full content of nonabelian gauge invariance.
These C-number identities for gauge invariance (Cg-identities) are sufficient for the operator condition (2.12). In Appendix B, we explicitly derive all types of Cg-identities and isolate those types which require nontrivial proofs. These few nontrivial identities are proven in Chapter 6 by suitable normalization. Here all symmetries of the theory, in particular the global $SU(N)$-symmetry (Appendix A) and charge conjugation invariance (Chapter 5b), are needed .
\vskip1.0cm

{\Large\bf 3. Gauge Invariance in Second Order}
\vskip 1cm

We prove gauge invariance in the sense of (2.12) up to the second order of perturbation theory in order to illustrate several points by explicit calculations. In Subchapter (a) we first consider the one-loop terms. We give the explicit expressions by using the splitting method for massless distributions introduced in Appendix C. In Subchapter (b) we study the gauge invariance of the tree terms. We prove that the tree part of $T_2(x,y)$ with appropriate normalization is gauge invariant [27].
The specific coupling $T_1(x)$ - defining the theory - contains the three-gluon vertex in addition to the gluon-ghost and gluon-matter coupling. One may wonder why the first-order $T_1(x)$ does not contain a quadrilinear term proportional to $g^2$, the four-gluon vertex. As we shall see, this term is automatically generated in second order by gauge invariance. This mechanism is essential if one wants to introduce the condition of gauge invariance as commutator relation (2.12) separately in every order of perturbation theory.

\vskip0,5cm
{\bf (a) Loop Terms in Second Order}
\vskip0,5cm
$\bullet$ We calculate the causal d-distributions. By normally ordering the direct product
$$R'_2(x;y)=-T_1(y)T_1(x), \eqno (3.1)$$
we select all terms with two contractions:
$$R'_2 (x;y)=r_{AA}^{\prime\nu \mu }(x-y):A_\nu (x)A_\mu (y):
+r_{AF}^{\prime\nu \mu \kappa }(x-y):A_\nu (x)F_{\mu \kappa }(y):+ \eqno (3.2)$$
$$+r_{FA}^{\prime\mu \kappa \nu }(x-y):F_{\mu \kappa }(x)A_\nu (y):
+r_{FF}^{\prime\mu \kappa \nu \tau }(x-y):F_{\mu \kappa }(x)F_{\nu \tau }(y):+$$
$$+r_{u \tilde u}^{\prime\mu} (x-y):u(x)\d_\mu\tilde u(y):
+r_{\tilde u u}^{\prime\mu} (x-y):\d_\mu\tilde u(x)u(y):+... \quad \mbox{tree terms}$$
where the two free field operators have the same colour index and there is a sum over this index. $r'_{AA}$ has three contributions: the gluon-loop term $r_{AA}^{\prime 1}$ containing two times the gluon vertex $:AAF:$ of $T_1$, the ghost-loop term $r_{AA}^{\prime 2}$ containing two times the ghost vertex $:Au\d\tilde u:$.

and the matter-loop term $r_{AA}^{\prime 3}$ containing two times the gluon-matter vertex $:jA:$. For the matter-loop term we can use the result of the corresponding term in the abelian theory (see chapter 3.6 in [12]) with some minimal modifications 
concerning the colour structure. Note that all propagators are massive in this term.\\
%$$\eqno (3.3)$$
\begin{eqnarray*}
r_{AA}^{\prime 1\,\nu \mu }(x-y) & = & g^2N[2(\d^\nu\d^\mu D^-_0)(x-y) 
D^-_0(x-y) \\
 &  & -3(\d^\nu D^-_0)(x-y)(\d^\mu D^-_0)(x-y)],\\
r_{AA}^{\prime 2\,\nu \mu }(x-y) & = & g^2N(\d^\nu D^-_0)(x-y)(\d^\mu D^-_0)(x-y),\\
r_{AA}^{\prime 3\,\nu \mu }(x-y) & = & 2g^2[2(\d^\mu D^-_m)(x-y)(\d^\nu D^-_m)(x-y)\\
 &  & -(\d^\kappa D^-_m)(x-y)(\d_\kappa D^-_m)(x-y) g^{\nu\mu}\\
 &  & + m^2 D^-_m(x-y) D^-_m(x-y) g^{\nu\mu}]\\
r_{AF}^{\prime\nu\mu\kappa}(x-y) &=& {g^2 \over 2}N[g^{\nu\kappa}(\d^\mu D^-_0)(x-y)D^-_0(x-y)\\ 
 &  & -g^{\nu\mu}(\d^\kappa D^-_0)(x-y)D^-_0(x-y)],\\
r_{FF}^{\prime\mu \kappa \nu \tau }(x-y) & = & -{g^2 \over 2}Ng^{\mu\nu}g^{\kappa\tau}D^-_0(x-y)D^-_0(x-y),\\
r_{u \tilde u}^{\prime\mu} (x-y) & = & r_{\tilde u u}^{\prime\mu} (x-y)=g^2N \d^\mu D^-_0(x-y)D^-_0(x-y),
\end{eqnarray*}  
$$\eqno (3.3)$$ 

The corresponding numerical distributions of $A'_2(x;y)=-T_1(x)T_1(y)$ are obtained (in this special case) by replacing $D^-_0$ by $D^+_0$.\\
This leads to the causal numerical distributions of

$$D_2(x,y) \left| \begin{array}{ll} &\\ \mbox{\small{loops}} \end{array} \right. = R'_2(x,y)-A'_2(x,y) \eqno (3.4)$$

{\bf Remark:} It may seem surprising why the products $\d^aD^-_0\d^bD^-_0,\,\mid a\mid ,\mid b\mid =0,1,2,$ appearing in
(3.3) exist. The reason is that in the Fourier transformed
expression 
$${1\over (2\pi)^2}\int d^4q\,(-i(p-q))^a\hat D^-_0(p-q)(-iq)^b
\hat D^-_0(q) \eqno (3.5)$$
the intersection of the supports of the two $\hat D^-_0$ is a
compact set. This is a general feature of perturbative quantum field theory:

As is well-known, L. Schwartz's famous theorem [28] states that there is no multiplication of all tempered distributions that is associative, commutative and which generalizes the multiplication of ordinary continuous functions.

But a general concept of multiplication is not at all necessary. In quantum field theoretical constructions, only products of Fourier transforms of advanced (resp. retarded) tempered distributions appear in addition to well-defined tensor products - a fact which can be verified explicitly in equations (3.3).

It is possible, however, to define a natural, associative and commutative multiplication on the set of Fourier transforms of advanced (resp. retarded) distributions so that the product of these distributions is the Fourier transform of some advanced (resp. retarded) distribution.

In this context one should keep in mind the classical theorem that the Fourier transforms $F[d(p)](x)=d(x)$ of advanced (resp. retarded) distributions $d(p)$, supp $d(p) \in V^+$  (resp. $\in V^-$), are boundary values of functions $d^+(z)\quad$ (resp. $d^-(z)$) analytic in the tube $\bf R^{\it m} \it +iV^+$\\
(resp. $\bf R^{\it m}\it -iV^+)$) \rm [29].
\vskip0,3cm
$\bullet$ In order to calculate the Fourier transformation of the causal numerical distributions of $D_2=R'_2-A'_2$, we insert the explicit expression of $\hat D^-_0$ in (3.5). We see that we have to deal with the integrals 
$$I^\pm_{...}(p)\=d \int d^4q \delta ((p-q)^2)\delta (q^2) \theta (\pm (p_0-q_0))\theta (\pm q_0)\{1,\,q_\nu,\,q_\nu q_\mu\}.\eqno (3.6)$$
Starting from (3.3) and the corresponding $a'$-distributions, 
one can easily express the $\hat d$-\hyphenation{dis-tri-bu-tions}distributions in terms of the integrals (3.6):

\begin{eqnarray*}
\hat d_{AA}^{1\nu\mu}(p) &=& -{g^2N\over (2\pi)^4}[2(I^{+\nu\mu}(p)-I^{-\nu\mu}(p))-3(p^\nu I^{+\mu}(p)\\
 & & -I^{+\nu\mu}(p)-p^\nu I^{-\nu\mu}(p)+I^{-\nu\mu}(p))],\\ 
\hat d_{AA}^{2\nu\mu}(p) &=& -{2g^2N\over (2\pi)^4}[p^\nu I^{+\mu}(p)- I^{+\nu\mu}(p)-p^\nu I^{-\mu}(p)+I^{-\nu\mu}(p)],\\
\hat d_{AA}^{3\nu\mu}(p) &=& -{g^2\pi\over 3(2\pi)^4} [p^\mu p^\nu - p^2g^{\mu\nu}] (1+{2m^2\over p^2})(1-{4m^2\over p^2})^{1\over 2} \theta (p^2-4m^2)\,{\rm sgn}\,p_0,\\
\hat d_{AF}^{\nu\mu\kappa}(p) &=& {ig^2N\over 2(2\pi)^4}[g^{\nu\mu}(I^{+\kappa}-I^{-\kappa})-g^{\nu\kappa}(I^{+\mu}-I^{-\mu})],\\
\hat d_{FF}^{\mu\kappa\nu\tau}(p) &=& -{g^2N\over 2(2\pi)^4}[I^+-I^-]g^{\mu\nu}g^{\kappa\tau},\\
\hat d_{u\tilde u}^\mu(p) &=& -{ig^2N\over (2\pi)^4}[I^{+\mu}-I^{-\mu}]. \hbox{\hspace{8.3cm}(3.7)}
\end{eqnarray*}

$\bullet$ It remains to compute the integrals (3.6). Due to the two $\delta\theta$ factors they vanish, except for $p^2>0$ and $p^0>0$ for $I^+_{...}(p)$ rsp. $p^0<0$ for $I^-_{...}(p)$. Therefore, we may choose a special Lorentz frame with $p=(p^0,\vec 0)$. Especially for the scalar integral $I^+$ we obtain (assuming $p^0>0$)

$$I^+(p)=\int d^4q\,\,\delta (p_0^2-2p_0q_0)\theta (p_0-q_0){1\over 2|\vec q|}\delta (q_0-|\vec q|)=$$

$$=\int d\Omega d|\vec q|\,\,|\vec q|^2{1\over 2p_0}
\delta (|\vec q|-{p_0\over 2}){1\over 2|\vec q|}
\theta (p_0-|\vec q|)={\pi \over 2}.\eqno (3.8)$$
$I^-$ can be calculated analogously and the result is in an arbitrary
Lorentz frame
$$I^\pm (p)={\pi \over 2}\theta (p^2)\theta (\pm p^0).\eqno (3.9)$$

Computing the vector integral $I^+_\nu$ for $p=(p^0,\vec 0),\,p^0>0$, we have a nonvanishing contribution for $\nu=0$ only. By comparing it with (3.8), we obtain for $I^+_0$ an additional factor $q_0$ in the integrand which is set equal to ${p_0\over 2}$. This leads us to
$$I^\pm_\nu (p)={\pi \over 4}p_\nu \theta (p^2)\theta (\pm p^0)\eqno (3.10)$$
in an arbitrary Lorentz frame.

The covariant decomposition of the tensor integral $I^\pm_{\nu\mu}$ is
$$I^\pm_{\nu\mu}(p)=A^\pm (p^2)p_\nu p_\mu+B^\pm (p^2)g_{\nu\mu}.\eqno (3.11)$$

A glance at (3.6) shows $I^{\pm\nu}_\nu=0$. This implies
$$B^\pm (p^2)=-{p^2\over 4}A^\pm (p^2).\eqno (3.12)$$
Calculating $I^\pm_{\nu\mu}(p)p^\nu p^\mu$ for $p=(p^0,\vec 0),\,p^0>0$, an additional factor $(p_0q_0)^2$ appears in the integrand of (3.8). Therefore, we obtain (in an arbitrary Lorentz frame)
$$I^\pm_{\nu\mu}(p)p^\nu p^\mu ={\pi\over 8}p^4\theta (p^2)\theta (\pm p^0).\eqno (3.13)$$ 
Taking (3.11),(3.12) into consideration, we finally arrive at 
$$I^\pm_{\nu\mu}(p)=[p_\nu p_\mu-{p^2\over 4}g_{\nu\mu}]{\pi\over 6}\theta (p^2)\theta (\pm p^0).\eqno (3.14)$$
$\bullet$ We obtain the causal numerical distributions:
\begin{eqnarray*}
\hat d_{AA}^{1\,\nu \mu }(p) &=& -{g^2N\pi\over (2\pi)^4 24}[2p^\nu p^\mu -5p^2g^{\nu\mu}]\theta (p^2)\,{\rm sgn}\,p^0,\hbox{\hspace{5.1cm}(3.15)}\\
\hat d_{AA}^{2\,\nu \mu }(p) &=& -{g^2N\pi\over (2\pi)^4 24} [2p^\nu p^\mu +p^2g^{\nu\mu}]\theta (p^2)\,{\rm sgn}\,p^0,\\
\hat d_{AA}^{3\,\nu \mu }(p) &=& -{g^2\pi\over (2\pi)^4 3} [p^\nu p^\mu - p^2g^{\nu\mu}] (1+{2m^2\over p^2})(1-{4m^2\over p^2})^{1\over 2}\theta (p^2-4m^2)\,{\rm sgn}\,p^0,\\
\hat d_{AF}^{\nu \mu \kappa }(p) &=& -\hat d_{FA}^{\mu \kappa \nu }(p) =-{g^2N\pi i\over (2\pi)^4 8}[g^{\nu\kappa}p^\mu -g^{\nu\mu}p^\kappa]\theta (p^2)\,{\rm sgn}\,p^0,\\
\hat d_{FF}^{\mu \kappa \nu \tau }(p) &=& -{g^2N\pi\over (2\pi)^4 4}g^{\mu\nu} g^{\kappa\tau}\theta (p^2)\,{\rm sgn}\,p^0,\\
\hat d_{u \tilde u}^\mu (p) &=& \hat d_{\tilde u u}^\mu (p)=-{g^2N\pi i\over (2\pi)^4 4}p^\mu \theta (p^2)\,{\rm sgn}\,p^0.
\end{eqnarray*}

Note the transversality relations
$$p_\nu (\hat d_{AA}^{1\,\nu \mu }(p)+\hat d_{AA}^{2\,\nu \mu }(p)) =0, \quad \hat d^\mu_{u\tilde u}(p)=p^\mu \hat d_{u\tilde u}(p), \eqno (3.16)$$
$$p_\nu \hat d_{AF}^{\nu \mu \kappa }(p)=0, \quad p_\nu \big( \hat d_{AA}^{3\,\nu \mu }(p) \big)=0.$$
The first one would be violated without the ghost term $\hat d_{AA}^2$.

$\bullet$ Now we apply the splitting method for massless distributions introduced in Appendix C to the causal $d$-distributions
(3.15). They all have the form
$$\hat d(p)=P(p)\hat d_1(p) \quad \mbox{with} \quad \hat d_1(p)\=d \theta (p^2)\,{\rm sgn }\,p^0 \eqno (3.17)$$
and $P(p)$ is a covariant polynomial in $p$. 
It suffices to find a covariant retarded part $\hat r_1(p)$ of $\hat d_1(p)$. Then
$$\hat r(p)\=d P(p)\hat r_1(p)\eqno (3.18)$$
is a splitting solution of $\hat d(p)$ (3.16). 

Inserting (C.41),(3.17) in (C.46) and taking $\omega (\hat d_1)=0$ into consideration, we obtain for

$p\in V^+,\,q\rightarrow 0$ in such a way that $p-q\in V^+,\,q^2<0$
$$\hat r_1(p)=\lim_{q\to 0}\left\{{i\over 2\pi}\int {dt\over (t-i0)(1-t+i0)} \theta ((tp+q)^2)\,{\rm sgn}\,(tp^0+q^0)+P_q(p)\right\},\eqno (3.19)$$
where the polynomial $P_q(p)$ is constant (in $p$). The zeros of $(tp+q)^2$ are
$$t_{1,2}={1\over p^2}(-pq\pm \sqrt N) \qquad \mbox{with}\qquad N\=d (pq)^2-p^2q^2>(pq)^2. \eqno (3.20)$$

Remember $p^2>0,\,q^2<0$.
Note $t_1>0,\,t_2<0$. The integral in (3.19) may be
simplified to
$$\left\{{-i\over 2\pi}\int_{-\infty}^{t_2}dt+{i\over 2\pi}
\int_{t_1}^\infty dt\right\}\left\{{1\over t}-P{1\over t-1}-i\pi\delta
(t-1)\right\}=$$
$$={i\over 2\pi}(-{\rm log}\mid t_2\mid+{\rm log}\mid t_2-1\mid-
{\rm log}\mid t_1\mid+{\rm log}\mid t_1-1\mid -i\pi).\eqno(3.21)$$
Since $t_{1,2}\rightarrow 0$ for $q\rightarrow 0$, the logarithms
${\rm log}\mid t_{1,2}-1\mid$ vanish in this limit and there remains in (3.21)
$$-{i\over 2\pi}({\rm log}\mid t_1t_2\mid +i\pi)={i\over 2\pi}\Bigl(
{\rm log}{p^2\over |q^2|}-i\pi\Bigl).\eqno (3.22)$$
Adding the polynomial $P_q(p)={i\over 2\pi}{\rm log}\,
{|q^2|\over M^2}$ in (3.19), where $M>0$ is an arbitrary
scale parameter which has the dimension of a mass,
we obtain a (covariant) splitting solution of $d_1$
$$\hat r_1(p)={i\over 2\pi}\Bigl({\rm log}\,{p^2\over M^2}-i\pi\Bigl).
\eqno (3.23)$$
So far this holds for $\lim_{q\to 0}(p-q)=p\in V^+$.
By analytic continuation in ${\bf R}^4+iV^+$, we
obtain for $p\in{\bf R}^4$
$$\hat r_1(p)={i\over 2\pi}{\rm log}\,{-(p^2+ip^00)
\over M^2}.\eqno (3.24)$$
Note that scale invariance is violated after
distribution splitting. This is also a general feature of the causal construction.

$\bullet$ The transversality relations (3.15) are also fulfilled by the corresponding retarded distributions:
$$p_\nu (\hat r_{AA}^{1\,\nu \mu }(p)+\hat r_{AA}^{2\,\nu \mu }(p)) =0, \quad \hat r^\mu_{u\tilde u}(p)=p^\mu \hat r_{u\tilde u}(p), \eqno (3.25)$$
$$p_\nu \hat r_{AF}^{\nu \mu \kappa }(p)=0, \quad p_\nu \big( \hat r_{AA}^{3\,\nu \mu }(p) \big) =0.$$
The first one would be violated without the ghost term $\hat d_{AA}^2$.
Their validity is determined exclusively by the covariant polynomial $P(p)$ of (3.17). But the covariant structure is not changed in the splitting process (see 3.18).
One can easily make out that the transversality relations of the numerical distributions (3.25) are sufficient for the operator gauge invariance of $R_2(x,y)\mid_{\mbox{\small{loops}}}$:
$$[Q,R_2(x,y) \mid_{\mbox{\small{loops}}} ]= \mbox{div} \eqno(3.26)$$
Since $R'_2(x,y) \mid_{\mbox{\small{loops}}}$ in (3.2) is trivially gauge invariant, we arrive at a gauge invariant two-point distribution $T_2=R_2-R'_2:$
$$[Q,T_2(x,y) \mid_{\mbox{\small{loops}}} ]= \mbox{div} \eqno(3.27)$$
\vskip0,5cm
{\bf (b) Tree terms in Second Order}
\vskip0,5cm

We now consider the tree terms $D_2(x,y) \mid_{\mbox{\small{trees}}}$ where we are able to study gauge invariance in an indirect but effective way:

$\bullet$ We are interested in the commutator $[Q, D_2]$. The commutation does not affect the numerical distributions in $D_2$, it only changes the field operators without disturbing normal ordering. Consequently, in the splitting of $[Q,\,D_2]$ we have to split only those numerical distributions which also  appear in $D_2$. With the same convention of normalization in the splitting of these numerical distributions, we can calculate $[Q, R_2]$ directly by splitting $[Q, D_2]$. This procedure has the advantage that it preserves the divergence structure and shows immediately where gauge invariance may break down. We start from
$$[Q,\,D_2(x,y)]=[Q,\,[T_1(x),\,T_1(y)]]=i \d_\nu^x([T_{1/1}^\nu(x),T_1(y)])+i \d_\nu^y([T_1(x),T_{1/1}^\nu(y)]).\eqno(3.28)$$
The first term is a divergence with regard to $x$ and the second with regard to $y$. In fact, the second term is obtained from the first by interchanging $x$ and $y$ and multiplying it by (-1). The question is whether the same (divergence form) is true for the commutator $[Q,\,R_2(x,y)]$ obtained by causal splitting of (3.28).\\
 Since this commutator agrees with (3.28) on $\{(x-y)^2\ge 0, x^0-y^0>0\}$, gauge invariance can only be spoiled by local terms with support $x=y$. But such terms do arise in the process of distribution splitting of (3.28). First we must normally order the commutator (3.28). In this process we get tree and loop graphs (which we have already considered in the last subchapter) due to the relation\\ 
$$[:ABC:,\,:DEF:]=:[ABC,\,DEF]:+{\rm loops},\eqno(3.29)$$
which is true in our situation.

$\bullet$ We only have to examine the first term on the right of (3.28). Substituting (2.11), but leaving out the terms including the gluon-matter vertex 
$T_1^{matter}$, we get a first contribution
$$K_1(x,y)=-if_{abc}f_{a'b'c'}\d_\nu^x:[A_{\mu a}u_b\d^\nu A_c^\mu,\,A_{\mu'a'} A_{\nu'b'}\d^{\nu'}A_{c'}^{\mu'}]:\=d\d_\nu^xK_1^\nu,\eqno(3.30)$$
where, after calculating the commutator,
$$K_1^\nu=f_{abc}u_b(x):A_{\mu a}(x)\{f_{cb'c'}\d_x^\nu D(x-y)
A_{\nu'b'}\d_y^{\nu'}A_{c'}^\mu\eqno(3.31a)$$
$$+f_{a'cc'}\d_x^\nu D(x-y)A_{\mu'a'}\d^\mu A_{c'}^{\mu'}\eqno(3.31b)$$
$$+f_{a'b'c}\d_x^\nu\d_y^{\nu'}D(x-y)A_{a'}^\mu A_{\nu'b'}\}:\eqno(3.31c)$$
$$+f_{abc}u_b(x):\{f_{a'b'a}\d_y^{\nu'}D(x-y)A_{\mu a'}A_{\nu'b'}+\ldots\}\d^\nu A_c^\mu(x):.\eqno(3.31d)$$
The splitting of (3.30) must be performed as follows: We carry out the derivative $\d_\nu^x$ and then we split the causal $D$-distributions in each term
$$D(x-y)=\Dr (x-y)-\Da (x-y).\eqno(3.32)$$
We examine whether the resulting retarded distribution $R_1$ is again a divergence, that means, whether the derivative $\d_\nu^x$ can again be taken out after the splitting. This is not the case because 
$$\d_\nu\d^\nu \Dr (x-y)=\delta(x-y)\eqno(3.33)$$
in contrast to $\w D(x-y)=0$. Here is the origin of local terms:
$$R_1(x,y)=\d_\nu^x\Bigl\{f_{abc}u_b(x):A_{\mu a}(x)f_{cb'c'}
\d_x^\nu\Dr(x-y)A_{\nu'b'}\d_x^{\nu'}A_{c'}^\mu+\ldots\Bigl\}+$$
$$-f_{abc}u_b(x):A_{\mu a}(x)\Bigl\{f_{cb'c'}
\delta(x-y)A_{\nu'b'}\d^{\nu'}A_{c'}^\mu+$$
$$+f_{a'cc'}\delta(x-y)A_{\mu'a'}\d_y^\mu A_{c'}^{\mu'}+f_{a'b'c}
\d_y^{\nu'}\delta(x-y)A_{a'}^\mu A_{\nu'b'}\Bigl\}:.\eqno(3.34)$$
The local terms compensate those coming from the divergence,
if $\d_\nu^x$ operates on $\d_x^\nu\Dr$; (3.31d) does not give rise to such terms.

For gauge invariance these local terms must drop out. In the splitting of the second commutator in (3.28) with $x$ and $y$ interchanged, the retarded part contains the advanced Pauli-Jordan distributions (3.32) $-\Da (y-x)$. These give rise to local terms with the same sign as in (3.34), so that there is no compensation. But the local terms coming from the vertex $x$ cancel out separately. We write (3.34) as follows: 
$$R_1(x,y)=\d_\nu^xR_1^\nu+f_{abc}f_{a'b'c}u_b(x):\Bigl\{\d_x
^{\nu'}A_a^\mu(x)\delta(x-y)A_{\mu a'}(y)A_{\nu' b'}(y)+$$
$$+A_a^\mu(x)\d_x^{\nu'}\delta(x-y)A_{\mu a'}(y)A_{\nu' b'}(y)
\Bigl\}:,$$
by means of the Jacobi identity for the $f_{abc}$, or
$$R_1(x,y)=\d_\nu^xR_1^\nu+f_{abc}f_{a'b'c}:\Bigl\{\d_\nu^x\Bigl(u_b A_a^\mu\delta(x-y)A_{\mu a'}A_{b'}^\nu\Bigl)\eqno(3.35)$$
$$-\d_\nu^xu_bA_a^\mu\delta(x-y)A_{\mu a'}A_{b'}^\nu\Bigl\}:.\eqno(3.35a)$$

$\bullet$ The second commutator $K_2$ contributing to (3.28) on the tree level
$$K_2(x,y)=if_{abc}f_{a'b'c'}\d^x_\nu:[A_{\mu a}u_b\d^\mu
A_c^\nu,\,A_{\mu'a'}A_{\nu'b'}\d^{\nu'}A_{c'}^{\mu'}]:$$
does not lead to a local term because there is no second
derivative $\d_x^\nu$. But the third commutator
$$K_3(x,y)=if_{abc}f_{a'b'c'}\d^x_\nu:[A_{\mu a}u_b\d_x^\nu A_c^\mu, \,A_{\mu'a'}u_{b'}\d_y^{\mu'}\tu_{c'}]:\=d\d_\nu^xK_3^\nu\eqno(3.36)$$
has such a derivative and produces a $\delta$-term:
$$K_3^\nu=f_{abc}:\Bigl\{f_{cb'c'}u_bA_{\mu a}\d_x^\nu D(x-y)\d^\mu \tu_{c'}u_{b'}+$$
$$+f_{ab'c'}u_b\d^\nu A_c^\mu D(x-y)\d_\mu^y\tu_{c'}u_{b'}-f_{a'b'b}
A_{\mu'a'}u_{b'}\d_y^{\mu'}D(x-y)A_{\mu a}\d^\nu A_c^\mu\Bigl\}:,\eqno(3.37)$$
namely
$$R_3=\d_\nu^xR_3^\nu-f_{abc}f_{cb'c'}:u_bA_{\mu a}\delta(x-y)\d^\mu 
\tu_{c'}u_{b'}:.\eqno(3.38)$$
The last local term is again transformed by means of the Jacobi identity
$$L_3\=d f_{abc}f_{cb'c'}:u_bu_{b'}A_{\mu a}\d^\mu\tu_{c'}:
\delta(x-y)\eqno(3.39)$$
$$=-f_{b'ac}f_{cbc'}:u_bu_{b'}A_{\mu a}\d^\mu\tu_{c'}:\delta
-f_{bb'c}f_{cac'}\ldots \eqno(3.39a)$$
Interchanging $b\leftrightarrow b'$ in the first term, it agrees with (3.39) up to the minus sign. Hence,
$$L_3=-{1\over 2}f_{bb'c}f_{cac'}:u_bu_{b'}A_{\mu a}\d^\mu\tu_{c'}:
\delta(x-y).\eqno(3.40)$$

$\bullet$ The commutator

$$K_4=-if_{abc}f_{a'b'c'}\d_\nu^x:[A_{\mu a}u_b\d^\mu A_c^\nu,\,A_{\mu'a'}u_{b'}\d^{\mu'}\tu_{c'}]:\eqno (3.41)$$
does not lead to a local term, one further commutator ($K_5$) vanishes, but the final one
$$K_6(x,y)=-{i\over 2}f_{abc}f_{a'b'c'}\d_\nu^x:[u_au_b\d^\nu\tu_c,
\,A_{\mu' a'}u_{b'}\d^{\mu'}\tu_{c'}]:\=d\d_\nu^x K_6^\nu \eqno (3.42)$$
$$K_6^\nu=-{i\over 2}f_{abc}f_{a'b'c'}:\Bigl\{u_au_b\{\d^\nu\tu_c,\,u_{b'}\}
\d^{\mu'}\tu_{c'}$$
$$-u_au_{b'}\{u_b,\,\d^{\mu'}\tu_{c'}\}\d^\nu\tu_c-u_{b'}\{u_a,\,
\d^{\mu'}\tu_{c'}\}u_b\d^\nu\tu_c\Bigl\}:A_{\mu'a'}=$$
$$={1\over 2}f_{abc}f_{a'cc'}:u_au_b\d_x^\nu D(x-y)\d^{\mu'}\tu
_{c'}:A_{\mu'a'}+\ldots,$$
produces another local term 
$$R_6=\d_\nu^xR_6^\nu-{1\over 2}f_{abc}f_{a'cc'}:u_au_bA_{\mu a'}
\d^\mu\tu_{c'}:\delta(x-y).$$
Changing $a'\to a$, $a\to b'$, this cancels against $L_3$ (3.40).

$\bullet$ Summing up, the only breakdown of gauge invariance so far, is the local term (3.35a). This term is just the commutator of the usual four-gluon interaction with $Q$ up to a minus sign.

There is one tree graph with $\omega=0$ in $D_2 |_{\mbox{\small{trees}}}$, namely
$$D_2(x,y)=-f_{abc}f_{a'b'c'}:[A_{\mu a}A_{\nu b}\d^\nu A_c^\mu,\,
A_{\mu'a'}A_{\nu'b'}\d^{\nu'}A_{c'}^{\mu'}]: \eqno (3.43)$$
$$=-if_{abc}f_{a'b'c}:A_{\mu a}A_{\nu b}\d_x^\nu\d_y^{\nu'}D(x-y)
A_{a'}^\mu A_{\nu'b'}:+\ldots$$
Therefore the general splitting solution contains exactly one undetermined normalization term
$$L_1=-iCf_{abc}f_{a'b'c}:A_{\mu a}A_{\nu b}\delta(x-y)A_{a'}^\mu
A_{b'}^\nu :.\eqno(3.44)$$
Its commutator  with $Q$
$$[Q,\,L_1]=4Cf_{abc}f_{a'b'c}:\d_\mu u_aA_{\nu b}\delta(x-y)A_{a'} ^\mu A_{b'}^\nu:\eqno(3.45)$$
has the same form as (3.35a). Consequently, choosing $C=\eh$, the
unwanted local term in (3.35) and the corresponding one from the second term in (3.28) drop out.

The terms including $T_1^{matter}$, $T_{1/1}^{matter}$ have to be separately gauge invariant which is shown analogously. Note that there is no additional normalization term.
Summing up, with the normalization $C=\eh$ the tree terms $R_2(x,y)\mid_{\mbox{\small{trees}}}$ are gauge invariant.
We see that gauge invariance fixes the normalization of $T_2(x,y)\mid_{\mbox{\small{trees}}}$ uniquely.

\vskip 1.0cm
{\Large\bf 4. Normalizability of Nonabelian Gauge Theories}
\vskip 1cm

In the causal approach the question of the normalizability of a quantum field theory does not involve the proof of its finiteness. Divergences do not appear at all in our approach.

The problem of normalizability, however, means above all that we have to show that the number of the (of course finite) constants $C_a$ in (C.37) to be fixed by physical conditions stays the same in all orders $n$ of perturbation theory. This means that finitely many normalization conditions are sufficient to determine the S-matrix completely.

If the number of the constants (still to be fixed) increases with the order $n$ of perturbation theory, one usually talks of a nonnormalizable theory. Such a theory is of course well-defined and can be constructed inductively but it has less predictive power.

It should be stressed that in the causal approach the question of the normalizability is separated from other conceptual questions such as gauge invariance and unitarity.

In this chapter we prove that the Yang-Mills theory with fermionic matter fields is normalizable. The concept of the singular order of distributions introduced in Appendix C (Definition C.3) is a rigorous definition of the usual power-counting degree. It allows a simple proof of normalizability which contains the abelian case (QED) as a by-product.

Normalizability is equivalent to the fact that the singular order of a distribution with fixed external field operators is independent of the order $n$ of perturbation theory.

The singular order $\omega$ depends on the external field operators only so that there are only finitely many cases with nonnegative $\omega$, i.e. with free normalization terms. Therefore, the following theorem establishes the normalizability of the Yang-Mills theory.

\vskip 0.5cm
{\bf Theorem 4.1.}  In the $SU(N)$-theory the singular order $\omega$ of a distribution with $b$ external gluon, $g_u$ external ghost operators, $g_{\tilde{u}}$ anti-ghost operators $\tilde{u}$, $d$ derivatives on these external operators and f quark or anti-quark pairs, is given by the following simple expression:
$$\omega \le 4-b-g_u-g_{\tilde{u}} -d-3f \eqno (4.1)$$
This expression is obviously independent of the order $n$ of perturbation theory.
\vskip0,3cm
{\bf Proof:} In the inductive construction of $T_n$ from the $T_m, m \le n -1,$ one must consider tensor products of two distributions
$$T_r^1(x_1, \ldots x_r)T_v^2(y_1, \ldots y_v) \eqno (4.2)$$
with singular orders, say, $\omega_1$ and $\omega_2$ which fulfill (4.1) the induction hypothesis. This product is normally ordered.

$\bullet$ {\bf Case 1:} We assume that $l$ contractions between gluons or ghosts arise in this process. Then, taking translation invariance into account, the numerical part of the contracted expression is of the form
$$t_1(x_1 - x_r, \ldots x_{r-1} - x_r) \prod_{j=1}^l \partial^{a_j} D_0^+ (x_{r_j} - y_{v_j})t_2(y_1 - y_v, \ldots y_{v-1} - y_v) \eqno (4.3)$$
$$\=d t(\xi_1, \ldots \xi_{r-1}, \eta_1, \ldots, \eta_{v-1}, \eta), $$
where $a_j \in N_0^4, \mid a_j \mid = 0, 1, 2.$ Here, $\{x_{r_j}\}$, is a subset of $\{x_1, \ldots x_r\}$ and $\{y_{v_j}\}$ is a subset of $\{y_1, \ldots y_v\}$. We have introduced relative coordinates
$$\xi = x_j-x_r, \quad \eta_j = y_j-y_v, \quad \eta=x_r-y_v.$$
The contraction function is given by
$$D_0^+(x)=\frac{i}{(2\pi)^3} \int d^4p\delta^{(1)}(p^2)\Theta(p^0)e^{-ipx}, \eqno (4.4)$$
where $\delta^{(n)}$ denotes the n-dimensional $\delta$-distribution. We compute the Fourier transform (omitting powers of $i$ and $2 \pi$)
$$\hat{t}(p_1,\ldots p_{r-1},q_1,\ldots q_{v-1},q)=\int t(\xi,\eta)e^{ip\xi + iq\eta}d^{4_r-4} \xi d^{4v} \eta. \eqno (4.5)$$
Since products go over into convolutions, we get
$$\hat{t}(\ldots)=\int \prod^l_{j'=1}d\kappa_j,\delta^{(4)}\Big{(}q-\sum^l_{h=1} \kappa_h \Big{)}$$
$$x \hat{t}_1(\ldots p_{r_i}-\kappa_i \ldots ) \prod^l_{j=1} \kappa_j^{a_j} D_0^+(\kappa_j)\hat{t}_2(\ldots q_{v_i}+\kappa_i \ldots). \eqno (4.6)$$
Here, $\kappa_i$ does not appear in the argument of $\hat{t}_1$ (rsp. $\hat{t}_2$) if $r_i=r$ (rsp. if $v_i=v$). If e.g. $k=r_i=r_s$ we have $\hat{t}_1(\ldots p_k-\kappa_i-\kappa_s \ldots)$. Applying $\hat{t}$ to a test function $\varphi \in \cal S( \it \bf R^{\it 4 \it (r+v-1)})$, we obviously have
$$\langle \hat{t},\varphi \rangle = \int d^{4r-4}p' d^{4v-4} q'\hat{t}_1(p')t_2 (q')\psi (p',q'),$$
with
$$\psi=\int d^4q \prod_{j'} d\kappa_{j'}\delta^{(4)}\Big{(} q-\sum_h \kappa_h \Big{)} \varphi (\ldots p'_{r_i} + \kappa_i \ldots , \ldots q'_{v_i} -\kappa_i \ldots q)$$
$$\times \prod^l_{j=1} \kappa_j^{a_j} \hat{D}_0^+(\kappa_j). \eqno (4.7)$$
In order to determine the singular order of $\hat{t}$ in $p$-space, we have to consider the scaled distribution
$$\Big{\langle} \hat{t} \Big{(} \frac{p}{\delta}\Big{)},\varphi \Big{\rangle} = \delta^m \langle \hat{t}(p),\varphi(\delta p)\rangle= \delta^m \int d^{4r-4}p' d^{4v-4}q' \hat{t}_1(p')\hat{t}_2(q') \psi_\delta (p', q') \eqno (4.8)$$
where
$$\psi_\delta(p',q') = \int d^4q \prod_{j'} d \kappa_{j'} \delta^{(4)} \big{(} q- \sum_h \kappa_h \big{)}$$
$$\varphi(\ldots \delta (p'_{r_i} + \kappa_i) \ldots, \ldots \delta (q'_{v_i}-\kappa_i) \ldots \delta q) \prod_j \kappa_j^{a_j} \hat{D}_0^+ (\kappa_j) \eqno (4.9)$$
and $m = 4(r + v - 1).$ We introduce scaled variables $\tilde{\kappa}_j = \delta_{\kappa j}, \tilde{q} = \delta q$ and note that
$$\hat{D}_0^+ \Big{(} \frac{\tilde{\kappa}}{\delta} \Big{)} = \delta^{(1)} \Big{(} \frac{\tilde{\kappa}^2}{\delta^2} \Big{)} \Theta \Big{(} \frac{\tilde{\kappa}^0}{\delta} \Big{)} = \delta^2 D_0^+(\tilde{\kappa}), \quad \sum_{j=1}^{l} \mid a_j \mid = a \eqno (4.10)$$
This implies
$$\psi_\delta (p',q') = \frac{\delta^{2l-a}}{\delta^{4l}} \int d^4\tilde{q} \prod_{j'} d\tilde{\kappa}_{j'} \quad \delta^{(4)} \big{(} \tilde{q} - \sum_h \tilde{\kappa}_h \big {)}$$
$$\times \varphi(\ldots \delta p'_{r_i} + \tilde{\kappa}_i \ldots, \ldots \delta q'_{v_i} - \tilde{\kappa}_i \ldots, \tilde{q}) \prod_j \tilde{\kappa}_j^{a_j} D_0^+ (\tilde{\kappa}_j) \eqno (4.11)$$
$$ = \frac{1}{\delta^{2l+a}} \psi (\delta p', \delta q').$$
Using again scaled variables $\delta p' = \tilde{p}, \delta q' = \tilde{q}$, we find
$$\Big{\langle} \hat{t} \Big{(} \frac{p}{\delta}\Big{)},\varphi \Big{\rangle} = \frac{\delta^4}{\delta^{2l+a}} \int d^{4r-4} \tilde{p} d^{4v-4} \tilde{q} t_1 \Big{(} \frac{\tilde{p}}{\delta} \Big{)} t_2 \Big{(} \frac{\tilde{q}}{\delta} \Big{)} \psi (\tilde{p}, \tilde{q}). \eqno (4.12)$$
According to the induction hypothesis, $\hat{t}_1$ and $\hat{t}_2$ have singular orders $\omega_1, \omega_2$ with power counting functions $\rho_1 (\delta), \rho_2(\delta)$ , respectively. Then the following limit exists:
$$\lim_{\delta\rightarrow 0} \delta^{2l+a-4} \rho_1 (\delta) \rho_2 (\delta) \Big{\langle} \hat{t} \Big{(} \frac{p}{\delta}\Big{)},\varphi \Big{\rangle} = \langle \hat{t}_0(p), \varphi \rangle. \eqno (4.13)$$
Hence, the singular order of $\hat{t}(p)$ is
$$\omega \le \omega_1 + \omega_2 + 2l + a -4 \eqno (4.14)$$
It remains to check that this result satisfies (4.1) Substituting
$$\omega_j \le 4 - b_j - g_{uj} - g_{\tilde{u} j} - d_j-f_j,\quad j = 1,2 , \eqno (4.15)$$
we find
$$\omega \le 4 - (b_1 + b_2 - 2l_b) - (g_{u1} + g_{u2} - l_g) - (g_{\tilde{u}1} + g_{\tilde{u}2}- l_g) - (d_1 + d_2 - a) -3(f_1+f_2)$$
where $l_b$ (resp. $l_g$) is the number of gluon-contractions (ghost-contractions) and the $l = l_b + l_g$ contractions have totally $a$ derivatives. Since the first bracket is just the number of gluon operators after the $l_b$ gluon-contractions and the second bracket is the number of ghost operators $u$ after the $l_g$ ghost-contrations etc., (4.1.) is proven for the $r', a'$ and $d$ distributions in this case.

$\bullet$ {\bf Case 2:}  Now we assume that $l_f=l$ contractions between quark and anti-quark operators arise in (4.2). The proof is identical as in case 1 with a few modifications: In (4.3) we have to substitute 
$$\prod_{j=1}^l \d^{aj} D_0^+ \longrightarrow \prod_{j=1}^l \hat S^{(+)} \eqno (4.17)$$
The contraction function $S^+(x)=(i \gamma_\mu \d^\mu+m) D_m^+(x)$  (see (4.4)) has $\omega=-1$. Thus, we get a factor $\delta^{3l}$ instead of $\delta^{2l+a}$ in (4.11). The singular order of $\hat t (p)$ is (see (4.14)) 
$$\omega \le \omega_1+\omega_2+3l-4 \eqno (4.18)$$
Inserting for $\omega_1$ and $\omega_2$ the induction assumption (4.15), we arrive at
$$\omega\le 4-(b_1+b_2)-(g_{u1}+g_{u2}) - (g_{\tilde u 1}+g_{\tilde u 2}) - (d_1+d_2)-3(f_1+f_2-l_f).  \eqno (4.19)$$
This is (4.1) after $l_f$ quark contractions. Thus (4.1) is also proven for the $r',a'$ and $d$ distributions in case 2.\\
The final step of the inductive construction is the splitting of the causal distribution into a retarded and advanced part. In this process the singular order is not changed (see (C.32)). Hence, (4.1.) is true in general. This implies that there are only finitely many cases with nonnegative $\omega$ normalization. This implies the normalizability of Yang-Mills theories with fermionic matter fields. {\bf q.e.d.}
\vskip0.2cm
$\bullet$ Our analysis of normalizability of the Yang-Mills theory also contains the abelian case (QED) as a by-product. One can take over the proof by setting $g_{\tilde{u}}=g_u=d=0;$ $b$ is then the number of photons, $f$ the number of electron-positron pairs.

\vskip 1cm

{\Large\bf 5. Discrete Symmetries}
\vskip 1cm

In the present chapter we study the discrete symmetry transformations in the causal formalism . For this purpose we have to state unitary or anti-unitary transformations of the field operators in the Fock space so that the defining equations of the theory in the causal formalism are invariant. The equations are:

$\bullet$ the fundamental (anti-)commutations relations of the free field operators,

$\bullet$ their dynamical equations and

$\bullet$ the specific coupling of the theory $T_{n=1}$.

On the basis of the invariance of $T_{n=1}$, we can carry out an inductive proof of the invariance of the n-point distributions $T_n$ which formally implies the invariance of the S-matrix $S(g)$.

The implications of the invariance postulates are decisive for the gauge invariance proof in both the abelian and the nonabelian case (see [26] and Chapter 6). 
In Subchapter (a) we present the (anti-)unitary representations in the abelian theory, in Subchapter (b) we give these representations in the nonabelian theory. From charge conjugation invariance (equivalently also from time reversal invariance) we derive restrictions for the $SU(N)$-tensor structure of the theory ($SU(N)$-Furry theorem). In conclusion we list the consequences of the famous $CPT$-theorem in the causal formulation.
Proofs are only carried out in an exemplary mode, except for the crucial inductive proofs.
\vskip0,5cm
{\bf (a) Abelian Theory}
\vskip0,5cm

$\bullet$ We look for unitary (anti-unitary) representations $U_i$ of the discrete symmetry transformations in the Fock space so that
$$ U_CS(g) U_C^{-1} = S(g) \quad \mbox{(     Invariance under Charge Conjugation)} \eqno (5.1)$$
$$ U_TS(g) U_T^{-1} = S^{-1}(g_T) \quad \mbox{(     Invariance under Time Reversal)} \eqno (5.2)$$
$$U_P S(g) U_P^{-1} = S(g_P) \quad \mbox{(     Invariance under Parity)} \eqno (5.3)$$
with  $g_p(x): = g(\Lambda_P^{-1}x)$, $g_T(x): = g (\Lambda_T^{-1}x)$, $(g \in \cal S)$.
Perturbatively, this means for the operator-valued n-point distribution (Notation: $x_T=\Lambda_T x,$    $x_P=\Lambda_p x)$:
$$ U_CT_n(x)U_C^{-1} = T_n (x), \quad U_TT_n(x)U_T^{-1} = \tilde{T}_n (x_T), \quad U_PT_n(x)U_P^{-1} = T_n (x_P).\eqno (5.4)$$
The following well-known \hyphenation{trans-for-ma-tions}transformations \hyphenation{ful-fill}fulfill the \hyphenation{re-quire-ments}requirements, above all, they leave the defining equations of the theory invariant [1]:

$$U_PA(\varphi)U_P^{-1} = A(\varphi^P), \quad \varphi_\mu^P(x) = \varphi^\mu(\Lambda_p^{-1}x), \quad U_P \quad \mbox{unitary} \eqno (5.5a)$$
$$U_TA(\varphi)U_T^{-1} = A(\varphi^T), \quad \varphi_\mu^T(x) = \varphi^\mu(\Lambda_T^{-1}x), \quad U_T \quad \mbox{anti-unitary}$$
$$U_CA(\varphi)U_C^{-1} = A(\varphi^C), \quad \varphi_\mu^C(x) = -\varphi^\mu (x), \quad U_C \quad \mbox{unitary}$$
$$\mbox{with} \qquad A(\varphi) = \int d^4 x \quad \varphi_\mu(x) A^\mu(x), \quad \varphi_\mu(x) \in \cal S (\bf R^{\it 4}\it)$$
(Note: Generally the integral has a purely symbolic meaning.)
This translates into:
$$U_PA_\mu(x) U_P^{-1} = A^\mu (x_P), \quad U_TA_\mu(x) U_T^{-1} = A^\mu (x_T), \quad U_CA_\mu(x) U_C^{-1} = -A_\mu (x). \eqno (5.5b)$$
We define the transformations in the fermionic sector of the Fock space through
$$U_T \psi(f) U_T^{-1} = \eta_T \psi (f_T), \quad f_T(x) = - \gamma^5 C^{-1} f(\Lambda_T^{-1}x), \quad U_T \quad \mbox{anti-unitary}, \quad \mid \eta_T \mid = 1 \eqno (5.6a)$$
$$U_P \psi(f) U_P^{-1} = \eta_P \psi (f_P), \quad f_P(x) = \gamma_0 f(\Lambda_P^{-1}x), \quad U_P \quad \mbox{unitary}, \quad \mid \eta_P \mid = 1$$
$$U_C \psi(f) U_C^{-1} = \eta_C \psi^+ (f_C); \quad f_C(x) = C\gamma^0 f^*(x), \quad U_C \quad \mbox{unitary}, \quad \mid \eta_C \mid = 1, \quad C= i\gamma^2 \gamma^0$$
$$\mbox{with}\quad \psi(f) = \int d^4 x f^+(x) \psi (x),\quad \psi^+(f) = \int d^4 x \psi^+(x) f(x), \quad f\in \cal S(\bf R^{\it 4}\it)^4$$
\newpage
(We use the Weyl-representation of the Clifford algebra over the four-dimensional (complex) Minkowski space.)

This implies for the \hyphenation{ope-rator-va-lued}operator-valued \hyphenation{dis-tri-bu-tions}distributions $\psi$ and $\bar{\psi}$:
$$U_T \psi(x) U_T^{-1} = \eta_T \gamma^5 C^{-1} \psi (x_T), \quad U_T \bar{\psi}(x) U_T^{-1} = \bar{\eta}_T \bar{\psi} (x_T) \gamma^5 C, \eqno (5.6b)$$
$$U_P \psi(x) U_P^{-1} = \eta_P \gamma_0 \psi (x_P), \quad U_P \bar{\psi}(x) U_P^{-1} = \bar{\eta}_P  \bar{\psi} (x_P) \gamma^0 $$
$$U_C \psi(x) U_C^{-1} = \eta_C ( \bar{\psi} (x) C^T )^T, \quad U_C \bar{\psi}(x) U_C^{-1} = \bar{\eta}_C ( C^{-1} \psi (x) )^T$$\\
\\
{\bf Comment }: One shows that the (anti-)unitary operators in Fock space are uniquely given up to the free observable phases $\eta_i$. Furthermore, one can show that $U_C$ commute with $U_T$ and $U_P$ if and only if $\eta_P$ is purely imaginary and $(\eta_C\eta_T)$ is real ([1],8.128). In fact,  any special choice of the free phases fulfilling 
$$ \eta_P^2= -1, \quad \eta_C^2\eta_T^2=1 \eqno(5.7)$$
leads to the following relations:

$$U_C^2=1,\quad U_T^2=U_P^2=U^v \eqno(5.8a)$$
$$U_C U_P=U_P U_C, \quad U_C U_T=U_T U_C \eqno(5.8b)$$
$$U_T U_P=U^v U_P U_T \eqno(5.8c)$$

with the unitary operator $U_v$ in Fock space called valency operator which is defined 
by $U^v f= - f$ if the number of spinor fields involved in the vector 
$f\in F$ is odd and by $U^v f= f$ if this number is even. On the 
spinor sector of $F$ we have $U^v= U(0,-\mbox{\bf 1})$ with $ -\mbox{\bf 1}\in SL(2,C)$. Note 
that $U^v T_n(x) U^v = T_n(x)$ because the n-point distributions $T_n$ are bilinear
in the spinor fields. In the following we set $\eta_P=i,\eta_C=\eta_T=1$. This choice fulfills (5.7).\\
\\

The transformations leave the constitutive equations of the theory invariant:

$\bullet$ Invariance of the fundamental commutator relations:
The parity transformation of the photon field (5.5) leads to
$$U_P \big{(}[A^\mu (x), A^\nu(y)] \big{)} U_P^{-1}=U_P \big{(}g^{\mu \nu} iD_0(x-y)\big{)} U_p^{-1} = g_{\mu \nu} i D_0 (x_p - y_p) \eqno (5.9)$$
Of particular interest is the case of time reversal. In this case the invariance postulate of the fundamental commutator relations already necessitates the anti-unitary implementation of the symmetry in the Fock space. An unitary implementation leads to a contradiction. For $U_T$ being anti-unitary, however, we have
$$\big{[}A^\mu (x_T), A^\nu (y_T)\big{]}_- = U_T \big{[}A_\mu (x), A_\nu (y) \big{]} U_T^{-1}\eqno (5.10)$$
$$= U_T ig_{\mu \nu} D_0(x-y) U_T^{-1} = (-i) g_{\mu \nu} D_0^* (x - y)= i g^{\mu \nu} D_0 (x_T - y_T) $$
The invariance of $\{\psi(x), \bar{\psi}(y)\} = \frac{1}{i} (i  \gamma_\mu \d^\mu + m) D(x)$ is shown analogously.
\newpage
$\bullet$ In the causal approach we have to show the invariance of the dynamical equations of the free field operators only, which is obvious.
$$\w A_\mu(x) = 0, \quad (i \gamma^\mu \d_\mu - m) \psi (x) = 0\eqno (5.11)$$
$\bullet$ The invariance of the specific coupling under P, C and T can also easily be verified.
$$T_{n=1}=ie A^\mu : \bar{\psi} (x) \gamma_\mu \psi(x):. \eqno (5.12)$$
Now we can prove the invariance of the n-point distributions $T_n$ inductively (5.4).\\
 This formally leads to the respective invariance of the S-matrix $S(g)$ (5.1) - (5.3). Note that in perturbation theory the S-matrix is constructed as a formal power series.\vskip 0,5cm
{\bf Proof of C-Invariance}
\vskip 0,3cm

$T_1$ and also $\tilde{T}_1 = -T_1$ are obviously C-invariant (5.12) Now all m-point distributions are assumed to be invariant with $m < n$:
$$U_C T_m (x) U_C^{-1} = T_m (x), \quad U_C \tilde{T}_m (x) U_C^{-1} = \tilde{T}_m (x) \quad m < n \eqno (5.13)$$
Thus we arrive at the invariance of all tensor products of these distributions including $R'_n, A'_n$ and $D_n$:
$$U_C A'_n(x) U_C^{-1} = U_C \sum_{P_2} \tilde{T}_{n_1} (x) T_{n-n_1} (Y, x_n) U_C^{-1} = A'_n (x) \eqno (5.14)$$
$$U_C R'_n(x) U_C^{-1} = U_C \sum_{P_2} T_{n-n_1} (Y, x_n) \tilde{T}(x) U_C^{-1} = R'_n (x)$$
$$U_C D_n(x) U_C^{-1} = U_C (R'_n(x) -A'_n(x)) U_C^{-1} = D_n(x)$$
Now let $R_n$ be a given splitting solution of $D_n$. Then $U_CR_nU_C^{-1}$ is also one.
$$\tilde{R}_n: = \frac{1}{2} [R_n + U_C R_n U_C^{-1}] \eqno (5.15)$$
is obviously an invariant splitting solution of $D_n$, $U_C\tilde{R}_n U_C^{-1} = R_n,$ because of (5.8a). From this follows that $T'=\tilde{R} - R'$ is invariant and one can easily see that this property survives the final symmetrization of $T'$:
$$T_n (x_1, \ldots, x_n) = \sum_{\pi} \frac{1}{n!} \quad T'_n (x_{\pi_1}, \ldots, x_{\pi_n})\eqno (5.16)$$
$\tilde{T}_n$ can be described as the sum of tensor products of $T_n's$ shown to be invariant and is thus also invariant. {\bf q.e.d}

\vskip 0,5cm
{\bf Proof of P-Invariance}
\vskip0,3cm
The statement is obviously valid for $n=1$ (5.12).
If
$$U_P T_m (x) U_P^{-1} = T_m (x_P) \quad \wedge \quad U_P \tilde{T}_m(x) U_P^{-1} = \tilde{T}_m (x_P) \quad \mbox{for} \quad m < n \eqno (5.17)$$
then we have
$$U_PD_n(x) U_P^{-1} = D_n (x_P).$$
By symmetrization of any given splitting solution $R_n$ we arrive at an invariant splitting solution:
$$\tilde{R}_n (x):= \frac{1}{2}[ R_n (x) + U_PR_n(x_p) U_P^{-1}]$$.
$$U_P\tilde{R_n}(x) U_P^{-1}=U_P \frac{1}{2} \big{[}R_n (x) + U_PR_n(x_p) U_P^{-1}\big{]} U_P^{-1}= \tilde{R}_n (x_p) \eqno (5.18)$$
due to (5.8a).
With $T'_n = \tilde{R}_n - R'_n$ the symmetrized n-point distributions $T_n$ and $\tilde{T}_n$ are again also invariant. {\bf q.e.d.}

\vskip0,3cm
It is obvious that the proof is completely analogous to the proof of C-invariance. The inner P-transformation $x \longmapsto x_p$ also leaves the time coordinate unchanged which is crucial for the support properties of the distribution. We now move on to the inductive proof of the T-Invariance

\vskip 0,5cm
{\bf Proof of T-Invariance}
\vskip0,3cm
We first go back to the result of pseudo-unitarity of the S-matrix.
We have
$$T_n^k = \tilde{T_n}\quad\mbox{or}\quad (R_n^k = -R_n) \quad \forall n \quad (\Longleftrightarrow S^{-1}(g) = S^k(g)),\eqno (5.19)$$
with $k$ signifying the conjugation with regard to the sesquilinear form in Fock space

$< \cdot \mid \eta \hspace{0,02cm}\cdot >: F \times F \longrightarrow C \rm$
with the metric tensor $\eta = (-1)^{N_{A_0}}.\quad N_{A_0}$ is the number operator of the time-like photons and
$$< \cdot \mid \cdot >: F \times F \longmapsto C^+$$
signifies the (positive definite) scalar product with the usual adjunction + in the Fock-Hilbert space.
We have $A_\mu^k = A_\mu$. As is well-known, the unitarity
$$T_n^+ = \tilde{T}_n \quad \forall n\quad \big{(}\Longleftrightarrow S^{-1}(g) = S^+(g)\big{)}\eqno (5.20)$$
is only valid in the physical subspace of transversal photons.

Now we can show the T-invariance (5.4) (see [30]). This statement is valid for $n=1$ (5.12). We assume that
$$U_T T_m(x) U_T^{-1} = \tilde{T}_m(x_T)= T_m^k(x_T),\quad m < n \eqno (5.21)$$
This leads to
$$U_T R'_n (x_1, \ldots, x_n) U_T^{-1} = \sum_{P_2} T_{n-n_1}^k (\Lambda_T Y, \Lambda_T x_n) T_{n_1} (\Lambda_T X)= A'^k_n (\Lambda_T x_1, \ldots, \Lambda_T x_n) \eqno (5.22)$$
Analogously we have
$$U_T(A'_n (x_1, \ldots, x_n)) U_T^{-1} = R'^k_n (\Lambda_T x_1, \ldots, \Lambda_T x_n) \eqno (5.23)$$
Then we arrive at
$$U_T(D_n (x_1, \ldots, x_n)) U_T^{-1} = -D^k_n (\Lambda_T x_1, \ldots, \Lambda_T x_n)\eqno (5.24)$$
Let $R_n - A_n = D_n$ be a decomposition of $D_n$ after retarded or advanced support with regard to the variable $x_n$:
$$U_T D_n(x_1, \ldots, x_n) U^{-1}_T=\underbrace{U_T R_n(x_1, \ldots, x_n) U_T^{-1}}_{c\Gamma^+(x_n)} - \underbrace{U_T(A_n(x_1, \ldots, x_n)U_T^{-1}}_{c\Gamma^-(x_n)}= \eqno (5.25)$$
$$= -D_n^k (\Lambda_T x_1, \ldots, \Lambda_T x_n) =\underbrace{-R_n^k(\Lambda_T x_1, \ldots, \Lambda_T x_n)}_{c\Gamma^-(x_n)} + \underbrace{A_n^k(\Lambda_T x_1, \ldots, \Lambda_T x_n)}_{c\Gamma^+(x_n)}$$
$R_n^k(x_T)$ has retarded support with regard to $x_n,$  $ A_n^k(x_T)$ advanced support respectively $x_n$. Note that the conjugation $k$ does not affect the support properties. We can carry out the following identifications up to local terms:
$$U_TR_n(x)U_T^{-1} = A_n^k(x_T) + \mbox{local terms},\quad U_T A_n(x)U_T^{-1} = R_n^k(x_T) + \mbox{local terms}\eqno (5.26)$$
Taking (5.8a) into account, the symmetrized splitting solution
$$\tilde{R}_n(x): = \frac{1}{2} \big{[}R_n(x) + U_TA_n^k (x_T) U_T^{-1} \big{]},\quad \tilde{A}_n(x): = \frac{1}{2} \big{[}A_n(x) + U_TR_n^k (x_T) U_T^{-1} \big{]} \eqno (5.27)$$
$$\mbox{fulfills}\hspace{2cm}  U_T\tilde{R}_n(x) U_T^{-1} = \tilde{A}_n^k (x_T), \quad U_T\tilde{A}_n(x) U_T^{-1} = \tilde{R}_n^k (x_T). \eqno (5.28)$$
With (5.22) and (5.23) follows
$$U_TT_n(x) U_T^{-1} = U_T\big{(}\tilde{R}_n(x) - R'_n(x)\big{)} U_T^{-1}=\tilde{A}_n^k (x_T) - A'^k_n (x_T) = T_n^k (x_T) = \tilde{T}_n (x_T). \eqno (5.29)$$

{\bf q.e.d.}
\vskip0,3cm
$\bullet$ The invariance postulates under discrete symmetry transformations imply additional conditions on the local distributions not yet fixed in the splitting process by causality and translational invariance. These conditions are decisive for the proof of abelian gauge invariance (see[26]).\\
$\bullet$ We now have to settle the question of the compatibility of the individual invariance conditions. But one can see at once that the crucial symmetrizations (5.15), (5.18), (5.27) can be carried out consecutively without destroying the symmetry previously reached because of (5.8 b,c).

With this statement we finish the presentation of the abelian case and move on to the nonabelian case where we only have to deal with the most essential points. The inductive proofs are to be taken over without any changes.

\vskip 0.5cm
{\bf (b) Nonabelian Case}
\vskip0,3cm
Again we look for (anti-)unitary representations of the discrete symmetry transformations in Fock space which leave the fundamental equations invariant (see (5.1), (5.2), (5.3))
\vskip 0,5cm
{\bf The Solution for Parity:}
\vskip0,3cm
The following unitary transformations in Fock space fulfill the requirements:
$$U_PA_\mu^a(x) U_p^{-1} = A_a^\mu (x_p), \quad U_Pu_a(x) U_p^{-1} = u_a (x_p),\quad U_P\tilde{u}_a(x) U_p^{-1} = \tilde{u}_a (x_p) \eqno (5.30)$$
We take the transformations of the fermionic matter field operators directly from the abelian case (5.6).
$$U_p \psi_\alpha (x) U_p^{-1} = i \gamma^0\psi_\alpha(x_p),\quad U_p \bar{\psi}_\alpha (x) U_p^{-1} = (-i) \bar{\psi}_\alpha(x_p) \gamma^0 \qquad (\eta_p=i)\eqno (5.31)$$
The P-invariance of the fundamental (anti-)commutation relations (2.3),(2.4), the dynamical equations (2.5) and the specific coupling $T_1$ (2.2) can be verified without any difficulties.

From $U_PT_1(x) U_P^{-1} = T_1(x_p)$ one again proves via induction:
$$U_PT_n(x) U_P^{-1} = T_n(x_p) \qquad \forall n \quad\big{(} \Longleftrightarrow U_pS(g) U_p^{-1} = S(g_p)\big{)} \eqno (5.32)$$
The proof of the abelian case can be taken over without any changes.
\vskip0.5cm
{\bf The Solution for Time Reversal:}
\vskip0,3cm
Preliminaries:

1) As in the abelian case, the invariance postulate of the fundamental (anti-)commutation relations necessitates the anti-unitary implementation of this symmetry in the Fock space.

2) (Again) one can introduce a sesquilinear form in the Fock-Hilbert space in addition to the canonical positive definite scalar product. We describe the corresponding conjugation through its action on the field operators. (We give a precise definition in Chapter 7.)
$$A_\mu^k = A_\mu, \quad \psi^+ = \psi^k, \quad u_a^k = u_a, \quad \tilde{u}_a^k = \tilde{u}_a(-1) \eqno (5.33)$$
$$\mbox{Then we have}\qquad T_1^k = \tilde{T}_1.$$
In Chapter 7 b we prove the so-called pseudo-unitarity via induction:
$$T_n^k(x) = \tilde{T}_n(x) \qquad \forall n \qquad \big{(} \Longleftrightarrow S^k(g) = S^{-1}(g) \big{)} \eqno (5.34)$$
Now we are prepared to construct the anti-unitary representation of the time reversal transformation in the Fock space. 
The transformation has to leave the specific coupling $T_1$ invariant

($T_1 = T_1^A+T_1^v+T_1^{matter}),(2.2))$ :
$$U_TT_1(x) U_r^{-1} = \tilde{T} (x_T) = T_1^k(x_T) \eqno (5.35)$$
First we look at the gauge boson-matter vertex $T_1^{matter}= :i\frac{g}{2} j_\mu^a A_a^\mu$ with $j_\mu^a =: \bar{\psi}_\alpha \lambda_{\alpha \beta}^a \gamma^\mu \psi_\beta:.$

We draw on the T-transformations of the spinors from the abelian case (5.6):
$$U_T\psi_\alpha (x) U_T^{-1} = \gamma^5C^{-1} \psi(x_T),\quad U_T \bar{\psi}_\alpha(x) U_T^{-1} = \bar{\psi}(x_T) \gamma^5C \quad (\eta_T = 1) \eqno (5.36)$$
Because $U_T$ is anti-unitary it follows  
$$U_T j_\mu^a(x) U_T^{-1} = -U_{ab} j_b^\mu (x_T) \quad \mbox{with} \quad \lambda'_a := U_{ab} \lambda_b = -\lambda_a^*. \eqno (5.37)$$
(5.37) and the postulate (5.35) lead to the required transformation of the gluon field operator
$$U_TA_\mu^a(x) U_T^{-1} = -U^{ab} A_b^\mu(x_T) \eqno (5.38)$$
From this follows for the gauge boson vertex
$$U_TT_1^A(x) U_T^{-1} = (T_1^A)^k(x_T) \eqno (5.39)$$
because of $U_{aa'}U_{bb'}U_{cc'}f_{a'b'c'} = f_{abc}$ (see Appendix A). We also set
$$U_Tu_a(x) U_T^{-1} = -U_{ab} u^b(x_T),\quad U_T\tilde{u}_a(x) U_T^{-1} = -U_{ab} \tilde{u}^b(x_T) \eqno (5.40)$$
then we get
$$U_TT_1^v(x) U_T^{-1} = (T_1^v)^k (x_T) \eqno (5.41)$$
and arrive at the required result (5.28).
Via induction this leads to 
$$U_TT_n(x) U_T^{-1} = T_n^k(x_T) = \tilde{T}_n(x_T) \quad \forall n \quad \big{(} \Longleftrightarrow U_TS(g)U_T^{-1} = S^{-1}(g_T) = S^k(g_T)) \eqno (5.42)$$
Once again the corresponding inductive proof of the abelian case is to be taken over without any changes.
\vskip0,5cm
{\bf The Solution for Charge Conjugation}
\vskip0,2cm
We define the unitary transformations
$$U_cA_a^\mu(x) U_c^{-1} = U_{ab} A_b^\mu(x),\quad U_cu_a(x) U_c^{-1} = U_{ab} u_b(x) \quad U_c\tilde{u}_a(x) U_c^{-1} = U_{ab} \tilde{u}_b(x) \eqno (5.43)$$
with $\lambda'_a = U_{ab} \lambda_b = -\lambda_a^*$. Because of $f' =f$ (see Appendix A), we have
$$U_c(T_1^A + T_1^v) U_c^{-1} =T_1^A + T_1^v \eqno (5.44)$$
The postulate
$$U_cT_1^{matter} (x) U_c^{-1} = T_1^{matter}(x) \eqno (5.45)$$
leads with (5.43) to the following required transformation behaviour for the matter current:
$$U_cj_a^\mu U_c^{-1} = U_{ab}j_b^\mu \eqno (5.46)$$
The C-transformations of the spinors introduced in the abelian theory (5.8)
$$U_C \psi_\alpha(x) U_C^{-1} = C \bar{\psi}_\alpha ^T (x) ,\quad U_C \tilde{\psi}_\alpha^T(x) U_C^{-1} = ( C^{-1} \psi_\alpha (x) )^T  \quad (\eta_c = 1) \eqno (5.47)$$
fulfill the requirement (5.46 ) (since $(-C)(\gamma^\mu)^T C^{-1} = \gamma^\mu$ in the Weyl-representation): 
$$U_cj_a^\mu U_c^{-1}=U_c(:\bar{\psi}_\alpha^i \lambda_{\alpha \beta}^a \gamma_{ij}^\mu \psi_\beta^j:) U_C^{-1}=:(C^{-1} \psi_\alpha)^i \lambda_{\alpha \beta}^a \gamma_{ij}^\mu \big{(}\bar{\psi}_\beta (-C)\big{)}^j: $$
$$=:\big{(}\bar{\psi}_\beta (-C)\big{)}^j (\lambda^{aT})_{\beta \alpha} (\gamma^{\mu T})_{ji} (C^{-1}\psi_\alpha)^i :(-1)=:\bar{\psi}_\beta^j \lambda_{\beta \alpha}^b \gamma_{ji}^\mu \psi_\alpha^i : U_{ab}= U_{ab} j_b^\mu$$

The unitary transformation $U_C$ thus leaves the specific coupling of the theory $T_1$ invariant as well as the fundamental (anti-)commutation relations and the dynamical equations, as one can easily verify. Via the same inductive proof given in the abelian case we get
$$U_CT_nU_C^{-1} = T_n \quad \forall n \quad \big{(} \Longleftrightarrow U_CS(g) U_C^{-1} = S(g) \big{)}\eqno (5.48)$$

{\bf Comments:}

$\bullet$ The invariance postulates under discrete symmetry transformations yield conditions for the free local normalization terms. Their compatibility again results from the explicit constructions in the inductive proofs (see abelian case).

$\bullet$ Moreover, from the C-invariance (5.48), we can deduce general statements about the tensor structure of the n-point distributions with regard to $SU(N)$. 
We normal order $T_n$ and make a covariant expansion of $T_n$ with regard to $SU(N)$:
$$T_n = \sum_{k,a} C_a^{SU(N)} t_k : \Omega_a^k:\eqno (5.49)$$
where $C_a^{SU(N)}$ are the numerically invariants of $SU(N)$ (see Appendix A) and $t_k:\Omega_a^k:$ the usual separation between the normal ordered operator and the numerical distribution $t_k$.
Because of (5.48), we can exclude several terms in the sum.
The invariance (5.48) has to be valid for every independent covariant structure $C_a^{SU(N)}$ and for every independent operator part $\Omega_a^k$ separately:

For two-leg-distributions there exists only one independent numerically invariant $"\delta_{ab}"$.
Because of 
$$\delta' = \delta \quad [U_{aa'}U_{bb'} \delta_{a'b'} = \delta_{ab}],\eqno (5.50)$$
there is no restriction in this case.

For three-leg distributions there are already two independent numerically invariants, namely $d_{abc}$ and $f_{abc}$.
We have
$$d' = -d\quad [U_{aa'} U_{bb'} U_{cc'} d_{a'b'c'} = -d_{abc}],\quad f' = f \quad [U_{aa'} U_{bb'} U_{cc'} f_{a'b'c'} = f_{abc}]\eqno (5.51)$$
from which follows (with 5.48) that we can exclude distributions with $SU(N)$-structure $d_{abc}$, while the terms with tensor-structure $f_{abc}$ are compatible with (5.48).
In general, every numerically invariant $SU(N)$-tensor can be represented as a product of the fundamental tensors $\delta, f$ and $d$.
$$C = \prod_{i=1}^{n_1} \delta^{(i)} \prod_{j=1}^{n_2} f^{(j)} \prod_{l=1}^{n_3} d^{(l)} \eqno (5.52)$$
Because of $f'=f, d'=-d, \delta' = \delta$, we can infer the following general statement (For a complete proof see [31].):
\vskip 0,5cm
{\bf SU(N)-Furry theorem}\hspace{1,5cm} In a theory (without $\gamma_5$) which is invariant under charge conjugation  - $U_CT_nU_C^{-1} = T_n,\quad \forall n$ - all distributions with an odd number of d-tensors ($n_3$ odd in (5.52)) in the covariant expansion with regard to $SU(N)$ disappear.
\vskip0,3cm
$\bullet$ In conclusion, we take down the anti-unitary CPT-symmetry transformations:
$$U_{CPT} A_\mu^a(x) U_{CPT}^{-1} = -A_\mu^a (-x),\quad U_{CPT} u_a(x) U_{CPT}^{-1} = -u_a(-x),\quad U_{CPT} \tilde{u}_a(x) U_{CPT}^{-1} = -\tilde{u}_a (-x)$$
$$U_{CPT} \psi_\alpha(x) U_{CPT}^{-1} = (-i) ( \bar{\psi}_\alpha (-x) \gamma^0 \gamma^5 )^T,\quad U_{CPT} \bar{\psi}_\alpha (x) U_{CPT}^{-1} = ( \gamma^5 \gamma_0 \psi_\alpha (-x) )^T (-i)\eqno (5.53)$$
From (5.48) (5.35) (5.32) we get
$$U_{CPT}T_n(x) U_{CPT}^{-1} = \tilde{T}_n(-x),\quad \big{(} \Longleftrightarrow S(g) = S^{-1} (g^{CPT}),\qquad g^{CPT}(x) = g(-x)\quad \big{)} \eqno (5.54)$$
According to the famous CPT-theorem (see [1]), a local quantum field theory, which is $L_+^\uparrow$-invariant and whose specific coupling $T_1$ is pseudo-hermitian, is automatically CPT-invariant. With the help of the Klein transformation, the CPT theorem can also be established for theories with anomalous commutation relations. In the causal approach, the CPT-theorem yields the following trivial conclusions:

$\bullet$ The CPT-invariance postulate does not result in a condition for the free local normalization terms.

$\bullet$ The conditions which are derivable for instance from the C-invariance postulate also have to result from the PT-invariance postulates.

\vskip1cm

{\Large\bf 6. Proof of Nonabelian C-Number Gauge Invariance}
\vskip1.0cm
The proof is carried out in two steps: We express the operator gauge invariance (2.12) by a set of identities between C-number distributions analogous to the Slavnov-Taylor identities. The latter are usually written less explicitly as operator identities and involve interacting fields and not the asymptotic free fields. In order to avoid missunderstandings, we shall use the notation "C-number identities for gauge invariance" or "Cg-identities". We see that the set of Cg-identities explicitly derived in Appendix B, are sufficient for the operator gauge invariance. In order to prove these Cg-identities, the inductive step needs some modifications (Subchapter (a)). We show in great detail that C-number gauge invariance can be preserved in the process of distribution by suitable normalization (Subchapter (b)) [32].
\vskip0.5cm
{\bf (a) General Inductive Step}
\vskip0,5cm
The Cg-identities are based upon the fact that operator gauge invariance
$$[Q,T_n(x_1,\ldots x_n)]=i\sum_{l=1}^n\d_\mu^lT^\mu_{n/l}(
x_1,\ldots x_n)\eqno(6.1)$$
must hold for every combination of normally ordered external field operators separately. However, there is the following subtlety. Consider the $l=1$ term on the r.h.s. of (6.1)
$$\d_\mu^1 T^\mu_{n/1}=(\d_\mu^1 t_{uA}^{\mu 1\nu})u(x_1)A_\nu(x_2) +t_{uA}^{\mu 1\nu}\d_\mu^1u(x_1)A_\nu(x_2)+\ldots\eqno(6.2)$$
If $t_{uA}^{\mu 1\nu}$ contains a contribution with a factor $\delta(x_1-x_3)$, then the terms with different field operators may compensate, due to the identity
$$[u(x_1)-u(x_3)]\d_\mu^1\delta(x_1-x_3)+\delta(x_1-x_3)\d_\mu^1
u(x_1)=0.\eqno(6.3)$$
(Note that the contribution with $u(x_3)$ comes from the $l=3$ term in (6.1).) This gives rise to a certain ambiguity in the definitions of the numerical distributions in (6.1). In order to eliminate this, we choose the convention of {\it only}
applying Wick's theorem to the products $\tilde T_n(X)T_{n-k}(Y,x_n)$ in $A'_n$ (doing nothing else), and similarly for $R'_n$. 

Because of the ambiguities in translating the operator equation (2.12) resp.(6.1) into the Cg-identities, we need some more input for the inductive step. Instead of proving (6.1), we prove the Cg-identities that imply (6.1).

The proof of the Cg-identities is now straightforward. As induction hypothesis we assume the Cg-identities for the $(t,t^l)$-distributions of $(T_k, T_{k/l})$ $(l\le k)$ and for
the $(\tilde t,\tilde t^l)$-distributions of $(\tilde T_k,\tilde T_{k/l})$ $(l\le k)$ in all lower orders $1\le k<n$ of perturbation theory. The induction hypothesis can be expressed in
the following form: Writing $T_k$, $\tilde T_k$, $T_{k/l}$ and $\tilde T_{k/l}$ in normally ordered form, and applying $[Q,\ldots]$, resp. $\sum_l\d_l$, we have 
$$[Q,T_k]=\sum_jt^Q_{kj}:\Omega_j:,\quad [Q,\tilde T_k]=\sum_j\tilde t^Q_{kj}:\Omega_j:,\eqno(6.4)$$
$$\sum_l\d_lT_{k/l}=\sum_j\tau_{kj}:\Omega_j:,\quad\sum_l\d_l\tilde T_{k/l}=\sum_j\tilde\tau_{kj}:\Omega_j:,\eqno(6.5)$$
with the Cg-identities
$$t^Q_{kj}=\tau_{kj},\quad\tilde t^Q_{kj}=\tilde\tau_{kj}\eqno(6.6)$$
being true for all $1\le k<n$ and for all all combinations of field
operators $\Omega_j$.

We insert (6.4) into 
$$[Q,\,A'_n(X,Y;x_n)]=\sum_{X,Y}\{[Q,\,\tilde T_k(X)]T_{n-k}(Y,x_n)
+\tilde T_k(X)[Q,\,T_{n-k}(Y,x_n)]\},\eqno (6.7)$$
and (6.5) into 
$$\sum_l\d_lA'_{n/l}(X,Y;x_n)=\sum_{X,Y}\Bigl(\sum_{l,x_l\in X}\d_l
\tilde T_{k/l}(X)\Bigl)T_{n-k}(Y,x_n)$$
$$+\sum_{X,Y}\tilde T_k(X)\Bigl(\sum_{l,x_l\in\{Y,x_n\}}\d_lT_{n-k/l}
(Y,x_n)\Bigl),\eqno (6.8)$$
Applying Wick's theorem to (6.7) and (6.8), we obtain the identical operator decomposition  for $[Q,A'_n]$ and for $\sum_l\d_lA'_{n/l}$. 
This is correct because the two following operations of $A'_n$ commute: the application of Wick's theorem to $A'_n$ and the differentiation of $A_n$. Thus the operator decomposition of $\sum_l\d_lA'_{n/l}$ is independent of the order of these two operations.

Analogously, the operator decomposition of $[Q,A'_n]$ is the same, if we first apply Wick's theorem and commute $A'_n$ with the charge $Q$ or if we do the two operations in the reversed order.

Because of (6.6) the numerical distributions in $[Q,A'_n]$ and in $\sum_l\d_lA'_{n/l}$ agree. Hence, the Cg-identities hold for $k=n$. In the same way one proves the Cg-identities for the $r'$-distributions of $R'$, $R'_{n/l}$. Therefore, they hold for the $d=r'-a'$ distributions, too. After distribution splitting the Cg-identities hold for the $r$ and, therefore, for the $t'=r-r'$ distributions up to local terms. In the symmetrization of $T'_n$ (resp. $T'_{n/l}$), $t'$ (resp. $t^{\prime l}$) is replaced by $t$ (resp. $t^l$), which is a linear combination of the various $t'$ (resp. $t^{\prime l}$) distributions where the splitting vertex $x_n$ can be an external or internal vertex. Since these different types of $(t',t^{\prime l})$-distributions separately satisfy the Cg-identities up to local terms, it follows that the $(t,t^l)$-distributions may violate the Cg-identities by local terms only.

In order to complete the inductive step, one has to prove the Cg-identities for the $(\tilde t,\tilde t^l)$-distributions corresponding to $(\tilde T_n, \tilde T_{n/l})$. With the
definition 
$$R_n''(x_1,\ldots x_n)=\sum_{X,Y}T_k(X)\tilde T_{n-k}(Y,x_n),\eqno(6.9)$$
$\tilde T_n$ is given by the symmetrization of
$$\tilde T'_n=-R_n-R''_n,\eqno(6.10)$$
and similarly for $R''_{n/l}$, $\tilde T'_{n/l}$ and $\tilde T_{n/l}$. This leads us to the following procedure: We define new retarded distributions by 
$$r^*\=d t+r',\quad r^{\ast l}\=d t^l+r^{\prime l}.\eqno(6.11)$$
Here we can make use of the correctly normalized $t$ resp. $t^l$. The retarded distributions (6.11) are splitting solutions of $(d,d^l)$ and satisfy the Cg-identities because $(t,t^l)$ and $(r',r^{\prime l})$ fulfill them. The Cg-identites for the $(r'',r^{\prime\prime l})$-distributions of $(R''_n,R''_{n/l})$ can be proven in the same way as above for the $a'$-distribution. Following (6.10), we now define 
$$\tilde t'\=d -r^*-r'',\quad\tilde t^{\prime l}\=d -r^{\ast l}-
r^{\prime\prime l},\eqno(6.12)$$
which satisfy the Cg-identities. As above, this property remains true after symmetrization. This finishes the inductive step apart from distribution splitting.

\vskip0.5cm

{\bf (b) Distribution Splitting}
\vskip0.5cm
It remains to prove that the Cg-identities are preserved under the operation of distribution splitting. In Appendix B, we explicitly derived all types of Cg-identities and isolated those few Cg-identities which require nontrivial proofs; these Cg-identities have $\kappa \ge -1$, where $\kappa$ is defined by
$$\kappa = 4 - b - g_u - g_{\tilde u} - d -3f. \eqno (6.13)$$
$b, g_u, g_{\tilde{u}}, d, f$ are determined by the field operator combination $\Omega$ which belongs to the considered Cg-identity as in (4.1).

The idea of the proof is quite straightforward: In the process of distribution splitting, gauge invariance can only be violated by local terms (anomalies).
We show that one can remove these local terms by choosing a suitable normalization. Here all symmetries of the theory, in particular the $SU(N)$-symmetry (Appendix A) and charge conjugation invariance (Chapter 5b), are needed to restrict the possible anomalies.

Our explicit procedure is the following:\\
$\bullet$ First we determine the general form of the anomaly $a$, i.e. the possible violation of the Cg-identity. In momentum space $\hat a$ is a Lorentz covariant polynomial in p of degree $\kappa + 1$ and $SU(N)$-covariant (Appendix A). Moreover, the polynomial is invariant with regard to permutations of the inner vertices and must have the same permutation symmetries as the corresponding external operator $\Omega$ because it is multiplied by the latter.\\
$\bullet$In the next step we further restrict the general ansatz of the anomaly by using additional symmetry properties of the t-distributions in the considered Cg-identity. These additional symmetries are introduced by inserting other Cg-identities as discussed in Appendix B. In this step we draw on the assumption that the divergences with regard to inner vertices vanish in the following limit:
$$ \sum_{l=r+1}^n p_{l \alpha} \hat t_\Omega^{\alpha l \ldots} (p_1, p_2, \ldots p_{n-1}) \rightarrow 0 \eqno (6.14)$$
$$\mbox{for}\qquad (p_1 + p_2 + \ldots + p_r) \rightarrow 0,\quad p_{r+1} \rightarrow 0, \ldots p_{n-1} \rightarrow 0, $$
$$p_n \=d -(p_1 + p_2 + \ldots + p_{n-1}).$$
where we consider r-leg distributions in the n-th order of perturbation theory in momentum space.

In configuration space the limit (6.14) means that the inner coordinates are integrated with $g(x) = 1$ , which corresponds to the adiabatic limit in regard to the inner coordinates. However, since $\hat t_\Omega^{\alpha l}$ is a distribution and not a continuous function, the assumption needs the following specification: We always approach the limit $p_l = 0$ from totally space-like points. In this region $\hat r'_\Omega, \hat a'_\Omega, d_\Omega$ vanish and, consequently, $\hat t_\Omega = \hat r_\Omega$ is analytic (see [13]). Therefore, the limit (6.14) can be understood in the sense of functions.

If the external momenta \hspace{0.7cm}  $p_1, \ldots p_{r-1}, p_r = -(p_1 + \ldots p_{r-1})$, \hspace{0.7cm} approach the mass shell $\quad p_1^2 = 0, l = 1, \ldots r$, infrared divergences appear [16]. In order to avoid them, we always assume $\mid p_l^2\mid \ge \varepsilon > 0, \forall l = 1, \ldots r$. This is sufficient for the proof because we only have to investigate a polynomial, namely the anomaly $\hat a$.

In order to shorten the notation, we simply set $p_1 + p_2 + \ldots + p_r= 0, p_{r+1} = 0, \ldots p_{n-1} = 0$, signifying the limit (6.14) in momentum space.\\
$\bullet$In the last step we remove the thus restricted anomaly $a$ by finite renormalizations of the $t$-distributions in the Cg-identity. These renormalization terms must be covariant local distributions with the same singular order and the same (permutation) symmetry as the corresponding $t$-distributions. If a certain distribution $t$ appears in several Cg-identities, the different normali- zations must be compatible.
\\
\\
Now we prove the few nontrivial Cg-identities, isolated in Appendix B, by this procedure:

\vskip0,5cm
$\bullet$ {\bf Proof of 2-Leg Cg-Identities}
\vskip0,3cm
For the Cg-identities (B.17), (B.18) we define the possible anomalies 
$$a_1^\nu\=d \d_\alpha^1 t_{AA}^{\alpha\nu}
+{1\over 2}\d_\alpha^2[t_{uA}^{\alpha 2\nu}-t_{uA}^{\nu 2\alpha}]
+\sum_{l=3}^n \d_\alpha^l t_{uA}^{\alpha l\nu},\eqno (6.15)$$
$$a_2^{\mu\nu}\=d \d_\alpha^1 t_{AF}^{\alpha\mu\nu}
+{1\over 2}[\d_2^\mu t_{u\tilde u}^\nu-\d_2^\nu t_{u\tilde u}^\mu]
+{1\over 4}[t_{uA}^{\mu 2\nu}-t_{uA}^{\nu 2\mu}]
+\sum_{l=3}^n \d_\alpha^l t_{uF}^{\alpha l\mu\nu}.\eqno (6.16)$$
Taking covariance and the singular order of the distributions
on the right side into account, we conclude that $a_1,a_2$
have the form
$$a_1^\nu(x_1,...,x_n)=[\sum_{i,j,k=1}^n C_{ijk}\d_i^\gamma
\d_{j\gamma}\d_k^\nu+\sum_{k=1}^n D_k\d_k^\nu]\delta
(x_1-x_n,...,x_{n-1}-x_n),\eqno (6.17)$$
$$a_2^{\mu\nu}(x_1,...,x_n)=[\sum_{i,j=1}^n K_{ij}\d_i^\mu
\d_j^\nu+Lg^{\mu\nu}]\delta (x_1-x_n,...,x_{n-1}-x_n).\eqno (6.18)$$
The constants $ C_{ijk},D_k,K_{ij}$ and $L$ are further restricted
by the permutation symmetry of the $t$'s in their inner vertices
and the antisymmetry in $\mu\nu$ of (6.16).
Using additionally 
$$\sum_{k=1}^n \d_k^\nu \delta (x_1-x_n,...,x_{n-1}-x_n)=0,\eqno (6.19)$$
(6.17),(6.18) can be simplified to
$$a_1^\nu(x_1,...,x_n)=[C'_{111}\w_1 \d_1^\nu+C'_{122}\d_1^\gamma
\d_{2\gamma}\d_2^\nu+C'_{221}\w_2 \d_1^\nu+C'_{121}\d_1^\gamma
\d_{2\gamma}\d_1^\nu+$$ 
$$+C'_{112}\w_1 \d_2^\nu+C'_{222}\w_2 \d_2^\nu \quad +C_1\sum_{i=3}^n\w_i\d_1^\nu+C_2\sum_{i=3}^n\w_i\d_2^\nu
+C_3\sum_{i=3}^n\d_1^\gamma\d_{i\gamma}\d_i^\nu+$$
$$+C_4\sum_{i=3}^n\d_2^\gamma\d_{i\gamma}\d_i^\nu+C_5\sum_{i=3}^n\w_i\d_i^\nu+D'_1\d_1^\nu+D\sum_{i=3}^n\d_i^\nu]
\delta (x_1-x_n,...,x_{n-1}-x_n),\eqno (6.20)$$
$$a_2^{\mu\nu}(x_1,...,x_n)=K(\d_1^\mu\d_2^\nu-\d_2^\mu\d_1^\nu)
\delta (x_1-x_n,...,x_{n-1}-x_n).\eqno (6.21)$$
Due to (6.19), these
expressions for the anomalies can be written in different form.
Performing the {\it finite} renormalizations
$$t_{AA}^{\alpha\nu}\rightarrow t_{AA}^{\alpha\nu}-[D'_1 g^{\alpha
\nu}+C'_{122}(\d_2^\alpha\d_2^\nu+\d_1^\alpha\d_1^\nu)
+(C'_{221}-C'_{222})(\w_1+\w_2)g^{\alpha \nu}+$$
$$+C'_{211}\d_2^\alpha\d_1^\nu+
(C'_{112}-C'_{111}+C'_{122}+C'_{221}-C'_{222})\d_1^\alpha\d_2^\nu]
\delta^{(4(n-1))},$$
$$t_{uA}^{\alpha l\nu}\rightarrow t_{uA}^{\alpha l\nu}-[Dg^{\alpha
\nu}-C'_{222}\w_2g^{\alpha \nu}-(C'_{111}-C'_{122}-C'_{221}+C'_{222})\w_1
g^{\alpha \nu}+$$
$$+C_1\d_l^\alpha\d_1^\nu+C_2\d_l^\alpha\d_2^\nu+C_3\d_1^\alpha\d_l^\nu
+C_4\d_2^\alpha\d_l^\nu+C_5\d_l^\alpha\d_l^\nu]\delta^{(4(n-1))},\quad
l\ge 3,$$
$$t_{AF}^{\alpha\mu\nu}\rightarrow t_{AF}^{\alpha\mu\nu}-
K(g^{\alpha\mu}\d_2^\nu-g^{\alpha\nu}\d_2^\mu)\delta^{(4(n-1))},
$$
$$t_{u\tilde u}^\nu\rightarrow t_{u\tilde u}^\nu,$$
$$t_{uF}^{\alpha l\mu\nu}\rightarrow t_{uF}^{\alpha l\mu\nu},\quad l\ge
3, $$
$$t_{uA}^{\alpha 2\nu}\rightarrow {1\over 2}[t_{uA}^{\alpha 2\nu}
-t_{uA}^{\nu 2\alpha}],\eqno (6.22)$$
the anomalies (6.16),(6.15) disappear. We stress that the distributions
with one $Q$-vertex have for $\omega \geq 0$ their own normalization freedom which is, as long as we do not care about gauge invariance, independent of the normalization of the distributions of in the main theory. 

Obviously, the renormalizations (6.22) preserve covariance, the permutation symmetry in the inner vertices and all other symmetry properties of $t$-distributions, in particular $t_{AA}^{\mu \nu} (x_1,x_2,x_3,\ldots)=t_{AA}^{ \nu \mu} (x_2,x_1,x_3,\ldots)$. Note that (6.22) are not the only renormalizations leading to symmetrical, covariant
$t,\,t^l$-distributions which fulfill the Cg-identities (B.17), (B.18). The normalization of the $t$-distributions is not uniquely determined by the Cg-identities.\\

\vskip0,5cm
$\bullet$ {\bf Proof of the 3-Leg Cg-Identities}
\vskip0.5cm
Now we consider the 3-leg Cg-Indentities: There is no possible anomaly $a$ for the Cg-identity (B.22), fulfilling all symmmetry requirements.

We remove the anomaly of (B.21)
$$a^{\mu \nu \lambda} =\sum_{j=1}^3 B_j [ g^{\mu \nu} i \d^\lambda_j - g^{\mu \lambda} i \d_j^\nu ] \delta^{(4(n-1))} \eqno (6.23)$$
by means of the renormalization
$$t_{AAF}^{\alpha \mu \nu \lambda} \rightarrow t_{AAF}^{\alpha \mu \nu \lambda}-i(B_1 - B_3)[g^{\mu \nu} g^{\alpha \lambda} - g^{\alpha \lambda} g^{\alpha \nu}] \delta^{(4(n-1))},$$
$$t_{uAF}^{\alpha 2 \mu \nu \lambda} \rightarrow t_{uAF}^{\alpha 2 \mu \nu \lambda} -i(B_2 - B_3)[g^{\mu \nu} g^{\alpha \lambda} - g^{\mu \lambda} g^{\alpha \nu}] \delta^{(4(n-1))},$$
$$t_{uAF}^{\alpha l \mu \nu \lambda} \rightarrow t_{uAF}^{\alpha l \mu \nu \lambda} +i B_3 [g^{\mu \nu} g^{\alpha \lambda} - g^{\mu \lambda} g^{\alpha \nu}] \delta^{(4(n-1))} , \quad \forall l=4, \ldots n.  \eqno (6.24)$$ 
Note that these renormalization are admissible, i.e. they do not destroy the permutation symmetries of $t_{AAF}, t_{uAF}^2, t_{uAF}^l$, and all other Cg-identities are maintained, if the first renormalization (6.24) is performed in $t_{uAF}^1$, too.
\vskip0,3cm
The possible anomaly of (B.19) must be invariant under the transposition of $x_1$ and $x_2$, hence
$$a^\mu= i[ A_1(\d_1^\mu + \d_2^\mu)+A_2 \d_3^\mu] \delta^{(4(n-1))}.  \eqno (6.25)$$
Performing the renormalizations
$$t_{uu \tilde{u}}^{\alpha 3 \mu} \rightarrow t_{uu\tilde{u}}^{\alpha 3 \mu} -i(A_2 -A_1)g^{\alpha \mu} \delta^{(4(n-1))},$$
$$t_{uu \tilde{u}}^{\alpha l \mu} \rightarrow t_{uu\tilde{u}}^{\alpha l \mu} -iA_1 g^{\alpha \mu} \delta^{(4(n-1))} \eqno (6.26)$$
the anomaly (6.23) vanishes. These renormalizations do not affect any other Cg-identity.
\vskip0,3cm
It remains the Cg-identidy (B.20). For the possible anomaly $a^{\mu \nu}$ we now have
$$a^{\mu\nu}=[K_1(\d_1^\nu \d_3^\mu - \d_1^\mu \d_2^\nu)+K_2(\d_1^\nu \d_2^\mu - p_1^\mu p_3^\nu)$$
$$+K_3(\d_3^\mu \d_3^\nu - \d_2^\nu \d_2^\mu)+K_4 g^{\mu \nu}(\d_1 \d_3 - \d_1 \d_2)+K_5 g^{\mu \nu} (\d_3^2 - \d_2^2)] \delta^{(4(n-1))}, \eqno (6.27)$$
by taking the antisymmetry under simultaneous exchange of $x_2 \leftrightarrow x_3$ and $ \mu \leftrightarrow \nu$ into account. In order to remove (6.27), the following normalization terms of $t_{AAA}, t_{uAA}^3$, and $t_{uAA}^l,(l=4,\ldots n)$ are avaiable 
$$N_{AAA}^{\alpha  \mu \nu}=i[G \big(g^{\mu \nu} (\d_2^\alpha - \d_3^\alpha)+g^{\alpha \nu}(\d_3^\mu - \d_1^\mu)+g^{\alpha \mu}(\d_1^\nu - \d_2^\nu) \big )]\delta^{(4(n-1))} $$
$$N_{AAA}^{\alpha 3 \mu \nu}=i[D_1 (\d_1^\alpha g^{\mu \nu} - \d_1^\nu g^{\mu \alpha}) + D_2 (\d_2^\alpha g^{\mu \nu} - \d_2^\nu g^{\mu \alpha})+ D_3 (\d_3^\alpha g^{\mu \nu} - \d_3^\nu g^{\mu \alpha})]\delta^{(4(n-1))} $$
$$N_{uAA}^{\alpha l \mu \nu}=i[C_1 g^{\mu \nu} (\d_2^\alpha - \d_3^\alpha)+C_2(\d_1^\mu g^{\alpha \nu} - \d_1^\nu g^{\alpha \mu})+C_3(\d_2^\mu g^{\alpha \nu}- \d_3^\nu g^{\alpha \mu})+$$ 
$$+C_4(\d_3^\mu g^{\alpha \nu} - \d_2^\nu g^{\alpha \mu})+C_5(\d_l^\mu g^{\alpha \nu} - \d_l^\nu g^{\alpha \mu}) ] \delta^{(4(n-1))} , \quad l=4, \ldots n.  \eqno (6.28)$$
In order to eliminate the anomaly (6.27), we must find constants $G,D_1,D_2,D_3,C_1,\ldots C_5$, so that
$$\d_{1\alpha} N_{AAA}^{\alpha \mu \nu} - \d_{2 \alpha} N_{uAA}^{\alpha 3 \nu \mu}+ \d_{3\alpha} N_{uAA}^{\alpha 3 \mu \nu} +\sum_{l=4}^n \d_{l \alpha} N_{uAA}^{\alpha l \mu \nu} = -a^{\mu \nu}   \eqno (6.29)$$
holds for arbitrarily given $K_1,\ldots K_5$. (6.29) is equivalent to the linear system
$$K_1=i(G+C_2-C_4-D_1),$$
$$K_2=i(C_2-C_3),$$
$$K_3=i(C_3-C_4-D_3),$$
$$K_4=i(-G+D_1+C_1),$$
$$K_5=i(C_1+D_3), \eqno (6.30)$$
But this system has only a solution, if
$$K_1+K_4-K_2-K_3-K_5=0, \eqno (6.31)$$
that means the anomaly (6.27) must be further restricted.

Supposing (6.31) to be proven, then a solution of (6.30) is given by
$$C_3=-iK_2+C_2, \quad D_1 =iK_4 - C_1$$
$$D_3=iK_5-C_1, \quad C_4=-i(K_2+K_3+K_5)+C_1+C_2,  \eqno (6.32)$$
where $D_2,C_1,C_2$ and $C_5$ can be chosen arbitrarily

In order to prove (6.31), we start from 
$$a^{\mu \nu}(x_1,x_2,x_3,x_4, \ldots)=\d_{1\alpha} t_{AAA}^{\alpha \mu \nu } (x_1,x_2,x_3,x_4, \ldots )+$$ 
$$-\frac{1}{2}\d_{2\alpha}[t_{uAA}^{\alpha 3 \mu \nu}(x_1,x_3,x_2,x_4, \ldots )-(\alpha  \leftrightarrow \mu)]$$
$$+\frac{1}{2}\d_{3 \alpha}[t_{uAA}^{\alpha 3 \mu \nu}(x_1,x_2,x_3,x_4, \ldots) - (\alpha \leftrightarrow \nu)]+ \sum_{l=4}^n \d_{l \alpha} t_{uAA}^{\alpha l \mu \nu}(x_1,x_2,x_3,x_4, \ldots ) $$
$$+g[\delta(x_1-x_2)-\delta(x_1-x_3)]t_{AA}^{\mu \nu}(x_2,x_3,x_4, \ldots)$$
$$-\frac{g}{2}\delta(x_2-x_3) [t_{uA}^{\mu 2 \nu}(x_1,x_2,x_4, \ldots )-(\mu \leftrightarrow \nu)]. \eqno (6.33)$$
Differentiating with  $\d_{2\mu} \d_{3\nu}$ and taking the limit (6.14), we obtain 
$$\d_{2\mu}\d_{3\nu}a^{\mu \nu}(x_1,x_2,x_3) = \d_{1\alpha} \d_{2\mu} \d_{3\nu} t_{AAA}^{\alpha \mu \nu}(x_1,x_2,x_3). \eqno (6.34)$$
Remember that the limit (6.14) means in configuration space that the inner coordinates are integrated with $g(x)=1$.
The right side has an additional symmetry: It is antisymmetric with regard to $x_1 \leftrightarrow x_2$ (and $x_1 \leftrightarrow x_3)$. Inserting (6.27) into the left side, we get
$$-\d_{2\mu}\d_{3\nu}a^{\mu \nu}(x_1,x_2,x_3) = [(K_1+K_4)\d_2 \d_3(\d_1\d_2 - \d_1\d_3)$$
$$+K_2((\d_1\d_2)\d_3^2-(\d_1\d_3)\d_2^2)+(K_3+K_5)\d_2\d_3(\d_2^2 - \d_3^2)]
\delta^{8} \eqno (6.35)$$
Because of (6.34), the right side must be antisymmetrical with regard to $x_1 \leftrightarrow x_2$. It is to verify  that this is equivalent to the condition (6.31). The additional symmetry (6.34) is precisely the information we need in order to complete the proof of the last 3-leg Cg-identity (B.20).
\vskip0,5cm

$\bullet$ {\bf Proof of the 4-Leg Cg-Identities}
\vskip0,5cm
Only the Cg-identities (B 23), (B 24) and (B 25) need a proof.

We start by defining a suitable basis for colour tensors of rank 4 (see Appendix A)
$$ C_{abcd} = f_{ahe} f_{bef} f_{cfg} f_{dgh}$$
$$=\delta_{ab} \delta_{cd} + \delta_{ad} \delta_{bc} + \frac{N}{4} (d_{abr} d_{cdr} - d_{acr} d_{dbr} + d_{adr} d_{bcr}). \eqno (6.36)$$
Note the relation
$$C_{abcd} = C_{dabc}, \quad C_{abcd} = C_{adcb}. \eqno (6.37)$$
There are three linarly independent permutations of $C_{abcd}$, which we call
$$C_{abcd}^1 = C_{abcd}, \quad C_{abcd}^2 = C_{acdb}, \quad C_{abcd}^3 = C_{adbc}. \eqno (6.38)$$

Using
$$f_{abr} f_{cdr} = \frac{2}{N} (C_{abcd}^2 - C_{abdc}^1), \eqno (6.39)$$
it is obvious (by drawing all possible diagrams) that, in 4-th order perturbation theory (n=4), all 4-leg distributions are in the linear span $[C^1, C^2, C^3]$. That means: Additionally to the Furry theorem (see Chapter 5), the colour space is reduced in this case from 6 dimensions for $SU(N)$, $N \ge 4$, or 5 dimensions for $SU(3)$, respectively, to 3 dimensions.

But the colour space for $SU(N)$, $n \ge 6$, is at least five-dimensional.

A symmetrical basis is given by
$$C^1, C^2, C^3, D^1, D^2, \eqno (6.40)$$
where
$$D_{abcd}^1 = d_{abr} d_{cdr} - d_{adr} d_{bcr},\quad D_{abcd}^2 = d_{acr} d_{bdr} - d_{abr} d_{cdr}. \eqno (6.41)$$
In order to cover the different possibilities with a single proof, we proceed in the following way: We shall give the proof for dimension 5 of the colour space $(SU(3), n \ge 6)$, by working with the basis (6.40). There will be a complete separation of the two sectors $[C^1, C^2, C^3]$ and $[D^1, D^2]$. The proof therefore holds in each sector separately. Especially it holds in $[C^1, C^2, C^3]$. This is the proof for $SU(N), N \ge 3$ in order $n = 4$. By setting $d_{abc} = 0$ in $C_{abcd}^1 = C_{abcd}$, i.e. by replacing $C^k$ by $G^k (k = 1,2,3)$,
$$G_{abcd}^1 = \delta_{ab} \delta_{cd} + \delta_{ad} \delta_{bc}, \quad G^2 = C^2 \mid_{d=0}, \quad G_3 = C^3 \mid_{d=0}, \eqno (6.42)$$
we obtain the proof for $SU(2)$ in arbitrary order $n \ge 4$. In fact $G^1, G^2, G^3$ have the same properties under permutations of $a, b, c, d$ as $C^1, C^2, C^3$. Moreover, note that equation (6.39) is essentially not changed in the $SU(2)$ case because
$$\varepsilon_{abr} \varepsilon_{cdr} = G_{abcd}^2 - G_{abcd}^1. \eqno (6.43)$$
There remains the case of dimension 6 $(SU(N)$ with $N \ge 4$ in order $n \ge 6$). Here we define:
$$E_{abcd}^1 = d_{abr} d_{cdr} - d_{acr} d_{bdr} + d_{adr} d_{bcr}$$
$$E_{abcd}^2 = d_{acr} d_{bdr} - d_{adr} d_{bcr} + d_{abr} d_{cdr} \eqno (6.44)$$
$$E_{abcd}^3 = d_{adr} d_{bcr} - d_{abr} d_{cdr} + d_{acr} d_{bdr},$$
i.e. $E^1, E^2, E^3$ are obtained by setting $\delta = 0$ in $C^1, C^2, C^3$ and omitting a factor $N/4$. Therefore, $E^1, E^2, E^3$ have the same behaviour under permutations of $a, b, c,$ as $C^1, C^2, C^3$. It is easy to verify that
$$C^1, C^2, C^3, E^1, E^2, E^3 \eqno (6.45)$$
are linearly independent, therefore, we may take (6.45) as a basis. The 6-dimensional proof is then a doubling of the 3-dimensional proof: The reasoning given for the sector $[C^1, C^2, C^3]$ can be repeated in the sector $[E^1, E^2, E^3]$ in exactly the same way, but with new constants in the ansatz for the anomaly and the normalization polynomials. The only difference is that there are no degenerate terms in $[E^1, E^2, E^3]$, in contrast to the sector $[C^1, C^2, C^3]$.
\vskip0,5cm
The identity (B.24) corresponds to external field operators: $\Omega :=: u^a(1)u^b(2) \partial_\mu \tilde u^c(3) A_\nu^d(4):$. The possible anomaly with the same permutation symmetry (antisymmetric in a, b), is
$$a^{\mu \nu} = g^{\mu \nu} \big [B(\delta_{ac} \delta_{bd} - \delta_{ad} \delta_{bc}) + E(d_{acr} d_{bdr} - d_{adr} d_{bcr}) \big ] \eqno (6.46)$$

with arbitrary constants $B, E.$ Due to the relations
$$\delta_{ac} \delta_{bd} - \delta_{ad} \delta_{bc} = \big(C^2 - C^1 - \frac {N} {2} (D^1 + D^2) \big)_{abcd} \eqno (6.47)$$
$$d_{acr} d_{bdr} - d_{adr} d_{bcr} = (D^1 + D^2)_{abcd} \eqno (6.48)$$
the anomaly (6.46) is an element of the 5-dimensional colour space. The nondegenerate terms in (B.24) have singular order $\omega = -1$ and, therefore, no freedom of normalization. The gauge invariant renormalizations of the 3-leg distributions in the $\delta$-degenerate terms drop out. Consequently, we have no possibility to remove (6.46) by renormalization! Therefore, we must use other properties of the theory.

Let us consider the limit (6.14) of the equation (B.24) = (6.46) in configuration space. Then, $\d_3=-(\d_1+\d_2+\d_4)$ and  $\sum_{l=5}^n \d_{l\alpha} t_{uu \tilde uA}^{\alpha l \mu \nu}$ vanishes, due to our assumption (6.14). Moreover, differentiating the equation (B.24) with $\d_{4 \nu}$, $t_{uu\tilde{u}A}^4$ drops out. It remains, using the Jacobi identity,
$$-\big [t_{1abcd}^\mu (x_1,x_2,x_3,x_4) - \big ( (a,x_1) \leftrightarrow (b, x_2) \big ) \big ] + i \d_3^\mu \d_{4\nu} \overline{t}_{uu\tilde{u}Aabcd}^{3 \nu} (x_1, x_2, x_3,  x_4)+$$
$$+g \big [ f_{adr}f_{bcr} \big\{ i\d_{4\nu} [ t_{Au\tilde{u}}^{\nu\mu} (x_1,x_2,x_3) \delta(x_1-x_4) + t_{Au\tilde{u}}^{\nu\mu}(x_4,x_1,x_3) \delta(x_2-x_3) ]+$$ 
$$- [i \d_{4\nu} t_{Au\tilde{u}}^{\nu\mu}(x_4,x_1,x_3) \delta(x_1-x_2) + i\d_4^\mu \overline{t}_{uu\tilde{u}}^3 (x_1,x_2,x_3) \delta(x_3-x_4) ] \big\}+$$ $$-f_{bdr}f_{acr} \{ x_1 \leftrightarrow x_2 \} \big ]= $$
$$= i\d_4^\mu \big\{ B (\delta_{ac}\delta_{bd} - \delta_{ad}\delta_{bc} ) + E(d_{acr}d_{bdr} - d_{adr}d_{bcr} \big\} \delta^{12},  \eqno (6.49)$$
where
$$t_{1abcd}^\mu (x_1,x_2,x_3,x_4) \=d i\d_{2\alpha} \d_{4\nu} t_{uA\tilde{u}Aabcd}^{\alpha\mu\nu} (x_1,x_2,x_3,x_4) \eqno (6.50)$$
Note the symmetry relation
$$t_{1abcd}^\mu (x_1,x_2,x_3,x_4) = t_{1adcb}^\mu (x_1,x_4,x_3,x_2) \eqno (6.51)$$
Subtracting from (6.49) the equations (6.49) with once $(x_1,a) \leftrightarrow (x_4,d)$ exchanged and once $(x_2,b) \leftrightarrow (x_4,d)$ exchanged, the $t_1$-terms cancel by means of (6.51). Using again the Jacobi identity, we obtain
$$-(\d_1+\d_2+\d_4)^\mu t_{2abcd} (x_1,x_2,x_3,x_4)+$$
$$+ g \big [f_{adr}f_{bcr} \big\{ t_3^\mu (x_1,x_2,x_3) \delta(x_1-x_4) - t_3^\mu (x_1,x_4,x_3) \delta(x_1-x_2) + t_3^\mu (x_1,x_4,x_3) \delta(x_2-x_3) +$$ 
$$- t_3^\mu (x_1,x_2,x_3) \delta(x_3-x_4)- i\d_4^\mu [\overline{t}_{uu\tilde{u}} (x_1,x_2,x_3) \delta(x_3-x_4)] + i\d_2^\mu [\overline{t}_{uu\tilde{u}} (x_1,x_4,x_3) \delta(x_3-x_2)] \big\}+$$ 
$$ \qquad -f_{bdr}f_{acr} \{x_1 \leftrightarrow x_2 \} \big ]$$
$$=i\d_4^\mu \delta^{12} \big\{ B( \delta_{ac}\delta_{bd} - \delta_{ad}\delta_{bc}) + E( d_{acr}d_{bdr} - d_{adr}d_{bcr}) \big\}$$
$$-i\d_1^\mu \delta^{12} \big\{ B( \delta_{dc}\delta_{ba} - \delta_{da}\delta_{bc}) + E( d_{dcr}d_{bar} - d_{dar}d_{bcr}) \big\}$$
$$-i\d_2^\mu \delta^{12} \big\{ B( \delta_{ac}\delta_{db} - \delta_{ab}\delta_{dc}) + E( d_{acr}d_{dbr} - d_{abr}d_{dcr}) \big\}\eqno (6.52)$$
where 
$$t_{2abcd} (x_1,x_2,x_3,x_4) \=d$$
$$i\d_{4\nu} \overline{t}_{uu\tilde{u}Aabcd}^{3\nu} (x_1,x_2,x_3,x_4) - i\d_{1\nu} \overline{t}_{uu\tilde{u}Adbca}^{3\nu} (x_4,x_2,x_3,x_1) - i\d_{2\nu} \overline{t}_{uu\tilde{u}Aadcb}^{3\nu} (x_1,x_4,x_3,x_2),\eqno (6.53)$$
$$t_3^\mu (x_1,x_2,x_3) \=d i\d_1^\nu t_{Au\tilde{u}}^{\nu\mu} (x_1,x_2,x_3) + i\d_2^\nu t_{Au\tilde{u}}^{\nu\mu} (x_2,x_1,x_3)\eqno (6.54)$$
Due to our assumption (6.14), the 3-leg Cg-identity (B.19) takes in the limit (6.14) the simple form
$$t_3^\mu (x_1,x_2,x_3) [= t_3^\mu (x_2,x_1,x_3)]
 = i(\d_1+\d_2)^\mu \overline{t}_{uu\tilde{u}}^3 (x_1,x_2,x_3)+ $$
$$ - ig  \big [ t_{u\tilde{u}}^\mu (x_1,x_3) \delta(x_1-x_2) - t_{u\tilde{u}}^\mu (x_2,x_3) \delta(x_1-x_3)  - t_{u\tilde{u}}^\mu (x_1,x_3) \delta(x_2-x_3) \big ]. \eqno (6.55)$$
Inserting this in (6.52), the left side gets a differential operator $(\d_1+\d_2+\d_4)^\mu$ and the 2-leg distributions $t_{u\tilde{u}}$ cancel
$$(\d_1+\d_2+\d_4)^\mu t_{4abcd} (x_1,x_2,x_3,x_4) = \d_4^\mu \{ \ldots \} - \d_1^\mu \{ \ldots \} - \d_2^\mu \{ \ldots \}  \eqno (6.56)$$
where 
$$t_{4abcd} (x_1,x_2,x_3,x_4) \=d i t_{2abcd} (x_1,x_2,x_3,x_4)+$$
$$ g  \big [ f_{adr}f_{bcr} \big\{ \overline{t}_{uu\tilde{u}}^3 (x_1,x_2,x_3) \delta(x_1-x_4) - \overline{t}_{uu\tilde{u}}^3 (x_1,x_4,x_3) \delta(x_1-x_2) +$$ 
$$+ \overline{t}_{uu\tilde{u}}^3 (x_1,x_4,x_3) \delta(x_3-x_2) -  \overline{t}_{uu\tilde{u}}^3 (x_1,x_2,x_3) \delta(x_4-x_3) \big \}
 - f_{bdr}f_{adr} \{ x_1 \leftrightarrow x_2 \} \big ]  \eqno (6.57)$$
Now we switch over to momentum space and choose a fixed point $( \bar{p}_1,\bar{p}_2,\bar{p}_4 ) \in {\bf R}^{12}$ such that $\bar{p}_1,\bar{p}_2,\bar{p}_4 \in {\bf R}^4$ are linearly independent and that (6.56) holds pointwise (i.e. in the sense of functions) at $( \bar{p}_1,\bar{p}_2,\bar{p}_4 )$. Then, there exist $g_1,g_2 \in {\bf R}^4$ with
$$g_1\bar{p}_1 \neq 0, \quad g_2\bar{p}_2\neq 0,\quad g_i\bar{p}_j = 0 \quad \forall i \neq j, \quad i \in \{1,2\}, \quad j \in \{1,2,3\}. \eqno (6.58)$$
Contracting (6.56) in momentum space first with $g_{1\mu}$ and afterwards with $g_{2\nu}$, we obtain
$$(g_1\bar{p}_1) \hat{t}_{4abcd} (\bar{p}_1,\bar{p}_2,\bar{p}_4) = - (g_1\bar{p}_1) \big \{ B (\delta_{ab}\delta_{cd} - \delta_{ad}\delta_{bc}) + E (d_{abr}d_{cdr} - d_{adr}d_{bcr})\big\} \eqno (6.59)$$
$$(g_2\bar{p}_2) \hat{t}_{4abcd} (\bar{p}_1,\bar{p}_2,\bar{p}_4) = - (g_2\bar{p}_2) \big \{ B (\delta_{ac}\delta_{bd} - \delta_{ab}\delta_{cd}) + E (d_{acr}d_{bdr} - d_{abr}d_{cdr})\big\} \eqno (6.60)$$
These two equations imply 
$$B=0, \quad E=0, \eqno (6.61)$$
i.e. there is no anomaly for (B.24).
\vskip0,3cm
The possible anomaly of (B.25) has exactly the same form as the general normalization term of 
$$t_{uAAA}^{\kappa 2 \lambda \mu \nu} = \eh[t_{uAAA}^{\kappa 2 \lambda \mu \nu} -(\kappa \leftrightarrow \lambda)], \eqno (6.62)$$
namely
$$N_{uAAAabcd}^{\kappa 2 \lambda \mu \nu} = ( g^{\kappa \mu} g^{\lambda \nu} - g^{\kappa \nu} g^{\lambda \mu} ) $$
$$\times [ N (C^2-C^1)_{abcd} + K(D^1 + D^2)_{abcd} ] \delta^{(4(n-1))} , \quad N,K= \mbox{const.}  \eqno (6.63)$$
Let us remember that the other nondegenerate terms in (B.25) have no freedom of normalization. Therefore, we remove a possible anomaly by a renormalization (6.63) of $t_{uAAA}^2$. This fixes the normalization of $t_{uAAA}^2$.
\vskip0,3cm
We now turn to (B.23). The possible anomaly $a$ is a covariant local distribution  of degree $\kappa +1=1$ which must be invariant under all permutations of $\{(b,\nu,x_2),(c,\kappa,x_3),(d,\lambda,x_4) \}$. After a lengthy calculation one obtains
$$a^{\nu\kappa\lambda} =i[ \quad M_1 [ C^1g^{\nu\lambda}\d_1^\kappa + C^2g^{\nu\kappa}\d_1^\lambda + C^3g^{\lambda\kappa}\d_1^\nu ]  $$
$$+M_2 [ C^1\d_3^\kappa g^{\nu\lambda}+\ldots ] +M_3[C^1(\d_2+\d_4)^\kappa g^{\nu\lambda}+\ldots]$$
$$+L_1[C^1(g^{\nu\kappa} \d_1^\lambda + g^{\lambda\kappa} \d_1^\nu)+\ldots] + L_2[C^1(g^{\nu\kappa} \d_3^\lambda + g^{\lambda\kappa} \d_3^\nu)+\ldots]$$
$$+L_3[C^1(g^{\nu\kappa} \d_2^\lambda + g^{\lambda\kappa} \d_4^\nu)+\ldots] + L_4[C^1(g^{\lambda\kappa} \d_2^\nu + g^{\nu\kappa} \d_4^\lambda)+\ldots]$$
$$+F_1[D^1(\d_1^\nu g^{\kappa\lambda} + \d_1^\kappa g^{\lambda\nu} - 2\d_1^\lambda g^{\nu\kappa}) + D^2(-\d_1^\nu g^{\kappa\lambda} + 2\d_1^\kappa g^{\nu\lambda} - \d_1^\lambda g^{\nu\kappa})]$$
$$+F_2[D^1(\d_2^\nu g^{\kappa\lambda} + \d_3^\kappa g^{\lambda\nu} - 2\d_4^\lambda g^{\nu\kappa}) + D^2(-\d_2^\nu g^{\kappa\lambda} - \d_4^\kappa g^{\nu\kappa} +2\d_3^\kappa g^{\nu\kappa})]$$
$$+F_3[D^1(\d_3^\nu g^{\kappa\lambda} - \d_4^\kappa g^{\lambda\nu} - 2\d_4^\lambda g^{\nu\lambda} + \d_2^\kappa g^{\nu\lambda})$$
$$+D^2(\d_3^\nu g^{\kappa\lambda} - \d_2^\lambda g^{\nu\kappa} - \d_4^\nu g^{\nu\lambda} +\d_3^\lambda g{\nu\kappa})]$$
$$+F_4[D^1(\d_4^\kappa g^{\nu\lambda} - p\d2^\lambda g^{\kappa\nu} + \d_4^\nu g^{\lambda\kappa} - \d_3^\lambda g^{\nu\lambda})$$
$$+D^2 ( \d_4^\kappa g^{\nu\lambda} - \d_3^\nu g^{\kappa\lambda} - \d_3^\lambda g^{\nu\kappa} +\d_2^\kappa g^{\nu\lambda} ) ]\quad ] \delta^{(4(n-1))}, \eqno (6.64)$$
where we have not written the colour indices of $C^1,C^2,C^3,D^1,D^2$, and all $M$'s, $L$'s and $F$'s are constants. The dots stand for 2 terms, obtained from the written one by cyclic permutation
$$(C^1,\kappa,x_3) \rightarrow (C^2,\lambda,x_4) \rightarrow (C^3,\nu,x_2) \rightarrow (C^1,\kappa,x_3). \eqno (6.65)$$
In 4-th order perturbation theory we only have three independent variables and the colour space $[C^1,C^2,C^3]$. There are only  $M_2,M_3,L_2,L_3,L_4, \neq 0$

We now turn to the normalization terms. For the nondegenerate terms they are
$$N_{AAAA}^{\alpha\nu\kappa\lambda}= [ \quad A_1 [C^1g^{\alpha\kappa} g^{\nu\lambda}+\ldots]$$
$$+A_2[C^1(g^{\alpha\lambda} g^{\nu\kappa} + g^{\alpha\nu} g^{\kappa\lambda}+\ldots]$$
$$+U[D^1(g^{\alpha\nu} g^{\kappa\lambda} + g^{\alpha\kappa} g^{\nu\lambda} - 2g^{\alpha\lambda} g^{\nu\kappa})$$
$$+D^2(2g^{\alpha\kappa} g^{\nu\lambda} + g^{\alpha\lambda} g^{\nu\kappa} - g^{\alpha\nu} g^{\kappa\lambda})]\quad ] \delta^{(4(n-1))} \eqno (6.66)$$
$$N_{uAAA}^{\alpha l \nu\kappa\lambda} = [ \quad B_1 [C^1 g^{\alpha\kappa} g^{\nu\lambda} + \ldots ]$$
$$+B_2[C^1(g^{\alpha\lambda} g^{\nu\kappa} + g^{\alpha\nu} g^{\kappa\lambda})+\ldots]$$
$$+V[D^1(g^{\alpha\nu} g^{\kappa\lambda} + g^{\alpha\kappa} g^{\nu\lambda} - 2g^{\alpha\lambda} g^{\nu\kappa})$$
$$D^2(2g^{\alpha\kappa} g^{\nu\lambda} - g^{\alpha\lambda} g^{\nu\kappa} - g^{\alpha\nu} g^{\kappa\lambda})] \quad ] \delta^{(4(n-1))}. \eqno (6.67)$$
Here, $l \ge 5$ and $A_1,\ldots V$ are constants. Again, the dots in (6.66) and (6.67) stand for two terms, obtained by cyclic permutation (6.65). The normalization of $t_{uAAA}^2$ is already fixed by (6.63). Inserting (6.67)(6.66) into the Cg-identity (B.23), we arrive at
$$\d_{1\alpha} N_{AAAA}^{\alpha\nu\kappa\lambda}  + \sum_{l=5}^n \d_{l \alpha} N_{uAAA}^{\alpha l \nu\kappa\lambda} = $$
$$=[ A_1 [ C^1 \d_1^\kappa g^{\nu\lambda} + \ldots ] + A_1 [C^1(\d_1^\lambda g^{\nu\kappa} + \d_1^\nu g^{\kappa\lambda})+\ldots ]$$
$$-B_1[C^1(\d_1+\d_2+\d_3+\d_4)^\kappa g^{\nu\lambda} +\dots]$$
$$-B_2[C^1(\d_1+\d_2+\d_3+\d_4)^\lambda g^{\nu\kappa} + (\d_1+\d_2+\d_3+\d_4)^\nu g^{\kappa\lambda}+\ldots]$$
$$+U[D^1(p_1^\nu g^{\kappa\lambda} + \d_1^\kappa g^{\lambda\nu} - 2\d_1^\lambda g^{\nu\kappa})$$
$$+D^2(-\d_1^\nu g^{\kappa\lambda} + 2\d_1^\kappa g^{\nu\lambda} - \d_1^\lambda g^{\nu\kappa}]$$
$$-V[D^1((\d_1+\d_2+\d_3+\d_4)^\nu g^{\kappa\lambda} + (\d_1+\d_2+\d_3+\d_4)^\kappa g^{\nu\lambda}$$
$$ - 2(\d_1+\d_2+\d_3+\d_4)^\lambda g^{\nu\kappa})$$
$$+D^2 (2 (\d_1+\d_2+\d_3+\d_4)^\kappa g^{\nu\lambda} - (\d_1+\d_2+\d_3+\d_4)^\lambda g^{\nu\kappa}$$
$$ - (\d_1+\d_2+\d_3+\d_4)^\nu g^{\kappa\lambda})]]\delta^{(4(n-1))} \eqno (6.68)$$

In order $n=4,\quad N_{uAAA}^l (l \ge 5)$ and $[D^1,D^2]$ are absent, moreover $\d_1 =-(\d_2+\d_3+\d_4)$. Therefore we have to set $B_1=B_2=U=V=0$ to obtain the $n=4$ case.

No independent additional freedom of normalization in (B.23) comes from the $\delta$-degenerate terms. Thus, only (6.68) is at our disposal to remove the anomaly (6.64), and it is obvious that we do not succeed. Therefore, the ansatz of the anomaly must be further restricted.

In fact, one proves after a lengthy calculation, inserting in (B.23) the 3-leg Cg-identities (B.19), (B.20), (B.21) and the 4-leg Cg-identities (B.24), (B.25) in the limit (6.14), the following additional symmetry:
$$\d_{3 \kappa} \d_{4 \lambda} a_{abcd}^{\nu\kappa\lambda} (x_1,x_2,x_3,x_4) -\big( (x_1,a) \leftrightarrow (x_3,c) \big) = 0. \eqno (6.69)$$

Substituting the ansatz (6.64) for the anomaly in the symmetry condition (6.69), we obtain by a straightforward calculation, that (6.69) is equivalent to
$$L_2=L_3=L_4 \=d: L, \quad M_2=M_3 \=d:M, \quad F_2=F_3 \=d:F, \quad F_4=F_2+F_3=2F. \eqno (6.70)$$
Now we perform renormalization (6.66),(6.67) with the following choice of the constants
$$A_1=M-M_1, \quad A_2=L-L_1, \quad U=F-F_1, $$
$$B_1=M, \quad B_2=L,\quad V=F,\eqno (6.71)$$
Then the anomaly (6.64) (with restriction  (6.69)) vanishes. In 4-th order perturbation theory the restrictions (6.69) are obtained in the same way, but the assumption (6.14) is not needed, because the anomaly is removed by choosing
$$A_1=M,\quad A_2=L.\eqno (6.72)$$

\vskip0,5cm
$\bullet$ {\bf Proof of the 5-Leg Cg-Identities}
\vskip0,3cm
There is only one operator combination with five external legs and $|\kappa| \ge -1$, namely
$$:\Omega:=:A_\mu^a u^b A_\nu^c A_\lambda^d A_pê:. \eqno (6.73)$$
The anomaly is a Lorentz and $SU(N)$ invariant constant so that its numerical part is of the following form
$$a_{abcde}^{\mu \nu \lambda \rho} = \sum_{j=1}^{22} \sum_{i=1}^3 C_{ij} L_i^{\mu \nu \lambda p} K_{abcd}^j \quad \delta^{(4(n-1))} , \quad C_{ij}=const. \eqno (6.74)$$
\newpage
Here the $L_i$'s are defined by 
$$L_1^{\mu \nu \lambda \rho} =g^{\mu \nu} g^{\lambda \rho}, \quad L_2^{\mu \nu \lambda \rho} =g^{\mu \lambda } g^{\nu \rho}, \quad L_3^{\mu \nu \lambda \rho} =g^{\mu \rho} g^{\nu \lambda},\eqno (6.75)$$
and the $K^j,j=1,\ldots 22$ are the basic (A.20) of the invariant 5th rank $SU(N)$-tensors $(N \ge 4)$ in the even sector, derived in the Appendix B. The index $b$ plays a distinct role in (A.20), this is the reason why the ghost field has this colour index (6.73). The odd sector is excluded in case of 5 external legs, because of the Furry theorem in Fock space (see Chapter 5, 5.52). The 16-dimensional basic for $SU(3)$ and the 10-dimensional basic for $SU(2)$ are contained in the 22-dimensional basis, we are using. Then it is easy to see that our proof holds for these special cases, too.

The anomaly is further restricted: It must be invariant under all permutations $\pi \in S_4$ of
$$(a,\mu),(c,\nu),(d,\lambda),(e,\rho) \eqno (6.76)$$
We are going to prove that this forces the anomaly to vanish due to the one $f$-tensors of the basic (A.20). In order to do this, we show that 
$$a_{abcde}^{\mu \nu \lambda \rho}=\sum_{i,j} C_{ij} \frac{1}{4!} \sum_{\pi \in S_4} L_i^{\pi(\mu) \pi(\nu) \pi(\lambda) \pi(\rho)} K_{b\pi(a)\pi(c)\pi(d)\pi(e)}^j \delta^{(4(n-1))}\eqno (6.77)$$
the terms compensate in pairs (or in pairs of pairs) for every fixed $(i,j)$, if the sum over $\pi \in S_4$ is carried out. This relies on the fact that $K^1,\ldots K^{22}$ is odd under certain permutations. However, the Lorentz indices are also permuted. Since
$$L_k^{\mu\nu\lambda\rho} = L_k^{\nu\mu\rho\lambda} = L_k^{\lambda\rho\mu\nu} =L_k^{\rho\lambda\nu\mu},\quad k=1,2,3, \eqno (6.78)$$
this does not seriously complicate the proof.

We prove that 
$$\sum_{\pi \in S_4} L_i^{\pi(\mu) \pi(\nu) \pi(\lambda) \pi(p)} K_{b\pi(a)\pi(c)\pi(d)\pi(e)}^j  \eqno (6.79)$$
Considering the 22 values of $j$, we   have to distinguish the following 4 cases, depending on the position of the nonpermuted index $b$.

$Case 1:$   $K_b^j\ldots = \delta .. f_b ..(j=2,3,4,8,9,10)$ in (A.20). Here we have
$$\delta_{de} f_{bac} L_i^{\mu \nu \lambda \rho} + \delta_{ed} f_{bca} L_i^{\nu \mu \rho \lambda}=0  \eqno (6.80)$$
where the second term is obtained from the first one by exchanging $(a,\mu) \leftrightarrow (c, \nu)$ and 

$(d,\lambda) \leftrightarrow (e,\rho)$.

$Case 2:$   $K_b^j\ldots = \delta .. f_b ..(j=1,5,6,7)$ in (A.20). Here we have the following compensations
$$\delta_{be} f_{acd} L_1^{\mu \nu \lambda p} + \delta_{eb} f_{cad} L_1^{\nu \mu \lambda p} = 0  \eqno (6.81)$$
$$\delta_{be} f_{acd} L_2^{\mu \nu \lambda p} + \delta_{eb} f_{dca} L_2^{\lambda \nu \mu p} = 0  \eqno (6.82)$$
$$\delta_{be} f_{acd} L_3^{\mu \nu \lambda p} + \delta_{be} f_{adc} L_3^{\mu \lambda \nu p} = 0  \eqno (6.83)$$
The second terms in (6.81-83) are generated by the permutations $(a,\mu) \leftrightarrow (c,\nu)$, $(a,\mu) \leftrightarrow (d,\lambda)$ or $(c,\nu) \leftrightarrow (d,\lambda)$, respectively.

$Case 3:$   $K_b^j\ldots = d_{..r} d_{.rs} f_{bs.} (j=14,15,16,20,21,22)$ in. We take the following terms together:
$$[d_{acr}d_{dsr}f_{bse}L_k^{\mu \nu \lambda \rho} + d_{car}d_{ers}f_{bsd}L_k^{\nu \mu \rho \lambda}]+[(a,\mu) \leftrightarrow (e,\rho),(c,\nu) \leftrightarrow (d,\lambda)]. \eqno (6.84)$$
Using the identity
$$d_{drs}f_{bse}=d_{ers}f_{bds} + d_{eds}f_{brs},$$
This is equal to 
$$=d_{acr}d_{eds}f_{brs}L_k^{\mu \nu \lambda \rho} +[(a,\mu) \leftrightarrow (e,\rho),(c,\nu) \leftrightarrow (d,\lambda)].  \eqno (6.85)$$
$Case 4:$   $K_b^j\ldots = d_{b.r} d_{.rs} f_{.s.} (j=11,12,13,17,18,19)$ in (A.20). In the expression

$d_{bar}d_{crs}f_{dse}L_k^{\mu \nu \lambda \rho}$, we use the identity
$$d_{bar}d_{crs}=d_{bsr}d_{car} - f_{bcr} f_{sar} + \frac{2}{N} (\delta_{bs}\delta_{ca} - \delta_{ba}\delta_{cs}). \eqno (6.86)$$
The two $\delta$-terms give contributions that have already been considered in case 1 and 2. The first $dd$-term cancels out in the following combinations:
$$d_{car}d_{brs}f_{dse}L_k^{\mu \nu \lambda \rho}+[(a,\mu) \leftrightarrow (c,\nu),(d,\lambda) \leftrightarrow (e,\rho)]=0   \eqno (6.87)$$
Concerning the $ff$-term, we consider
$$[f_{bcr}f_{ars}f_{dse} L_k^{\mu \nu \lambda \rho}+f_{bar}f_{crs}f_{esd} L_k^{\lambda \rho \mu \nu}]+$$
$$+[(a,\mu) \leftrightarrow (d,\lambda),(c,\nu) \leftrightarrow (e,\rho)]. \eqno (6.88)$$
Using now the Jacobi identity
$$f_{bcr}f_{ars}=f_{acr}f_{rsb} + f_{crs}f_{bar},$$
this is equal to 
$$=f_{acr}f_{rsb}f_{dse}L_k^{\mu \nu \lambda \rho}+[(a,\mu) \leftrightarrow (d,\lambda),(c,\nu) \leftrightarrow (e,\rho)]=0. \eqno (6.89)$$
This completes the proof of the 5-Leg Cg-Identities

\vskip0.5cm
$\bullet$ {\bf Proof of the Identities with External Fermion Pairs}
\vskip0,5cm
There remain the cases with an external fermion pair to be considered. Similar to the abelian theory, gauge invariance here relies on C-invariance (see Chapter 5b) 
$$U_C T_n U_C^{-1}=T_n, \quad U_C T_{n/l} U_C^{-1} = T_{n/l}, \quad l \le n \eqno (6.90)$$
The Cg-identities with external fermions (B.29) (B.30) correspond to field operators

$:\Omega:=: \bar{\psi}^\alpha (x_i) \psi^\beta(x_j): u^a(x_n)$ and $:\bar{\psi}^\alpha(x_i) \psi^\beta (x_j):  u^a(x_n) A^b_\nu (x_m)$. The proof of these identities follows the corresponding one used in the abelian case ([26], case III and case IV). Now we have to use the nonabelian version of C-invariance in order to restrict the anomaly. One only has to make some modifications concerning the colour structure and the additional assumption (6.14). The identities (B.29) and (B.30) are only proven between Dirac spinors in this way. But note that operator gauge invariance is sufficient for the unitarity.

\vskip 1cm
{\Large\bf 7. Proof of the Unitarity in Nonabelian Gauge Theories}
\vskip 1cm
The proof of the operator gauge invariance now puts us in a position to show the unitarity of the S-matrix $S(g)$ in the physical subspace of the Fock space. This is the most important and most subtle property of the S-matrix. The subtlety comes from the well-known fact that, because of the gauge structure, the gauge boson sector of the Fock space contains more elements than are physically distinguishable.

We proceed in four steps: In Subchapter (a) we introduce a representation of the varions field operators in a positive definite Fock-Hilbert space. We avoid the usual treatment in a Pseudo-Hilbert space with an indefinite metric.
In Subchapter (b) we prove inductively the pseudo-hermiticity of the n-point distribution $T_n$ with regard to a sesquilinear form in the Fock space. This guarantees the pseudo-unitarity of the S-matrix.
In Subchapter (c) we give a definition of the physical subspace, using the methods introduced by Kugo and Ojima [33].
In Subchapter (d) we finally prove inductively the perturbative unitarity in the physical subspace with the help of the operator gauge invariance in the causal construction.

Note that in the causal approach all examinations concerning the unitarity of the S-matrix $S(g)$ are mathematically well-defined, even in a massless theory because in the causal formulation the physical infrared problem is naturally separated by adiabatic switching of the S-matrix $S(g)$ with a test function $g \in \cal S$ and also absent before the limit $g \longrightarrow 1$ is taken.
      
\vskip0.3cm
{\bf (a) Preliminaries}
\vskip0,3cm
In this subchapter we introduce a concrete representation of the various field operators in a positive definite Fock-Hilbert space. We avoid working in a vector space with an indefinite metric. As is well-known, the realization of the gauge boson field on a positive definite Hilbert space $F$ is not possible in a manifestly Lorentz covariant way: The zeroth component of the gauge boson field must be skew-hermitean, in contrast to the hermitean spatial components:
$$A^0(x) =(2\pi)^{-\frac{3}{2}} \int \frac{d^3k}{\sqrt{2w}} \big{(} a^0(\stackrel{\rightharpoonup}{k}) e^{-ikx} -a^0(\stackrel{\rightharpoonup}{k})^+ e^{+ikx}\big{)} \eqno (7.1)$$
$$A^j(x) =(2\pi)^{-\frac{3}{2}} \int \frac{d^3k}{\sqrt{2w}} \big{(} a^j(\stackrel{\rightharpoonup}{k}) e^{-ikx} +a^j(\stackrel{\rightharpoonup}{k})^+ e^{ikx}\big{)} \eqno (7.2)$$
where $w = \mid \stackrel{\rightharpoonup}{k} \mid$ and the $a(\stackrel{\rightharpoonup}{k}), a(\stackrel{\rightharpoonup}{k})^+$ are the absorption and emission operators satisfying the usual bosonic commutation relations:
$$\big{[} a^0(\stackrel{\rightharpoonup}{p}), a^{0+} (\stackrel{\rightharpoonup}{p})\big{]}_- = \delta (\stackrel{\rightharpoonup}{p} - \stackrel{\rightharpoonup}{p}') \eqno (7.3)$$
$$\big{[} a^j(\stackrel{\rightharpoonup}{p}), a^{i+} (\stackrel{\rightharpoonup}{p})\big{]}_- = \delta_{ij} \delta (\stackrel{\rightharpoonup}{p} - \stackrel{\rightharpoonup}{p}') \eqno (7.4)$$
We have omitted the colour indices which are irrelevant in this section.
In addition to the (positive definite) scalar product
$$<\cdot \mid \cdot>: \quad F \times F \longrightarrow \mbox{C}^+ \eqno (7.5)$$
we now introduce a sesquilinear form in $F$ (initially only in the gauge boson sector of $F$)
$$<\cdot \mid \eta_A \cdot >: \quad F \times F \longrightarrow \mbox{C} \eqno (7.6)$$
with the metric tensor $\eta_A^+ = \eta_A^{-1} = \eta_A$ which is in the one-particle Hilbert space $H_1^A$ in $F$ given by
$$\eta_A \left| \begin{array}{ll} & =\\ H_1^A  \end{array} \right. \left( \begin{array}{rccr} -1 0 0 0\\ 0 1 0 0 \\ 0 0 1 0\\ 0 0 0 1 \end{array} \right) \mbox{.} \eqno (7.7)$$
This means
$$\eta_Aa_0 \eta_A = -a_o,\quad \eta_Aa_j\eta_A = a_j \eqno (7.8)$$
The corresponding expansion on the whole gauge boson sector of $F$ is given by
$$\eta_A = (-1)^{N^{A_0}} \eqno (7.9)$$
where $N_{A_0}$ is the particle number operator of the scalar gauge bosons.
The corresponding conjugation $k$ for any operator $\hat{O}$ is given by
$$\hat{O}^k = \eta \hat{O}^+ \eta \eqno (7.10)$$
with "+" describing the hermitean adjunction with regard to the Hilbert scalar product.
One finds that the gauge boson field is at least pseudo-hermitean
$$A_\mu^k = A_\mu \eqno (7.11)$$
Furthermore, we have
$$A_\mu(x) =(2\pi)^{-\frac{3}{2}} \int \frac{d^3k}{\sqrt{2w}} \big{(} a_\mu(\stackrel{\rightharpoonup}{k}) e^{-ikx} +a_\mu(\stackrel{\rightharpoonup}{k})^k e^{ikx}\big{)}$$
$$\mbox{with} \qquad \big{[} a_\mu (\stackrel{\rightharpoonup}{k}), a_\nu (\stackrel{\rightharpoonup}{k'})^k \big{]} = g_{\mu \nu} \delta (\stackrel{\rightharpoonup}{k} - \stackrel{\rightharpoonup}{k}'), \quad a_\mu^k = \eta a_\mu^+ \eta \eqno (7.12)$$
One possible representation of the ghost field is
$$u(x) =(2\pi)^{-\frac{3}{2}} \int \frac{d^3p}{\sqrt{2w}} \big{(} c_2(\stackrel{\rightharpoonup}{p}) e^{-ipx} +c_1^+(\stackrel{\rightharpoonup}{p}) e^{ipx}\big{)} \eqno (7.13)$$
$$\tilde{u}(x) =(2\pi)^{-\frac{3}{2}} \int \frac{d^3p}{\sqrt{2w}} \big{(}-c_1(\stackrel{\rightharpoonup}{p}) e^{-ipx} +c_2^+(\stackrel{\rightharpoonup}{p}) e^{ipx}\big{)} \eqno (7.14)$$
Here $\omega = \mid \stackrel{\rightharpoonup}{p} \mid = p^0$ and the creation and annihilation operators satisfy the (anomalous) anti-commutation relations.
$$\big{\{} c_i (\stackrel{\rightharpoonup}{p}), c_j (\stackrel{\rightharpoonup}{q})^+ \big{\}}_+ = \delta_{ij} \delta (\stackrel{\rightharpoonup}{p} - \stackrel{\rightharpoonup}{q}) \eqno (7.15)$$
and all other anti-commutators vanish.
This implies the equations
$$\big{\{} u_a^\pm (x), \tilde{u_b}^\mp (y) \big{\}}_+ = (-i) \delta_{ab} D^\mp (x - y) \eqno (7.16)$$
We define a sesquilinear form also in the ghost sector of $F$ by giving the metric tensor in the one-particle subspace:
$$\eta_G \left| \begin{array}{ll} & =\\ H_1^G \end{array} \right. \left( \begin{array}{lccr} 0 & 1 \\ \\  1 & 0 \end{array} \right) \mbox{,} \eqno (7.17)$$
that means
$$\eta_Gc_2\eta_G = c_1 \quad \wedge \quad \eta_G c_1 \eta_G = c_2 \eqno (7.18)$$
The corresponding conjugation is then given by
$$c_2^k: = \eta_G c_2^+ \eta_G = c_1^+,\quad c_1^k: = \eta_G c_1^+ \eta_G = c_2^+ \eqno (7.19)$$
which implies
$$u^k = u \quad \wedge \quad \tilde{u}^k = -\tilde{u}. \eqno (7.20)$$
The ghost fields take the symmetric form
$$u(x) =(2\pi)^{-\frac{3}{2}} \int \frac{d^3p}{\sqrt{2\omega}} \big{(} c_2(\stackrel{\rightharpoonup}{p}) e^{-ipx} +c_2(\stackrel{\rightharpoonup}{p})^k e^{ipx}\big{)} \eqno (7.21)$$
$$\tilde{u}(x) =(2\pi)^{-\frac{3}{2}} \int \frac{d^3p}{\sqrt{2\omega}} \big{(} -c_1(\stackrel{\rightharpoonup}{p}) e^{-ipx} +c_1^k(p) e^{ipx}\big{)} \eqno (7.22)$$

\vskip0.5cm

{\bf (b) Pseudo-Unitarity of S(g)}
\vskip0,3cm
The 1-point distribution is skew-conjugate with regard to the sesquilinear form defined in Subchapter (a).
$$T_1^k = -T_1 = \tilde{T}_1 \eqno (7.23)$$
This holds for all n-point distributions $T_n(x)$ by induction:
\vskip 0,3cm
{\bf Theorem 7.1.}
$$T_n^k(x) = \tilde{T}_n(x) \qquad \forall n \qquad \mbox{(Pseudo-Hermiticity of $T_n$)} \eqno (7.24)$$
$$\big{(} \Longleftrightarrow S^k(g) = S^{-1}(g) \qquad \mbox{(Pseudo-Unitarity of $S(g)$} \big{)} \eqno (7.25)$$
Note that (7.25) is a statement about a formal power series.\\
{\bf Proof:}
The proof method is analogous to the one used in Chapter 5 for the discrete symmetries:
The statement (7.24) is valid for $n=1$. Let us assume that (7.24) holds for all $m \le n-1$. \\
Now let $R_n$ be a splitting solution of $D_n$. Taking (C.10a,b)
into account, this total retarded distribution
$$R_n(x_1, \ldots x_n) = \sum_{P_2^0} T_{n-n_1} (Y, x_n) \tilde{T}_{n_1} (X) = R'_n + T_n \eqno (7.26)$$
can also be written as follows
$$R_n = -\sum_{P_2^0} T_{n_1} (X) \tilde{T}_{n-n_1} (Y, x_n) = -(R'_n)^k - \tilde{T}_n. \eqno (7.27)$$ 
Analogously, one derives $A_n = A'_n - T_n = - (A'_n)^k - \tilde{T}_n$. It directly follows
$$D_n^k = -D_n. \eqno (7.28)$$
Now $R^s_n$ given by
$$R^s_n: = \frac{1}{2} [R_n-R_n^k] \eqno (7.29)$$
is a splitting solution with
$$(R^s_n)^k = - R^s_n \eqno (7.30)$$
Taking (7.27) and (7.30) into consideration, we get
$$T^s_n = R^s_n - R'_n , \quad \tilde{T}^s_n = (-R'_n)^k - R^s_n \eqno (7.31)$$
and finally
$$(\tilde{T}^s_n)^k = -R'_n - (R^s_n)^k = -R'_n + R^s_n = 
+T^s_n \eqno (7.32)$$
This proves the induction assumption (7.24) for $m = n$ and completes the proof of pseudo-unitarity. {\bf q.e.d.}

\vskip0,3cm
{\bf Comments:}\\
$\bullet$Again we are confronted with the question of the compatibility of the various normalization conditions on the n-point distribution $T_n$. In Chapter 6 we ended up with a $T_n$ which is $L_+^\uparrow$-covariant, C-,P- and T-invariant (Chapter 5) and in particular gauge invariant. Obviously, the discrete symmetry properties survive in the symmetrization in (7.29), if the (anti-)unitary discrete symmetry operators commute or anticommute with the  operator $\eta$ which defines the sesquilinear form in $F$.
In an exemplary mode, we show $[U_C, \eta]=0$ in the gauge boson sector of $F$:\\
An explicit unitary representation of $U_C$ in the gluon sector of $F$ is the
following:
$$ U_C= \prod^{N^2 -1}_{a=1} \theta_a^{\frac{1}{2} - \frac{1}{2} \tau_a} $$

with $\theta_a=(-1)^{N_a}$, $N_a$is the particle number operator of the gauge bosons with colour $a$. $\tau$ is given by $U_{aa'}=-\tau_a \delta_{aa'} \quad$ (see A.9). Now (5.43) holds: $U_C A_a U_C = U_{aa'} A_{a'}$.\\
Remember that on the gauge boson sector $\eta$ is given by $ \eta_A=(-1)^{N_{A_0}}$ (7.9) with the particle number operator $N_{A_0}$ of the scalar gauge bosons.
Using these explicit representations of the operators, it is obvious that $U_C$ commute with $\eta_A$.\\ 
Furthermore, all necessary (re-)normalizations - in order to arrive at a gauge invariant splitting solution in Chapter 6 - do not violate pseudo-unitarity nor any discrete symmetry.\\  
\\
$\bullet$The following equation is a direct consequence of pseudo-unitarity:
$$ReT_n = \frac{1}{2} (T_n + T_n^k) = \frac{1}{2} \big{[}\tilde{R}_n - R'_n + \tilde{R}_n^k - (R'_n)^k \big{]} = \frac{1}{2}(-R'_n - R'^k_n) \eqno (7.33)$$
The pseudo-unitarity, $\tilde{R}^k = \tilde{R} (-1)$, confines the normalization freedom to the imaginary part $ImT_n$.
This fact is the starting point of the well-known Cutkosky rules in pertubative QFT.\\

\vskip 0,5cm
{\bf (c) Definition of the Physical Subspace}
\vskip 0,5cm
We want to arrive at the explicit form of the charge operator $Q$. Therefore, we first compute
$$\partial_\mu A^\mu(x) = -i(2\pi)^{-3/2} \int d^3k \sqrt{\frac{\omega}{2}} \big{[}(a^0 + a_\|)e^{-ikx} +(a^{0+} -a_\|^+)e^{ikx} \big{]} \eqno (7.34)$$
Here we have introduced creation and annihilation operators for longitudinal bosons
$$a_\|(\stackrel{\rightharpoonup}{k}) = \frac{k_j}{\omega} a^j(\stackrel{\rightharpoonup}{k}), \eqno (7.35)$$
which also fulfill the ordinary commutation relations
$$[a_\|(\stackrel{\rightharpoonup}{k}), a_\|^+(\stackrel{\rightharpoonup}{k}')] = \delta(\stackrel{\rightharpoonup}{k} - \stackrel{\rightharpoonup}{k}'). \eqno (7.36)$$
This enables us to express the charge operator
$$Q = \int d^3x (\partial_\nu A^\nu \partial_0 u - \partial_0 \partial_\nu A^\nu u) \eqno (7.37)$$
in momentum space
$$Q = \int d^3k \quad \omega(\stackrel{\rightharpoonup}{k})[(a_\|(\stackrel{\rightharpoonup}{k})^+ - a^{0+} (\stackrel{\rightharpoonup}{k}))c_2 (\stackrel{\rightharpoonup}{k}) + c_1^+(\stackrel{\rightharpoonup}{k})(a_\| (\stackrel{\rightharpoonup}{k}) + a^0(\stackrel{\rightharpoonup}{k}))]. \eqno (7.38)$$
Note that $Q$ is pseudo-hermitesch
$$Q^k = Q \eqno (7.39)$$
Let us introduce the combinations
$$b_1(\stackrel{\rightharpoonup}{k}) = \frac{1}{\sqrt{2}} (a_\|(\stackrel{\rightharpoonup}{k}) + a_0(\stackrel{\rightharpoonup}{k})),\quad b_2(\stackrel{\rightharpoonup}{k}) = \frac{1}{\sqrt{2}} (a_\|(\stackrel{\rightharpoonup}{k}) - a_0(\stackrel{\rightharpoonup}{k})) \eqno (7.40)$$
Since
$$[b_i(\stackrel{\rightharpoonup}{k}), b_j^+(\stackrel{\rightharpoonup}{k})]_- = \delta_{ij} \delta(\stackrel{\rightharpoonup}{k} - \stackrel{\rightharpoonup}{k}') \eqno (7.41)$$
they describe different degrees of freedom and we have
$$Q = \sqrt{2} \int d^3k \quad \omega (\stackrel{\rightharpoonup}{k}) \big{[} b_2^+ (\stackrel{\rightharpoonup}{k}) c_2 (\stackrel{\rightharpoonup}{k}) + c_1^+ (\stackrel{\rightharpoonup}{k}) b_1 (\stackrel{\rightharpoonup}{k}) \big{]} = \sqrt{2} \int d^3k \quad \omega (\stackrel{\rightharpoonup}{k}) \big{[} b_1^k c_2 +  c_2^k b_1  \big{]} \eqno (7.42)$$
In order to find the operation of $Q$, it is therefore sufficient to consider the sector of unphysical particles generated by the creation operators $b_1^+, b_2^+, c_1^+$ and $c_2^+$.
Now let
$$N = \int d^3k \big{[} b_1^+ (\stackrel{\rightharpoonup}{k}) b_1 (\stackrel{\rightharpoonup}{k}) + b_2^+ (\stackrel{\rightharpoonup}{k}) b_2 (\stackrel{\rightharpoonup}{k}) + c_1^+ (\stackrel{\rightharpoonup}{k}) c_1 (\stackrel{\rightharpoonup}{k}) + c_2^+ (\stackrel{\rightharpoonup}{k}) c_2 (\stackrel{\rightharpoonup}{k}) \big{]} \eqno (7.43)$$
be the particle number operator of the unphysical particles. $N$ is a positive self-adjoint operator with discrete spectrum $n = 0, 1, 2, 3,$..  . $Q$ manifestly does not change the number of unphysical particles. This means that $N$ commutes with $Q$. Hence the eigenspaces of the operator $N$ for fixed $n$ are invariant under $Q$ and $Q$ commutes with the corresponding projection operators.

The nullspace Ker$N$ contains no unphysical particles (scalar or longitudinal vector bosons or ghosts).

This is the physical subspace of transversal gauge bosons $F_\bot$.  It follows from (7.42) that $F_\bot$ is a subspace of Ker$Q$.
We state the definitions:
$$\mbox{Ker}Q: = \{ \alpha \in F \mid Q \alpha = 0\} \eqno (7.44)$$
$$F_\bot: = \{\alpha \in F \mid Q \alpha = 0 \wedge N \alpha = 0\} \eqno (7.45)$$
$$F_0: = KerQ \cap \oplus_{n>0}Eig(N,n) \eqno (7.46)$$
($n\in$ {\bf N}; $Eig(N,n)$ is the eigenspace of the operator $N$ corresponding to the eigenvalue $n$)\\
We have
$$\mbox{Ker}Q = F_\bot \oplus F_0 \eqno (7.47)$$
We call the corresponding projection operators $P, P_\bot$ and $P_0$. 
We prove some properties of these subspaces:
\vskip 0.3cm
{\bf Proposition 7.1.}
$$F_0 = QF = \mbox{Range} Q \eqno (7.48)$$
{\bf Proof:} We introduce the operator
$$Q_1 = \int \frac{d^3k}{\sqrt{2\omega(\stackrel{\rightharpoonup}{k}})} \big{[} b_1^+ (\stackrel{\rightharpoonup}{k}) c_1(\stackrel{\rightharpoonup}{k}) + c_2^+ (\stackrel{\rightharpoonup}{k}) b_2 (\stackrel{\rightharpoonup}{k}) \big{]} \eqno (7.49)$$
It has a simple anti-commutator with $Q$
$$\{Q, Q_1\} = N \eqno (7.50)$$
This can easily be verified by means of the identity
$$\{AB, C\} = A \{B, C\} - [A, C] B = \{C, AB\} \eqno (7.51)$$
Since in $F_0 = P_0F$ the null space of $N$ is projected out, $N$ has a bounded inverse $N^{-1}$ on $F_0$.
$$P_0 = NN^{-1} P_0 = (Q Q_1 + Q_1Q)N^{-1}P_0 \eqno (7.52)$$
where (7.51) has been used. In the second term $Q$ can be commuted with $N^{-1}$ and is zero when applied to $P_0$. Hence
$$P_0f = QQ_1N^{-1} P_0f \=d Qg \in \mbox{Range} Q \eqno (7.53)$$
is in the range of $Q$ for all $f \in F$. Actually $QF=F_0$ because Range$Q \subset$ Ker$Q = PF$ (since $Q^2=0$) and because Range$Q$ is orthogonal to $P_\bot F = F_\bot$:
$$<P_\bot g, Qf > = <g, P_\bot Qf> = < g, QP_\bot f> = 0 \eqno (7.54)$$

{\bf q.e.d.}

\vskip 0,3cm
{\bf Proposition 7.2.}
\vskip0,2cm
Let $A$ be a gauge invariant operator in $F: \quad [Q,A] = 0$.
Then we have
$$\qquad (\alpha)\quad A(\mbox{Ker}Q) \subset \mbox{Ker}Q \eqno (7.55)$$
$$(\beta)\quad A(F_0) \subset F_0 \eqno (7.56)$$
$$\mbox{but} \qquad A(F_\bot) \not\subset F_\bot \qquad \mbox{in general} \eqno (7.57)$$
{\bf Proof:}
$(\alpha)$ Let $f \in$ Ker$Q$, then $Q(AF) = AQf = 0$.

$(\beta)$ Let $f \in F_o.\quad \exists\quad g \in F$ with $Qg=f$ because $F_0 \subset QF$.
If follows: $Af = AQg =$

$QAg: = Qh$ with $h \in F$ Because of $QF \subset F_o$, we have $Af \in F_0$.  {\bf q.e.d.}
\vskip 0,3cm
{\bf Note:}
The operator gauge invariance states $[Q, T_n]$ = div $\not= 0$, $\forall n$ (2.12). Therefore we are able to show the invariance of Ker $Q$ and $F_0$ under $T_n$ modulo a divergence (in the sense of vector analysis).
But this is sufficient to prove unitarity in the physical subspace $F_\bot$ because a divergence formally (infrared problem) vanishes in the adiabatic limit $g\longrightarrow 1$.
\vskip 0.5cm
{\bf (d) Proof of Physical Unitarity}
\vskip0,3cm
In order to prove the physical unitarity, one has to consider the following two propositions. They state that the unphysical degrees of freedom compensate each other in intermediate states up to a divergence:
\vskip 0,3cm
{\bf Proposition 7.3.} For arbitrary n-point distributions $T_n(X)$ we have
$$T_n(X)P = PT_n(X)P + div. \eqno (7.58)$$
where div denotes terms of divergence form as they occur in the condition of gauge invariance.
\vskip 0,3cm
{\bf Proof:} For arbitrary $f \in F$ we investigate the vectors
$$T_nPf = g. \eqno (7.59)$$
Operating with $Q$ on (7.59) and using gauge invariance, we get
$$QT_nPf = T_nQPf + div_1. \eqno (7.60)$$
This implies
$$Qg = div_1, \eqno (7.61)$$
because $Pf$ is in the kernel of $Q$. This is a linear inhomogeneous equation for $g$ and we know that it has a solution (7.59). This solution must be of the following form:
$$g = Ph + div_2. \eqno (7.62)$$
Here $Ph$ is a solution of the homogeneous equation and the solution of the inhomogeneous equation must also have divergence form. Operating with $P$ on (7.62), we get
$$Pg = Ph + div_3, \eqno (7.63)$$
which enables us to write $g$ in the form
$$g = Pg + div_2 - div_3. \eqno (7.64)$$
This completes the proof of Prop. (7.3).  {\bf q.e.d.}
\vskip 1cm
{\bf Proposition 7.4.} Let $P_\bot$ be the projection operator on $F_\bot$ then
$$P_\bot T_n(X)P_\bot = P_\bot T_n(X)P + div_1 \eqno (7.65)$$
where $P$ projects on Ker $Q$.
\vskip 0,5cm
{\bf Proof:} Let be $f \in F$. $Pf$ has a unique decomposition because of Ker$Q = F_0 \oplus F_+$:
$$Pf = P_0f + P_\bot f = f_0 + f_\bot. \eqno (7.66)$$
We have to prove that $T_n f_0 = h + div$ with $h \in F_0.$
Because of $F_0 = QF$, there is a $g \in F$ with $Qg=f_0$. With the help of gauge invariance, we then arrive at
$$T_n f_0 = T_n Qg = QT_ng + div = h + div, \quad \hbox{with} \quad h \in F_0 \qquad \mbox{{\bf q.e.d.}}\eqno (7.67)$$
\vskip1cm
The proof of physical unitarity is now quite simple. According to (7.24), we have pseudo-unitarity
$$T_n^k(x) = \tilde{T}_n(x) \qquad \forall n,\quad \big{(} \Longleftrightarrow S(g)^k = S(g)^{-1} \big{)} $$
We want to prove a similar perturbative relation for the restriction of the S-matrix to the physical subspace,
$$P_\bot S(g) P_\bot = S_\bot (g) \eqno (7.68)$$
Its inverse is given by
$$\big{(} P_\bot S(g) P_\bot \big{)}^{-1} = \sum_{n} \frac{1}{n!} \int d^4x_1 \ldots \int d^4x_n \quad \tilde{T}_n^{P_\bot} (x_1, \ldots, x_n) g(x_1)\ldots g(x_n) \eqno (7.69)$$
where the n-point distributions are equal to the following sum over subsets of X $= \{x_1, \ldots, x_n\}$
$$\tilde{T}_n^{P_\bot}(X) = \sum_{r=1}^{n} (-)^r \sum_{Pr} P_\bot T_{n_1}(X_1) P_\bot \ldots P_\bot T_{n_r} (X_r) P_\bot. \eqno (7.70)$$
{\bf Theorem 7.2.} (Physical Unitarity)
$$\tilde {T}_n^{P_\bot} = P_\bot T_n^+ P_\bot + div \quad \forall n \eqno(7.71a)$$

$$\big{(} \Longleftrightarrow \quad (S_\bot)(g)^{-1} =S_\bot^+(g) + div(g) \qquad \mbox{where} \qquad S_\bot = P_\bot SP_\bot \big{)} \eqno(7.71b)$$
\vskip0,5cm
Note that (7.71b) is a statement about a formal power series.\\ 
{\bf Proof:}
$$\tilde {T}_n^{P_\bot}(X) = \sum_{r=1}^{n} (-1) \sum_{P_r} P_\bot T_{n_1}(X_1) P_\bot \ldots P_\bot T_{n_r} (X_r) P_\bot \eqno (7.72)$$
Using Proposition 7.4., all internal projection operators $P_\bot$ can be changed into $P$
$$\tilde {T}_n^{P_\bot}(X) = \sum_{r=1}^{n} (-1)^r \sum_{P_r} P_\bot T_{n_1}(X_1) P \ldots P T_{n_r} (X_r) P_\bot + div_1. \eqno (7.73)$$
With the help of Proposition 7.3. they can be transformed away:
$$\tilde {T}_n^{P_\bot}(X) = \sum_{r=1}^{n} (-1) \sum_{P_r} P_\bot T_{n_1}(x_1) \ldots T_{n_r} (x_r) P_\bot + div_2 \eqno (7.74)$$
$$= P_\bot \tilde{T}_n(x) P_\bot + div_2$$
By means of pseudo-unitarity (7.24), we finally arrive at the desired perturbative unitarity
$$\tilde {T}_n^{P_\bot}(X) = P_\bot T_n(X)^k P_\bot + div_2 = P_\bot T_n(X)^+ P_\bot + div_2 \eqno (7.75)$$
In the last step we have used the fact that the conjugation "k" agrees with the adjoint "+" on $F_\bot$. {\bf q.e.d.}

\vskip1cm
{\Large\bf Appendix A $\quad$ SU(N) - Group Theory}
\vskip 1cm
For the purposes of completeness, we quote the most important group theoretical facts, in particular we list the linearly independent, numerically invariant tensors which transform according to the r-times tensor product of the adjoint representation. We prove their independence.

$\bullet$ Let $T_a(a=1,\ldots ,N^2-1)$ be the $SU(N)$-generators which close a Lie algebra:
$$[T_a,T_b]=i f_{abc} T_c,\quad T_r[T_a]=0, \eqno (A.1)$$
$f_{abc}$ being real and totally antisymmetric, normalized in such a way that
$$f_{abc} f_{dbc}=N \delta_{ad} \eqno (A.2)$$
The adjoint representation is given by
$$(T_a)_{bc}=-i f_{abc} \eqno (A.3)$$
The fundamental (lowest-dimensional cogradient) representation is given by
$$T_a=\frac{\lambda_a}{2} \eqno (A.4)$$
where $\lambda_a$ are hermitian, traceless NxN-matrices generalizing the well-known Gell-Mann matrices of $SU(3)$ and satisfying
$$[\lambda_a,\lambda_b]_-=i2f_{abc}\lambda_c,\quad \{\lambda_a,\lambda_b\}_+=\frac{4}{N}\delta_{ab} {\bf 1}+2 d_{abc} \lambda_c \eqno (A.5)$$
We choose $(\lambda_a)_{a=1}^{N^2-1}$ so that
$$f_{abc} \in {\bf R}, \quad d_{abc} \in {\bf R} \quad \forall_{a,b,c,} \eqno( A.6)$$
$\bullet$ We define an equivalent representation by
$$\lambda'=U\lambda,\quad \lambda'_a=-\lambda_a^*=U_{ab}\lambda_b,\quad U=U^{-1}=U^T. \eqno (A.7)$$
\vskip0,2cm
(A.5) leads to 
\begin{eqnarray*}
\mbox{\hspace{2,3cm}}f'&=&f,\quad f_{a'b'c'}=U_{a'a}U_{b'b}U_{c'c} f_{abc} \mbox{\hspace{4,8cm}(A.8)}\\
d'&=&-d, \quad (-d_{a'b'c'})=U_{a'a}U_{b'b}U_{c'c} d_{abc}
\end{eqnarray*}
$$\mbox{we find} \quad \lambda_a^*=(\lambda_a^T)^+=(\lambda_a^+)^T=\lambda_a^T=\pm \lambda_a,\quad \mbox{therefore}\quad U_{ab}=\pm \delta_{ab}. \eqno (A.9)$$
$\bullet$ Suppose that $D_{ij\ldots k}^{(\alpha)}\big{(}1\le\alpha\le\beta(r)\big{)}$ are $\beta(r)$ numerically invariant tensors of rank $r$, that is, they are tensors which transform according to the r-times tensor product of the adjoint representation of $SU(N)$, i.e.
$$if_{pit}D_{tj \ldots k}^{(\alpha)}+if_{pjt}D_{it\ldots k}^{(\alpha)}+\ldots+if_{pkt}D_{ij\ldots t}^{\alpha}+\ldots=0 \eqno (A.10)$$
and they are also just sets of numbers which are the same in all $SU(N)$ frames. The number $\beta(r)$ of linearly independent, numerically invariant tensors of rank $r$ equals the multiplicity of the one-dimensional representation in the Clebsch-Gordon decomposition of the following tensor representations
$$\bigotimes\limits_{j=1}^r Ad_j(SU(N)).\eqno (A.11)$$
and can be evaluated by actually performing the reduction of this representation by the method of Yang tableaux, but note that there might be additional dependences between these tensors; in fact, there are not any as long as $N > r$ [34].
$$\eqno (A.12)$$
\begin{tabular}{lllllllll}
$N>$ 4:&$\beta$(0)=  1 &$N=4$:& $\beta$(0)=1 & $N=3$:& $\beta$(0)=  1 & $N=2$:& $\beta$(0)=  1\\
           &$\beta$(1)=  0 &     & $\beta$(1)=0 &    &$\beta$(1)=  0 &     & $\beta$(1)=  0\\
           &$\beta$(2)=  1 &     & $\beta$(2)=1 &    &$\beta$(2)=  1 &     & $\beta$(2)=  1\\
           &$\beta$(3)=  2 &     & $\beta$(3)=2 &    &$\beta$(3)=  2 &     & $\beta$(3)=  1\\
           &$\beta$(4)=  9 &     & $\beta$(4)=9 &    &$\beta$(4)=  8 &     & $\beta$(4)=  3\\
           &$\beta$(5)= 44 &     & $\beta$(5)=43&    & $\beta$(5)= 32 &     & $\beta$(5)= 10\\
\end{tabular}\\
\\
We only found $\beta(5)=43$ invariant tensors for $N=4$ and $r=5$ because there is an additional linear dependence for $N=4$ as it is assured by an explicit calculation of the corresponding determinant (see (A.23)).

$\bullet$ In order to prove the independence of a given set of tensors $D_{ij\ldots k}^{(\alpha)}$,  we define
$$Q^{\alpha\beta}=Q^{\beta\alpha}=\sum_{ij \ldots k=1}^{N^2-1}D_{ij \ldots k}^{(\alpha)}D_{ij \ldots k}^{(\beta)}. \eqno (A.13)$$
It is obvious that the $\beta(r)$ tensors $D_{ij \ldots k}^{(\alpha)}$ are linearly independent if and only if $detQ\neq0$. The case $\fbox{r=2}$ is trival: There is only one numerically invariant tensor $\delta_{ab}$.

In the case $\fbox{r=3}$, there are two numerically invariant tensors for $N\ge3$:
$$D_{abc}^1=f_{abc},\quad D_{abc}^2=d_{abc} \eqno (A.14)$$
The equations of invariance (A.10) leads in the case of these two tensors to well-known relations, in particular to the Jakobi identity:
$$f_{ijp}f_{kep} + f_{iep}f_{jkp} + f_{ikp}f_{ejp} = 0, \quad f_{ijp}d_{kep} + f_{iep}d_{jkp} + f_{ikp}d_{ejp} = 0 \eqno (A.15)$$
We can calculate
$$detQ=(N^2-1)^2(N^2-4)\neq 0 \quad \mbox{for}\quad N\ge 3 \eqno (A.16)$$
For $N=2 \quad d_{abc}=0$, therefore $\beta(2)=1$.
In case $\fbox{r=4}$, we know that there are $\beta (4)=9$ independent linear tensors for $N \ge 4$ (A.12). We consider the following set which consists of some of the simplest tensors we can build:
\begin{eqnarray*}
\mbox{\hspace{3,2cm}}D_{abcd}^{(\alpha)} &=&\{\delta_{ab}\delta_{cd},\quad \delta_{ad}\delta_{bc},\quad \delta_{ac}\delta_{bd}, \mbox{\hspace{3,5cm}(A.17)}\\
&&d_{abe} d_{cde},\quad d_{ade} d_{cbe},\quad d_{ace} d_{dbe},\\
&&d_{abe} f_{cde},\quad d_{ade} f_{bce},\quad d_{ace} f_{dbe}\}
\end{eqnarray*}
For the determinant of the corresponding matrix $Q$ (A.13) we get:
$$detQ=(\frac{N^4}{4})(N^2-1)^9(N^2-4)^6(N^2-9)\eqno (A.18)$$ 
Therefore (A.17) is a basis for $N \ge 4$. The case $N=3$ needs an extra consideration. The 9 tensors (A.17) are linearly dependent due to the well-known relation
$$\delta_{ij}\delta_{kl}+\delta_{ik}\delta_{jl}+\delta_{il}\delta_{jk}=3(d_{ijm} d_{kem}+d_{ikm}d_{jlm}+d_{ilm}d_{jkm}) \eqno (A.19)$$
Hence one gets a 8-dimensional basis for $N=3$, if one of the 3 dd-tensors in (A.17) is left out. 
For $N=2 \quad d_{abc}=0$ that means $\beta(4)=3$.
We see from (A.18) that $N=2$ and $3$ are the only exceptional cases.

Case $\fbox{r=5}$ : For the fifth rank tensors the procedure is in principle exactly the same, but much more involved.
Because of (A.8), there is no nontrivial relation between the invariant tensors with an even number of d's and the invariant tensors with an odd number of d's. Therefore we have two types of fifth rank tensors we can independently deal with.

Generalizing the method of Dittner [35] to the general $SU(N)$, we get an ansatz for the independent set of fifth rank invariant tensors by manipulating Jacobi relations like (A.15) (in case $N=3$ also (A.19)). It is not sure that these relations represent the only dependences between fifth rank tensors. Therefore we compute the corresponding determinants again [36].
Our ansatz in the even sector is the following:
\begin{eqnarray*}
\mbox{\hspace{1cm}}D_{abcde}^{(\alpha)(+)} &=&\{\delta_{ab} f_{cde},\quad \delta_{ac} f_{bde},\quad \delta_{ad} f_{bce},\quad \delta_{ae} f_{bcd},\quad \delta_{bc} f_{ade},\mbox{\hspace{1,5cm}(A.20)}\\
&&\delta_{bd} f_{ace},\quad \delta_{be} f_{acd},\quad \delta_{cd} f_{abe},\quad \delta_{ce} f_{abd},\quad \delta_{de} f_{abc},\\
&&d_{abm} d_{cmk} f_{dke},\quad d_{abm} d_{dmk} f_{cke},\quad d_{abm} d_{emk} f_{dkc},\\
&&d_{dem} d_{cmk} f_{akb},\quad d_{cem} d_{dmk} f_{akb},\quad d_{dcm} d_{emk} f_{akb},\\
&&d_{cbm} d_{amk} f_{dke},\quad d_{cbm} d_{dmk} f_{ake},\quad d_{cbm} d_{emk} f_{dka},\\
&&d_{aem} d_{cmk} f_{dkb},\quad d_{aem} d_{dmk} f_{ckb},\quad d_{acm} d_{emk} f_{dkb}\}
\end{eqnarray*}

The corresponding determinant shows that we have an independent set in the even sector for $N \ge 4$:
$$detQ^{(+)}=(\frac{N^{20}}{128})(N^2-1)^{22}(N^2-4)^{16}(N^2-9)^4 \eqno (A.21)$$
For $N=3$ we have to leave out the 16., 18., 19., 20., 21. and 22. tensor in (A.20) in order to arrive at an independent set. 
For $N=2$, we only have the first 10 tensors in (A.20) because $d=0$. They are obviously independent.\\
%Analogously one can show that 22 tensors in the odd sector are independent
%for $N>4$ [36], but these tensors with an odd number of d's are excluded in a nonabelian gauge theory (without $\gamma_5$) because of the Furry theorem in Fock space (see Chapter 5, 5.52). Thus, an explcit representation of an independent set in the even sector is irrelevant for our purposes.

Now we consider the tensors with an odd number of d's. Analogously to the even sector, we get the ansatz:
\begin{eqnarray*}
\mbox{\hspace{1cm}}D_{abcde}^{(\alpha)(-)} &=&\{\delta_{ab} d_{cde},\quad \delta_{ac} d_{bde},\quad \delta_{ad} d_{bce},\quad \delta_{ae} d_{bcd},\quad \delta_{bc} d_{ade},\mbox{\hspace{1,5cm}(A.22)}\\
&&\delta_{bd} d_{ace},\quad \delta_{be} d_{acd},\quad \delta_{cd} d_{abe},\quad \delta_{ce} d_{abd},\quad \delta_{de} d_{abc},\\
&&d_{abm} d_{cmk} d_{dke},\quad d_{abm} d_{dmk} d_{cke},\quad d_{abm} d_{emk} d_{dkc},\\
&&f_{abm} d_{cmk} f_{dke},\quad f_{abm} d_{dmk} f_{cke},\quad f_{abm} d_{emk} f_{dkc},\\
&&d_{cbm} d_{amk} d_{dke},\quad d_{cbm} d_{dmk} d_{ake},\quad d_{dbm} d_{emk} d_{akc},\\
&&f_{cbm} d_{amk} f_{dke},\quad f_{cbm} d_{dmk} f_{ake},\quad f_{dbm} d_{emk} f_{akc}\}
\end{eqnarray*}

In order to prove the linear independence of these 22 tensors for $N > 4$, we calculate the determinant $Q$ (A.13):
$$detQ^{(-)}=(\frac{N^8}{256})(N^2-1)^{22}(N^2-4)^{22}(N^2-9)^5 (N^2 - 16) \eqno (A.23)$$
The factor $(N^2 - 16)$ indicates an additional dependence between the 22 tensors for $N=4$ (see(A.12)). We have to leave out the 19.tensor in (A.22) in order to arrive at an independent set. For $N=3$ we have to leave out the 13., 17., 19., 20.,21. and 22. tensor in (A.22).

\vskip1.0cm

{\Large\bf Appendix B $\quad$ Derivation of the Cg-Identities}
\vskip1cm
We derive all types of the C-number identities which express gauge invariance of the nonabelian theory analogously to the Slavnov-Taylor identities. These numerical Cg-identities are sufficient for the operator gauge invariance (2.12) with the choice $\alpha = 0$ (2.14). We isolate those few Cg-identities which require nontrivial proofs. (This analysis was already carried out in great detail in reference [37].) From this large set of identities we can derive 5 summed identities which can directly compared with the Slavnov-Taylor identities. We explicitly show at one-loop level that our Cg-identities imply the well-known relations between the Z-factors.
\vskip 0.5cm
$\bullet$ {\bf Lorentz-structure of some distributions with one $Q$-vertex}
\vskip 0.5cm
Our convention of denoting operator-valued distributions is the following
$$t_{AB\ldots ab\ldots}^{\alpha 2}(x_1,x_2,\ldots):A^a(x_1)B^b(x_2)\ldots:
\eqno(B.1)$$
means a distribution with external field operators (legs) $A^a$ and
$B^b$, $a$ and $b$ are colour indices. The subscripts $\alpha 2$ show
that this term belongs to $T^\alpha_{n/2}(x_1,x_2,\ldots)$ with 
$Q$-vertex at the second
argument of the numerical distribution $t$.
An immediate consequence of this notation is the relation
$$t_{AB\ldots ab\ldots}^{\alpha 1}(x_1,x_2,\cdots)=\pm t_{BA\ldots ba
\ldots}^{\alpha 2}(x_2,x_1,\cdots),\eqno(B.2)$$
where we have a minus sign, if $A, B$ are both Fermi operators and a
plus sign in all other cases.
The Lorentz structure is preserved in the
process of distribution splitting. For example, if the $d$-distribution
$d^{\mu\nu}\sim g^{\mu\nu}$, the retarded distribution must also be $r^{\mu\nu}\sim g^{\mu\nu}$, or if $d^{\mu\nu}=-d^{\nu\mu}$, we
antisymmetrize a (covariant) splitting solution so that $r^{\mu\nu}=-r^{\nu\mu}$. 
As a consequence, the Lorentz structure of $T_{1/1}^\alpha$ goes over to $T_{n/l}^\alpha$.  If the $Q$-vertex is
$$T_{1/1}^{u\alpha}(x_1)\sim :u(x_1)u(x_1)\d^\alpha\tilde u(x_1):,\eqno(B.3)$$
the terms with $\d^\alpha\tilde u(x_1)$ not contracted have the
following form:
$$T_{n/1}^\alpha(\xnn)=t_{\tilde
uuu\ldots}^{\alpha1\mu}(x_1,x_2,x_3,\ldots): \d_\mu
\tilde u(x_1)u(x_2)u(x_3)\ldots:+\ldots$$
where
$$t_{\tilde uuu\ldots}^{\alpha 1\mu}=g^{\alpha\mu}\bar t^1_{\tilde uuu\ldots} $$
Next we assume that the $Q$-vertex is of the following type
$$T_{1/1}^{A\alpha}(x_2)\sim :A_\mu(x_2)u(x_2)F^{\alpha\mu}(x_2):.
\eqno(B.4)$$
Considering the terms with $F^{\alpha\mu}(x_2)$ or $A_\mu(x_2)$ not contracted, respectively,
$$T_{n/1}(\xnn)=t_{uF\ldots}^{\alpha 2\nu\mu}(x_1,x_2,\ldots) 
:u(x_1)F_{\nu\mu}(x_2)\ldots:$$
$$+t_{uA\ldots}^{\alpha 2\mu}:u(x_1)A_\mu(x_2)\ldots:+\ldots\eqno(B.5)$$
we obtain by $F^{\alpha\mu}=-F^{\mu\alpha}$
$$t_{uF\ldots}^{\alpha 2\nu\mu}=-g^{\alpha\nu}\bar t_{uF\ldots}
^{2\mu}+g^{\alpha\mu}\bar t_{uF\ldots}^{2\nu},\quad t_{uA\ldots}^{\alpha 2\mu}=-t_{uA\ldots}^{\mu 2\alpha}.\eqno(B.6)$$
\vskip 0.5cm
$\bullet$ {\bf Degenerate Diagrams}
\vskip 0.5cm
The distributions $D_n,D_{n/l},R_n,R_{n/l},A_n,A_{n/l}$ contain {\it connected} diagrams only, due to their causal supports. The disconnected diagrams appear in $R'_n,R'_{n/l},A'_n,A'_{n/l}$ and are, therefore, automatically gauge invariant due to gauge invariance in lower orders.

For 2-leg diagrams the external legs must be attached to {\it different} vertices. This is not the case, if we consider diagrams with 3 or more legs. We call a connected diagram {\it degenerate}, if it has at least one vertex with two external legs; otherwise the connected diagram is called {\it nondegenerate}. Let $x$ be the degenerate vertex with two external fields, say $B_1$, $B_2$. Then, due to its causal support, the $D_n$-distribution is given by
$$D_n(x,y,z,\ldots ;x_n)=:B_1^a(x)B_2^b(x)B_3^c(z)\ldots :f_{abd}$$
$$[\lap_\r(x-y)r_{n-1}^{dc\ldots}(y-x_n,z-x_n,\ldots)-\lap_\a(x-y) 
a_{n-1}^{dc\ldots}(y-x_n,z-x_n,\ldots)]+\ldots\eqno(B.7)$$
where $\lap_{\r /\a}=D_{\r /\a}, \d D_{\r /\a}$, or (B.9) below, and $r_{n-1}$, $a_{n-1}$ are the retarded and advanced distributions of the subdiagram of order $n-1$, which contain the vertices $y,z,\ldots x_n$.

Following the inductive construction, it is easy to prove that
$$[Q,\,D_n]=\sum_{l=1}^n\d_{\alpha}^lD_{n/l}^\alpha,\eqno(B.8)$$
by using gauge invariance in lower orders. Collecting all terms in
(B.8) with a certain field operator combination $:\Omega:$,  degenerate ($:\Omega:=:B_1(x)B_2(x)B_3(z)\ldots:$) and nondegenerate diagrams ($:\Omega:=B_1(x)B_2(y)B_3(z)\ldots :$) cannot get mixed up, except the degenerate diagram contains a factor $\delta(x-y)$. Terms with derivatives of $\delta(x-y)$ do not appear (see above). If there is such a factor $\delta(x-y)$ in a degenerate term, we call it $\delta$-degenerate, and if there is no such $\delta(x-y)$, we call it truly degenerate. A $\delta$-degenerate term in $[Q,\,D_n]$ must be $\delta$-degenerate already in $D_n$. The $\delta(x-y)$ originates from the 4-gluon interaction, i.e. the normalization term $-\eh g^{\mu\nu}
\delta(x-y)$ of the propagator:
$$\lap_{\r /\a}(x-y)=\d^\mu\d^\nu D_{\r /\a}(x-y)-\eh g^{\mu\nu}\delta (x-y).\eqno(B.9)$$
This normalization, which is fixed in second order by gauge invariance (see Chapter 3(b)), goes over to tree-like diagrams in higher orders (see also Sect.4b of [38]). A $\delta$-degenerate term in $\sum_l\d_\alpha^l D_{n/l}^\alpha$ is either $\delta$-degenerate already in $D_{n/l}^\alpha$, due to the 4-gluon interaction (second term in (B.9)), or the $\delta$-distribution is generated by the divergence $\d_\alpha^l$
$$d^{\alpha l}(x,y,\ldots)\sim\d^\alpha D_{\r/\a}(x-y)\to\cases{
\d_\alpha^x d^{\alpha l}(x,y\ldots)\sim\delta(x-y)+\ldots
\quad{\rm(B.10a)}\cr\d_\alpha^y d^{\alpha l}(x,y\ldots)\sim
-\delta(x-y)+\ldots\quad{\rm (B.10b)}\cr}$$
In (B.10a) the $Q$-vertex is $x_l=x$ and in (B.10b) it is $x_l=y$.
Terms with $\d^\mu\delta(x-y)$ are produced in $\sum_l\d_\alpha^lD_{n/l} ^\alpha$ in two different ways (let $x_l=x$)
$$d^{\alpha l\mu}(x,y,\ldots)\sim\d^\mu\d^\alpha D_{\r/\a}(x-y)\to
\d_\alpha^xd^{\alpha l\mu}\sim\d^\mu\delta(x-y)+\ldots\eqno(B.11)$$
$$d^{\alpha l\mu}(x,y,\ldots)\sim g^{\mu\alpha}\delta(x-y)
\quad(\hbox{4-gluon interaction})\quad \to\d_\alpha^x d^{\alpha l}\sim\d^\mu\delta(x-y).\eqno(B.12)$$
One can easily show that these two types of $\d\delta$-terms cancel.\\  
Moreover, one can show, by means of the Cg-identites for $t_{n-1}(y-x_n,
z-x_n,\ldots)$ or

$r_{n-1}(y-x_n,z-x_n,\ldots)$, respectively that the truly degenerate terms also cancel out (see [37]). The $\delta$-degenerate terms, however, do not cancel. Therefore, the latter have to be included in the nondegenerate Cg-identities. This must be done in the following symmetric way:
$$:B_1(x)B_2(x)B_3(z)\ldots:\delta(x-y)=\eh :B_1(x)B_2(y)B_3(z)
\ldots:\delta(x-y)\ldots$$
$$+\eh:B_1(y)B_2(x)B_3(z)\ldots:\delta(x-y)\ldots\eqno(B.13)$$
Otherwise, we would get a contradiction between the Cg-identites
corresponding to $:\Omega:$ $=$ $:B_1(x)B_2(y)B_3(z)\ldots:$ and
$:\Omega:=:B_1(y)B_2(x)B_3(z)\ldots:$. These two identites must be
identical, up to exchange of $x$ and $y$.

\vskip 0.5cm
$\bullet$ {\bf The different types of Cg-identities}
\vskip 0,5cm
In the various diagrams contributing to $T_n$ and $T_{n/l}^\alpha$, we have the basic external field operators $A, F, u$ and $\d\tilde u$. Going over to $[Q,\,T_n]$, we get one external field operator $\d u$ or $\d F$ in each nonvanishing diagram. In $\sum\d_\alpha^l T_{n/l}^\alpha$, the derivative may act on the numerical distribution or on an external field operator. In the first case all external field operators are basic ones appearing also in the specific coupling $T_1$ i.e. $A, F, u$ or $\d\tilde u$. In the second case the term contains one of the following nonbasic external field operators:
$\d u$, $\d \d \tilde{u}, \d F$ or $\d A = {1\over 2}F+(\d A)_s$

After the operator decomposition of
$$[Q,\,T_n]=i\sum_{l=1}^n\d_\alpha^lT_{n/l}^\alpha,\eqno(B.14)$$
according to our convention, (6.7),(6.8) , we collect all terms with a particular combination of field operators $:\Omega$. Then
we get the following types of Cg-identities:

{\it Type I}: $\Omega$ contains one nonbasic field operator (i.e. $\d u, \d F, \d\d\tilde u$ or $(\d A)_s$). Then the derivative in (B.14) must act on a field operator. 

{\it Type II}: $\Omega$ consists of basic field operators only. Then the derivative in (B.14) acts on the numerical distribution, with the following exception: If $\Omega$ contains an operator $F(x_i)$, then the terms with $\eh F_{\alpha\mu}(x_i)$ in (B.14) must also be included.

In proving gauge invariance (B.14), the nontrivial step is to show that the Cg-identities can be preserved in the process of distribution splitting. In this operation they can be spoiled by terms with local support only.

The Cg-identities of type I are either identifications of numerical distributions of the theory with one $Q$-vertex with numerical distributions of the physical theory, and can easily be preserved in the process of distribution splitting because we
are free to normalize the extended theory properly;
or they concern the Lorentz structure only. In the latter case they hold true if the Lorentz structure is preserved in the process of distribution splitting, which is always assumed.

Therefore the Cg-identities of type II are the only ones which require nontrivial proofs. We consider a certain Cg-identity of type II which belongs to a certain combination of external field operators. Let
$$\kappa \=d 4-b-g_{\tilde{u}}-g_u-d-3f \eqno(B.15)$$
where $b, g_u, g_{\tilde u}, d, f$ are determined by the field operator combination as in (4.1). The divergence derivative $\partial_\mu^{l}$ in (B.14) operates on the numerical distribution in this case (type II) with the trivial, above mentioned exception. Thus, the derivated numerical distribution has a singular order $\omega \le \kappa + 1$. We conclude that the anomaly, i.e. the possible violation of the considered Cg-identity, is a polynomial of degree at most $\kappa + 1$ in the partial derivatives of $\delta^{n-1}(x)$. Therefore, only Cg-identities of type II which belong to field operator combinations with
$$\kappa \ge -1 \eqno(B.16)$$
need a nontrivial proof. Hence, Cg-identities with more than 5-legs, all 5-leg identities except one and even some 4-leg-Cg-identities are automatically fulfilled.

In the following we list all 2-, 3-, 4-leg-cg-identities of type II. The trivially fulfilled identities of type I are identifications of numerical distributions of the theory with one $Q$-vertex with numerical distributions of the physical theory or they concern the Lorentz structure of certain distributions. We insert these trivial type I-identities in the below listed type II-identities in order to eliminate the distributions with one $Q$-vertex as far as possible.
\vskip 0.5cm
$\bullet$ {\bf The Cg-identities for 2-leg distributions}
\vskip0,5cm
All 2-leg distributions contain the colour tensor $\delta_{ab}$ (see Appendix A). Therefore we define the numerical distributions without this factor $\delta_{ab}$. We list all nontrivial 2-leg-Cg-identities according to the above analysis:

For $:\Omega:=\delta_{ab}:u_a(x_1)A_\nu^b(x_2):$ with $\kappa=2 \ge -1$ we obtain
$$\d_\alpha^1 t_{AA}^{\alpha\nu}+{1\over 2}\d_\alpha^2
[t_{uA}^{\alpha 2\nu}-t_{uA}^{\nu 2\alpha}]
+\sum_{l=3}^n\d_\alpha^l t_{uA}^{\alpha l\nu}=0,\eqno(B.17)$$

For $:\Omega:=\delta_{ab}:u_a(x_1) F_{\mu \nu}^b (x_2):$ with $\kappa=1\ge(-1)$
$$\d_\alpha^1 t_{AF}^{\alpha\mu\nu}+{1\over 2}[\d_2^\mu
t_{u\tilde u}^{\nu}-\d_2^\nu t_{u\tilde u}^\mu]
+{1\over 4}[t_{uA}^{\mu 2\nu}-t_{uA}^{\nu 2\mu}]+
\sum_{l=3}^n\d_\alpha^l t_{uF}^{\alpha l\mu\nu}=0.\eqno(B.18)$$

\vskip0,5cm
$\bullet$ {\bf The Cg-identities for 3-leg distributions}
\vskip0,5cm
Since all 3-leg distributions contain the colour tensor $f_{abc}$, the numerical distribution are defined without this factor $f_{abc}$. 

For $:\Omega:=f_{abc}:u^a(x_1)u^b(x_2)\d_\mu\tilde u
^c(x_3):$ with $\kappa=0\ge -1$, we obtain
$$0=\d_\alpha^{x_1}t_{Au\tilde u}^{\alpha\mu}(x_1,x_2,x_3,\ldots)+
\d_\alpha^{x_2}t_{Au\tilde u}^{\alpha\mu}(x_2,x_1,x_3,\ldots)+\d_{x_3}
^\mu\bar t_{uu\tilde u}^3(x_1,x_2,x_3,\ldots)$$
$$+\sum_{l=4}^n\d_\alpha^{x_l}t_{uu\tilde u}^{\alpha l\mu}(x_1,x_2,x_3, \ldots)+g\delta(x_1-x_2)t_{u\tilde u}^\mu(x_2,x_3,\ldots)$$
$$-g\delta(x_1-x_3)t_{u\tilde u}^\mu(x_2,x_3,\ldots)-g\delta(x_2-x_3) t_{u\tilde u}^\mu(x_1,x_3,\ldots).\eqno(B.19)$$
For $:\Omega:=f_{abc}:u^a(x_1)A_\mu^b(x_2)A_\nu^c(x_3):$ with $\kappa=1$, we get
$$0=\d_\alpha^{x_1}t_{AAA}^{\alpha\mu\nu}(x_1,x_2,x_3,x_4,\ldots)-
{1\over 2}\d_\alpha^{x_2}\Bigl[t_{uAA}^{\alpha 3\nu\mu}(x_1,x_3,x_2,x_4, \ldots)-(\alpha\leftrightarrow\mu)\Bigl]$$
$$+{1\over 2}\d_\alpha^{x_3}\Bigl[t_{uAA}^{\alpha 3\mu\nu}(x_1,x_2,x_3, x_4,\ldots)-(\alpha\leftrightarrow\nu)\Bigl]+\sum_{l=4}^n\d_\alpha^{x_l}
t_{uAA}^{\alpha l\mu\nu}(x_1,x_2,x_3,x_4,\ldots)$$
$$+g[\delta(x_1-x_2)-\delta(x_1-x_3)]t_{AA}^{\mu\nu}(x_2,x_3,x_4,\ldots)$$
$$-g\delta(x_2-x_3){1\over 2}\Bigl[t_{uA}^{\mu 2\nu}(x_1,x_2,x_4,\ldots)-(\mu\leftrightarrow\nu)\Bigl].\eqno(B.20)$$
For $:\Omega:=f_{abc}:u^a(x_1)A_\mu^b(x_2)F_{\nu\lambda}
^c(x_3):$, with $\kappa=0$, we obtain
$$0={1\over 4}\Bigl[t_{uAA}^{\nu 3\mu\lambda}(x_1,x_2,x_3,\ldots)-(\nu \leftrightarrow\lambda)\Bigl]$$
$$+\d_\alpha^{x_1}t_{AAF}^{\alpha\mu\nu\lambda}(x_1,x_2,x_3,\ldots)+{1\over 2}\d_\alpha^{x_2}\Bigl[t_{uAF}^{\alpha
2\mu\nu\lambda}(x_1,x_2,x_3,\ldots)-(\alpha\leftrightarrow\mu)\Bigl]$$
$$+{1\over 2}\Bigl[\d_{x_3}^\nu t_{u\tilde uA}^{\lambda\mu}(x_1,x_3,x_2, \ldots)-(\nu\leftrightarrow\lambda)\Bigl]+\sum_{l=4}^n\d_\alpha^{x_l}t_{uAF}^{\alpha l\mu\nu\lambda}(x_1, x_2,x_3,\ldots)$$
$$+g[\delta(x_1-x_2)-\delta(x_1-x_3)]t_{AF}^{\mu\nu\lambda}(x_2,x_3,x_4,\ldots)$$
$$+{g\over 2}\Bigl[g^{\mu\nu}\delta(x_2-x_3)t_{u\tilde u}^\lambda(x_1, x_3,x_4,\ldots)-(\nu\leftrightarrow\lambda)\Bigl].\eqno(B.21)$$
For $:\Omega:=f_{abc}:u^a(x_1)F_{\mu\tau}^b(x_2)F_{\nu\lambda}^c(x_3):$ with $\kappa=-1$, we obtain
$$0={1\over 4}\Bigl[t_{uAF}^{\mu 2\tau\nu\lambda}(x_1,x_2,x_3\ldots)-(\mu\leftrightarrow\tau)\Bigl]-{1\over 4}\Bigl[t_{uAF}^{\nu 2\lambda \mu\tau}(x_1,x_3,x_2\ldots)-(\nu\leftrightarrow\lambda)\Bigl]$$
$$+\d_\alpha^{x_1}t_{AFF}^{\alpha\mu\tau\nu\lambda}(x_1,x_2,x_3,\ldots)+{1\over 2}\Bigl[\d_{x_2}^\tau t_{u\tilde u F}^{\mu\nu\lambda}(x_1,x_2,x_3\ldots)-(\mu\leftrightarrow\tau)\Bigl]$$
$$-{1\over 2}\Bigl[\d_{x_3}^\lambda t_{u\tilde uF}^{\nu\mu\tau}(x_1, x_3,x_2\ldots)-(\lambda\leftrightarrow\nu)\Bigl]$$
$$+\sum_{l=4}^n\d_\alpha^{x_l}t_{uFF}^{\alpha l\mu\tau\nu\lambda}(x_1,x_2,x_3\ldots)$$
$$+g[\delta(x_1-x_2)-\delta(x_1-x_3)]t_{FF}^{\mu\tau\nu\lambda}(x_2,
x_3,x_4\ldots).\eqno(B.22)$$

\vskip 0.5cm
$\bullet$ {\bf The Cg-identities for 4-leg distributions}
\vskip 0.5cm
We now write down all types-II Cg-identities for 4-leg distributions.

We obtain for $:\Omega:=:u^a(x_1) A_\nu^b(x_2) A_\kappa^c(x_3)$ $A_\lambda^d(x_4):$ with $\kappa=0$ using the abbreviation $l$ for $x_l$ in the arguments:
$$0 = \d_\alpha^{x_1}t_{AAAAabcd}^{\alpha\nu\kappa\lambda}(1,2,3,4,5,\ldots)+\eh\d_\alpha^{x_2}[t_{uAAAabcd}^{\alpha 2\nu\kappa\lambda}(1,2,3,4,5, \ldots)-(\alpha\leftrightarrow\nu)]$$
$$+\eh\d_\alpha^{x_3}[t_{uAAAacbd}^{\alpha 2\kappa\nu\lambda}(1,3,2,4,5, \ldots)-(\alpha\leftrightarrow\kappa)]+\eh\d_\alpha^{x_4}[t_{uAAAadbc}^ {\alpha 2\lambda\nu\kappa}(1,4,2,3,5,\ldots)-(\alpha\leftrightarrow\lambda)]$$
$$+\sum_{l=5}^n\d_\alpha^{x_l}t_{uAAAabcd}^{\alpha l\nu\kappa\lambda}(1,2,3,4,5, \ldots)+$$ 
$$+\Bigl\{gf_{abr}f_{cdr}\Bigl[\delta(1-2)t_{AAA}^{\nu\kappa\lambda} (2,3,4,5,\ldots)+\delta(3-4)\eh
[t_{uAA}^{\kappa 2\lambda\nu}(1,3,2,5,\ldots)-(\kappa\leftrightarrow \lambda)]\Bigl]\Bigl\}$$
$$+\Bigl\{(b,\nu, x_2)\to (c,\kappa,x_3)\to (d,\lambda,x_4)\to (b,\nu,x_2) \Bigl\},\eqno(B.23)$$
The invariance of $:\Omega:$ under permutations of $\{(b,\nu,x_2), (c,\kappa,x_3), (d,\lambda,x_4)\}$ is manifestly realized in (B.23).

For $:\Omega:=:u^a(x_1)u^b(x_2)\d_\mu\tilde u^c(x_3)A_\nu^d(x_4):$ with $\kappa=-1$, we obtain the following Cg-identity:
$$0= - \Bigl[\d_\alpha^{x_2}t_{uA\tilde uAabcd}^{\alpha\mu\nu}(1,2,3,4,5,\ldots) -\Bigl((a,x_1)\lra (b,x_2)\Bigl)\Bigl]$$
$$+\d_{x_3}^\mu\bar t_{uu\tilde uAabcd}^{3\nu}(1,2,3,4,5,\ldots)
+\d_\alpha^{x_4}\eh\Bigl[t_{uu\tilde uAabcd}^{\alpha 4\mu\nu}(1,2,3,4,5, \ldots)-(\alpha\leftrightarrow\nu)\Bigl]$$ 
$$+\sum_{l=5}^n\d_\alpha^{x_l}t_{uu\tilde uAabcd}^{\alpha l\mu\nu}(1,2,3,4,5, \ldots)$$
$$+g\{f_{adr}f_{bcr}[\delta(1-4)t_{Au\tilde u}^{\nu\mu}(4,2,3,5\ldots)$$
$$+\delta(2-3)t_{Au\tilde u}^{\nu\mu}(4,1,3,5\ldots)]\}-g\{(a,x_1)\lra (b,x_2)\}$$
$$+gf_{abr}f_{cdr}[\delta(1-2)t_{Au\tilde u}^{\nu\mu}(4,2,3,5\ldots)$$
$$+\delta(3-4)g^{\nu\mu}\bar t_{uu\tilde u}^3(1,2,4,5\ldots)],
\eqno(B.24)$$

For $:\Omega:=:u^a(x_1)F_{\kappa\lambda}^b(x_2)A_\mu ^c(x_3)A_\nu^d(x_4):$ with $\kappa=-1$, we obtain the following Cg-identity:
$$0=\d_\alpha^{x_1}t_{AFAAabcd}^{\alpha\kappa\lambda\mu\nu}(1,2,3,4,5,\ldots) +\eh[\d_{x_2}^\lambda t_{u\tilde uAAabcd}^{\kappa\mu\nu}(1,2,3,4,5,\ldots) -(\lambda\lra\kappa)]$$
$$+\eh\d_\alpha^{x_3}[t_{uFAAabcd}^{\alpha 3\kappa\lambda\mu\nu} (1,2,3,4,5,\ldots)-(\alpha\lra\mu)]$$
$$+\eh\d_\alpha^{x_4}[t_{uFAAabdc}^{\alpha 3\kappa\lambda\nu\mu}(1,2,4,3,5, \ldots)-(\alpha\leftrightarrow\nu)] +\sum_{l=5}^n\d_\alpha^{x_l}t_{uFAAabcd}^{\alpha l\kappa\lambda\mu\nu} (1,2,3,4,5,\ldots)$$
$$+{\scriptstyle{1\over 4}}[t_{uAAAabcd}^{\kappa 2\lambda\mu\nu}(1,2,3,4,5, \ldots)-(\kappa\leftrightarrow\lambda)]$$
$$+gf_{abr}f_{cdr}\delta(3-4)\eh [t_{uAF}^{\mu 2\nu\kappa\lambda} (1,3,2,5\ldots)-(\mu\lra\nu)]$$
$$+{g\over 2}\{f_{adr}f_{bcr}\delta(2-3)[g^{\mu\kappa}t_{u\tilde uA} ^{\lambda\nu}(1,3,4,5\ldots)-(\kappa\lra\lambda)]\}+{g\over 2}\{(c,\mu, 3)\lra (d,\nu, 4)\}$$
$$+g[f_{acr}f_{dbr}\delta(1-3)-f_{adr}f_{cbr}\delta(1-4)]t_{AAF}^ {\mu\nu\kappa\lambda}(3,4,2,5\ldots)$$
$$+gf_{abr}f_{cdr}\delta(1-2)t_{AAF}^{\mu\nu\kappa\lambda}(3,4,2,5\ldots)],\eqno(B.25)$$

For $:\Omega:=:u^a(x_1)A_\mu^b(x_2)F_{\kappa\lambda}^c(x_3)F_{\sigma 
\tau}^d(x_4):$ with $\kappa=-2<-1$(!), we obtain the following Cg-identity:
$$0=\d_\alpha^{x_1}t_{AAFFabcd}^{\alpha\mu\kappa\lambda\sigma\tau} (1,2,3,4,5,\ldots)+\eh\d^{x_2}_\alpha[t_{uAFFabcd}^{\alpha 2\mu\kappa \lambda\sigma\tau}(1,2,3,4,5,\ldots)-(\alpha\lra\mu)]$$
$$+\eh\{\d^\lambda_{x_3}t_{uA\tilde uFabcd}^{\mu\kappa \tau\rho}(1,2,3,4,5,\ldots)-(\kappa\lra\lambda)]\}$$
$$+\eh\{(c,\kappa,\lambda,x_3)\lra (d,\sigma,\tau,x_4)\} +\sum_{l=5}^n\d_\alpha^{x_l}t_{uAFFabcd}^{\alpha l\mu\kappa\lambda\sigma \tau}(1,2,3,4,5,\ldots)$$
$$+{\scriptstyle{1\over 4}}[t_{uAAFabcd}^{\kappa 3\mu\lambda\sigma\tau} (1,2,3,4,5,\ldots)-(\kappa\leftrightarrow\lambda)]+{\scriptstyle{1\over 4}} [(c,\kappa,\lambda,x_3)\lra (d,\sigma,\tau,x_4)]$$
$$+g\{f_{abr}f_{cdr}\delta(1-2)t_{AAF}^{\mu\kappa\lambda\sigma\tau} (2,3,4,5\ldots)$$
$$-{g\over 2}\{f_{adr}f_{bcr}\delta(2-3)[g^{\mu\kappa}t_{u\tilde uF} ^{\lambda\sigma\tau}(1,3,4,5\ldots)-(\kappa\lra\lambda)]\} $$
$$-{g\over 2}\{(c,\kappa,\lambda,x_3)\lra (d,\sigma,\tau,x_4)\}$$
$$+g[f_{adr}f_{bcr}\delta(1-4)-f_{acr}f_{bdr}\delta(1-3)]t_{AFF}^ {\mu\kappa\lambda\sigma\tau}(2,3,4,5\ldots),\eqno(B.26)$$

For $:\Omega:=:u^a(x_1)F_{\mu\nu}^b(x_2)F_{\kappa\lambda} 
^c(x_3)F_{\sigma\tau}^d(x_4):$ with $\kappa=-3<-1$(!), we obtain the following Cg-identity:
$$0=\d_\alpha^{x_1}t_{AFFFabcd}^{\alpha\mu\nu\kappa\lambda\sigma\tau} (1,2,3,4,5,\ldots)+\eh[\d_{x_2}^\nu t_{u\tilde uFFabcd}^{\mu\kappa \lambda\sigma\tau}(1,2,3,4,5,\ldots)-(\mu\lra\nu)]$$
$$+\eh[\d_{x_3}^\lambda t_{u\tilde u FFacbd}^{\kappa\mu\nu\sigma\tau} (1,3,2,4,5,\ldots)-(\kappa\lra\lambda)]$$
$$+\eh[\d_{x_4}^\tau t_{u\tilde u FFadbc}^{\sigma\mu\nu\kappa\lambda} (1,4,2,3,5,\ldots)-(\sigma\leftrightarrow\tau)] +\sum_{l=5}^n\d_\alpha^{x_l}t_{uFFFabcd}^{\alpha l\mu\nu\kappa\lambda \sigma\tau}(1,2,3,4,5,\ldots)$$
$$+{\scriptstyle{1\over 4}}[t_{uAFFabcd}^{\mu 2\nu\kappa\lambda\sigma\tau} (1,2,3,4,5,\ldots)-(\mu\leftrightarrow\nu)] +{\scriptstyle{1\over 4}}[t_{uAFFacbd}^{\kappa 2\lambda\mu\nu \sigma\tau} (1,3,2,4,5,\ldots)-(\kappa\leftrightarrow\lambda)]$$
$$+{\scriptstyle{1\over 4}}[t_{uAFFadbc}^{\sigma 2\tau\mu\nu\kappa\lambda} (1,4,2,3,5, \ldots)-(\sigma\leftrightarrow\tau)]$$
$$-g[f_{acr}f_{bdr}\delta(1-3)+f_{adr}f_{cbr}\delta(1-4) +f_{abr}f_{dcr}\delta(1-2)]t_{FFF}^{\mu\nu\kappa\lambda\sigma\tau} (2,3,4,5\ldots),\eqno(B.27)$$

For $:\Omega:=:u^a(x_1)u^b(x_2)\d_\mu\tilde u^c(x_3)F_{\lambda\kappa}^ d(x_4):$ with $\kappa=-2<-1$(!), we obtain the following Cg-identity:
$$0=\Bigl[\d_\alpha^{x_1}t_{Au\tilde uFabcd}^{\alpha\mu\lambda\kappa} (1,2,3,4,5,\ldots)-\Bigl((a,x_1)\lra (b,x_2)\Bigl)\Bigl]$$
$$+\d_{x_3}^\mu\bar t_{uu\tilde uFabcd}^{3\lambda\kappa}(1,2,3,4,5,\ldots) +\eh\Bigl[\d_{x_4}^\kappa t_{uu\tilde u\tilde u abcd}^{\mu\lambda} (1,2,3,4,5,\ldots)-(\kappa\leftrightarrow\lambda)\Bigl]$$ 
$$+\sum_{l=5}^n\d_\alpha^{x_l}t_{uu\tilde uFabcd}^{\alpha l\mu\lambda \kappa}(1,2,3,4,5,\ldots)$$
$$+{\scriptstyle{1\over 4}}[t_{uu\tilde uAabcd}^{\lambda 4\mu\kappa} (1,2,3,4,5,\ldots)-(\lambda\leftrightarrow\kappa)]$$
$$+gf_{abr}f_{cdr}\delta(1-2)t_{u\tilde uF}^{\mu\lambda\kappa} (2,3,4,5\ldots)$$
$$-g[f_{acr}f_{bdr}\delta(1-3)t_{u\tilde uF}^{\mu\lambda\kappa}(2,3,4,5 \ldots)-((a,x_1)\lra (b,x_2))]$$
$$+g[f_{adr}f_{bcr}\delta(1-4)t_{u\tilde uF}^{\mu\lambda\kappa}(2,3,4,5 \ldots)-((a,x_1)\lra (b,x_2))],\eqno(B.28)$$

$\bullet$ Summing up,

besides the one nontrivial $(\kappa \le -1)$ 5-leg-Cg-identity corresponding to the operator $:\Omega: = \, :u_a A_\delta ^b A_\mu ^c A_\kappa ^d A_\lambda ^e:$, we have isolated two 2-leg-Cg-identities (B.17), (B.18), four 3-leg Cg-identities (B19), (B20), (B21), (B22) and three 4-leg Cg-identities, (B23), (B24), (B25) which need a nontrivial proof. Their proof is given in Chapter 6. The last three 4-leg Cg-identities on our list (B26),(B27),(B28) have $\kappa < -1$ and are therefore automatically fulfilled.\\
\\
Furthermore, there are two identities with external fermionic matter fields with $\kappa \ge (-1)$, analogously to the abelian theory (see [26], case III and case IV) : \\
\\
For $:\Omega :=:\overline{\psi}_\alpha(x_i)  \psi_\beta(x_j): u^a(x_n)$ with $\kappa =0$, we obtain

$$ \sum_{1\le i,j \le n-1;i\not= j} \quad :\overline{\psi}_\alpha(x_i) \, \big{[} \quad \sum^n_{l=1} \d_\nu^l t_{\overline{\psi}\psi u}^{\nu l\alpha\beta a}(x_i,x_j, \ldots ,x_n) \, + \, i m \, \overline{t}^{1\alpha\beta a}_{\overline{\psi}\psi u}(x_i,x_j, \ldots ,x_n)$$
$$ - i m \, \overline{t}^{2\alpha\beta a}_{\overline{\psi}\psi u}(x_i,x_j, \ldots  ,x_n) \, + \, degenerate \, terms \quad \big{]} \, \psi_\beta(x_j): \quad = \,0. \eqno (B.29)$$
\\
\\
For $:\Omega:=:\overline{\psi}_\alpha(x_i)\psi_\beta(x_j): A^a_\mu(x_{n-1}) u^b(x_n)$ with $\kappa =-1$, we obtain
$$ \sum_{1 \le i,j \le n-2;i\not= j} \quad :\overline{\psi}_\alpha(x_i) \, \big{[} \quad \sum^n_{l=1} \d_\nu^l t^{\nu l\mu\alpha\beta ab}_{\overline{\psi}\psi Au}(x_i,x_j, \ldots ,x_{n-1},x_n)  \, + \, i m \, \overline{t}^{1\mu\alpha\beta ab}_{\overline{\psi}\psi Au}(x_i,x_j, \ldots ,x_{n-1},x_n)$$
$$ - i m \, \overline{t}^{2\mu\alpha\beta ab}_{\overline{\psi}\psi Au}(x_i,x_j, \ldots ,x_{n-1},x_n)  \, + \, degenerate \, terms \quad \big{]} \, \psi_\beta(x_j): \, =\,0. \eqno(B.30)$$

The notation  $\overline{t}$  is defined by the following equations:
$$t^{1\nu} = \gamma^\nu \overline{t}^1, \quad t^{2\nu} = \overline{t}^2 \gamma^\nu \eqno(B.31$$ 
These equations hold because the terms in (B.29) and (B.30) multiplied by a mass correspond to the vertex $T^{matter}_{1/1}$ (see (2.11)).

\vskip0,5cm
$\bullet$ {\bf The summed Cg-identities } \vskip0,5cm
The derived Cg-identities are sufficient for the operator gauge invariance with the choice $\alpha =0$ in (2.14). $\alpha =0$ corresponds to a special choice of the $Q$-vertex. One can eliminate all distributions with one $Q$-vertex besides the divergences in regard to the inner variables. One arrives at relations which almost only involve distributions of the orginal theory: In order to get the summed 2-leg identity, one has to insert (B.18) into (B.17). Besides the 3-leg identity (B.19) one attends another summed (3-leg) identity  by inserting (B.22) into (B.21), then (B.21) into (B.20). The 4-leg identities are treated analogously: Inserting (B.28) into (B.24), we get the first summed 4-leg identity and inserting (B.27) into (B.26), then (B.26) into (B.25) and finally (B.25) into (B.23), we arrive at the second summed 4-leg identity (see [24]).\\
These 5 (summed) identities can be directly compared with the Slavnov-Taylor identities. This will be done in detail in a forthcoming paper [41]. \\
\\
However, one  can immediately  derive for example the well-known relation between the Z-factors of the gluon vertex, the gluon propagator, the ghost vertex and the ghost propagator at one-loop level from these summed Cg-identities:\\
$\bullet$ In the first step we define the following summed 2-leg and 3-leg distributions (The dots $(\ldots)$ stand for the coordinates $(x_3,\ldots,x_{n-1})$ in the first two definitions and for the coordinates $(x_4,\ldots,x_n)$ in all the other definitions):
$$\Pi^{\kappa\nu}_{AA}(x_1,x_2,\ldots):= t_{AA}^{\kappa\nu}(x_1,x_2,\ldots)+$$ $$-2\d_{\lambda}^{x_2}t^{\kappa\lambda\nu}_{AF}(x_1,x_2,\ldots)-2\d_{\lambda}^{x_1}t^{\lambda\kappa\nu}_{FA}(x_1,x_2,\ldots)+4\d_{\lambda}^{x_1}\d_{\tau}^{x_2}t^{\lambda\kappa\tau\nu}_{FF}(x_1,x_2,\ldots) $$
$$\Pi^{\kappa l \nu}_{uA}(x_1,x_2,\ldots):= t^{\kappa l \nu}_{uA}(x_1,x_2,\ldots)-2\d^{x_2}_{\lambda}t^{\kappa l\lambda\nu}_{uF}(x_1,x_2,\ldots)$$
$$\Pi^{\mu\nu}_{u\overline{u}A}(x_1,x_2,x_3,\ldots):= t^{\mu\nu}_{u\overline{u}A}(x_1,x_2,x_3,\ldots) - 2 \d_{\kappa}^{x_3} t^{\mu\kappa\nu}_{u\overline{u}F}(x_1,x_2,x_3,\ldots)$$
$$\Pi^{\alpha\mu\nu}_{AAA}(x_1,x_2,x_3,\ldots):= t^{\alpha\mu\nu}_{AAA}(x_1,x_2,x_3,\ldots)+$$ $$+2 \delta(x_1-x_2)t^{\nu\alpha\mu}_{AF}(x_3,x_2,\ldots)-2 \delta(x_1-x_3)t^{\mu\alpha\nu}_{AF}(x_2,x_3,\ldots)+2 \delta(x_2-x_3)t^{\alpha\mu\nu}_{AF}(x_1,x_2,\ldots)+$$
$$-2\d^{x_1}_{\kappa} [ t^{\kappa\alpha\mu\nu}_{FAA}(x_1,x_2,x_3,\ldots)+2\delta(x_2-x_3)t^{\mu\nu\kappa\alpha}_{FF}(x_3,x_1,\ldots) ]+$$
$$-2\d^{x_2}_{\kappa} [ t^{\alpha\kappa\mu\nu}_{AFA}(x_1,x_2,x_3,\ldots)-2\delta(x_1-x_3)t^{\alpha\nu\kappa\mu}_{FF}(x_3,x_2,\ldots) ]+$$
$$-2\d^{x_3}_{\kappa} [ t^{\alpha\mu\kappa\nu}_{AAF}(x_1,x_2,x_3,\ldots)+2\delta(x_1-x_2)t^{\kappa\nu\alpha\mu}_{FF}(x_3,x_2,\ldots) ]+$$
$$+4\d_{\kappa}^{x_2}\d_{\lambda}^{x_3}t^{\alpha\kappa\mu\lambda\nu}_{AFF}(x_1,x_2,x_3,\ldots)+4\d_{\kappa}^{x_1}\d_{\lambda}^{x_2}t^{\kappa\alpha\lambda\mu\nu}_{FFA}(x_1,x_2,x_3,\ldots) -8 \d_\kappa^{x_1} \d_\lambda^{x_2} \d_\sigma^{x_3}
t_{FFF}^{\kappa\alpha\lambda\mu\sigma\nu}(x_1,x_2,x_3,...)$$  
$$\Pi^{\alpha l\mu\nu}_{uAA}(x_1,x_2,x_3,\ldots):=t_{uAA}^{\alpha l\mu\nu}(x_1,x_2,x_3,\ldots)+$$ $$-2\d^{x_3}_{\kappa}t^{\alpha l\mu\kappa\nu}_{uAF}(x_1,x_2,x_3,\ldots)-2\d^{x_2}_{\kappa}t^{\alpha l\kappa\mu\nu}_{uFA}(x_1,x_2,x_3,\ldots)+4\d^{x_2}_{\kappa}\d^{x_3}_{\lambda}t^{\alpha l\kappa\mu\lambda\nu}_{uFF}(x_1,x_2,x_3,\ldots)+$$ $$+2 \delta(x_2-x_3)t_{uF}^{\alpha (l-1)\mu\nu}(x_1,x_2,\ldots), \quad l>3 $$

These definitions are natural, because the defined distributions represent in each case the sum of all
distributions contributing to the same operator in the adiabatic limit where partial integrating is formally possible. (Note that also the four-gluon terms ( proportional to $\delta$, see (B.9)) contribute to the operator where
all external legs are attached to different vertices; for further details see [41])\\
$\bullet$ Inserting (B.18) into (B.17) and inserting (B.22) and (B.21) into (B.20) and using the new definitions, we already arrive at 2 of the 5 summed identities of gauge invariance:\\
$$ \d^{x_1}_{\kappa}\Pi^{\kappa\nu}_{AA}(x_1,x_2,x_3,\ldots ,x_{n-1})-\d^{\alpha}_{x_2} [\d^{\alpha}_{x_2}t_{u\overline{u}}^{\nu}(x_1,x_2,x_3,\ldots ,x_{n-1})-(\alpha\leftrightarrow\nu)]+$$
$$+\sum^n_{l=3} \d_{\kappa}^{x_l}\Pi^{\kappa l\nu}_{uA}(x_1,x_2,x_3,\ldots ,x_{n-1})=0 \eqno(B.32)$$

$$ \d^{x_1}_{\alpha}\Pi^{\alpha\mu\nu}_{AAA}(x_1,x_2,x_3,x_4,\ldots ,x_n)+$$ 
$$+[\bigl(\d^{x_2}_{\alpha}[\d^{\alpha}_{x_2}\Pi^{\mu\nu}_{u\overline{u}A}(x_1,x_2,x_3,x_4,\ldots ,x_n)-(\alpha\leftrightarrow\mu)]\bigl) - \bigl((x_2,\nu)\longleftrightarrow(x_3,\mu)\bigl) ]+$$
$$+g[\delta(x_1-x_2)-\delta(x_1-x_3)]\Pi^{\mu\nu}_{AA}(x_2,x_3,x_4,\ldots ,x_n)+$$ 
$$+g[\bigl(\d^{x_2}_{\alpha}[\delta(x_2-x_3)g^{\alpha\mu}t^{\nu}_{u\overline{u}}(x_1,x_2,x_4,\ldots ,x_n)-(\alpha\leftrightarrow\nu)]\bigl) - \bigl((x_2,\nu)\longleftrightarrow(x_3,\mu)\bigl) ]+$$
$$+\sum^{n}_{l=4}\d^{l}_{\alpha}\Pi^{\alpha l\mu\nu}_{uAA}(x_1,x_2,x_3,x_4,\ldots ,x_n) = 0 \eqno(B.33)$$
$\bullet$ One easily checks that the following (local) renormalisations of the self energy distributions  are compatible with the first summed identity of gauge invariance (B.32) (and also with Lorentz invariance, all the discrete symmetries and pseudo-unitarity) in the nth step of the inductive construction. Because we are interested in the comparison with the Slavnov-Taylor identities, we state only the relevant local normalisation terms which survive in the adiabatic limit in regard to the inner coordinates (see (6.14)):\\
$$ \Pi^{\mu\nu}_{AA} + C_{AA}^{n-1} [ \d^{\mu}_{x_1}\d^{\nu}_{x_1} - g^{\mu\nu}\d_{x_1}\d_{x_1} ] \delta^{n-2} $$
$$ t^{\nu}_{u\overline{u}} + C_{u\overline{u}}^{n-1} \d^{\nu}_{x_2} \delta^{n-2}$$
$\bullet$The possible renormalisations of the two vertices (compatible with Lorentz invariance, the discrete symmetries and pseudo-unitarity) are the following:\\
$$ \Pi_{AAA}^{\alpha\mu\nu} + C_{AAA}^{n} [ g^{\alpha\mu}(\d^{\nu}_{x_1}-\d^{\nu}_{x_2}) +g^{\alpha\nu}(\d^{\mu}_{x_3}-\d^{\mu}_{x_1}) +g^{\mu\nu}(\d^{\alpha}_{x_2}-\d^{\alpha}_{x_3}) ] \delta^{n-1} $$
$$ \Pi_{u\overline{u}A}^{\mu\nu} + C_{u\overline{u}A}^{n}  \delta^{n-1} g^{\mu\nu}$$
The second summed identity (B.33) implies the following relation between these four
normalisation constants in the nth step of the inductive construction:\\
$$g  C_{AA}^{n-1}+C_{u\overline{u}A}^{n}- g C_{u\overline{u}}^{n-1}-C_{AAA}^{n} = 0$$
Because of $Z_{i}:=1+C_{i}$ (Note our conventions in the ghost sector.), we directly get the well-known relation between the Z-factors at one-loop level:\\
$$\frac{Z_{AA}}{Z_{AAA}} = \frac{Z_{u\overline{u}}}{Z_{u\overline{u}A}} $$

The interpretation of this relation is slightly different in the causal approach: It represents the restrictions by gauge invariance on finite normalisation terms only. Of course we do not need any infinite part in the Z-factors to absorb divergences in the causal approach.\\ 
Using the two summed 4-leg identities or the identities with external fermion pairs (B.29) and (B.30), we can analogously deduce the corresponding relation of the Z-factor of the four-gluon vertex or of the matter vertex.\\

\newpage
%\vskip1.0cm
{\Large\bf Appendix C $\quad$ The Causal Approach to QFT}
\vskip1.0cm

For the purposes of completeness, we give a brief introduction to the Epstein-Glaser method in quantum field theory (for details see [13,12]). We present a solution to the crucial problem of distribution splitting
following a recently given formulation of this question (see [42]). We state some results which are decisive especially for the causal construction of massless theories (see [27]).\\
$\bullet \bullet $ {\bf The Method of Epstein and Glaser:} In the traditional Lagrangean approach to quantum field theory, the basic objects are the quantized interacting fields. The Greens functions and the time-ordered products constructed by the famous Feynman rules are not mathematically well-defined since they contain the product of operator-valued distributions with discontinuous step functions. This leads to the well-known ultraviolet divergences in perturbation theory. They have to be corrected by subsequent renormalization. In order to manipulate divergent integrals in the renormalization program, one has to introduce an intermediate regularization. 

Epstein and Glaser followed the Bogoliubov's formalism in order to keep apart the different difficulties encountered in perturbative quantum field theory and to show that the standard renormalization procedure, especially the intermediate regularization, is neither essential nor constitutive for the physical theory at all [13,12]. There is a related approach to perturbative quantum field theory given by Steinmann [39].

In contrast to the usual Lagrangean approach, Epstein and Glaser construct the  perturbative scattering matrix $S(g)$ directly in the well-defined Fock space of free fields $F$. They introduce the S-matrix without reference to Lagrangean formalism and do not use the problematic concept of a quantized interacting field.
$$S(g)\in B(F),\quad \it g\in \cal S \it (\bf R^{\it 4}) \eqno(C.1)$$
In order to obtain the explicit form of the S-matrix, they use certain physical conditions. Here the condition of causality plays the most important role:
\vskip0,3cm
$\bullet$ If the support of $g_{1}\epsilon\cal S$ in Minkowski space is earlier than the support of $g_{2}\epsilon\cal S$ in some Lorentz frame (supp$g_{1}<$ supp$g_{2}$), then the S-matrix fulfills the following functional equation:
$$S(g_1+g_2)=S(g_2)\cdot S(g_1)\quad \mbox{\bf[Causality (I)]} \eqno(C.2)$$
$\bullet \quad U(a, \Lambda)$ shall be the usual representation of the Poincar\'e group $P_+^\uparrow$ in the Fock space $F$. The condition of {\bf Poincar\'e invariance} of the S-matrix can be expressed as follows:
$$U(a,\bf 1\it )S(g)U(a,\bf 1\it )^{-1} = S(g_a) \quad \forall a \epsilon \bf R^{\it 4} $$
$$\mbox{where} \quad g_a(x)=g(x-a).\quad \hbox{\bf[Translational invariance (II)]} \eqno(C.3)$$
$$U(0,\Lambda)S(g)U(0,\Lambda)^{-1}=S(g_\Lambda), \quad \forall \Lambda \in L_+^\uparrow $$
	$$\mbox{where} \quad g_\Lambda(x)=g(\Lambda^{-1}x).\quad \hbox{\bf[Lorentz Invariance (III)]} \eqno(C.4)$$
$\bullet$ Epstein and Glaser search for a solution of the functional equation for the S-matrix of the following form (formal power series in $g\epsilon\cal S$)
$$S(g)=\bf 1 \it + \sum_{n=1}^\infty \frac{1}{n!} \int d^4x_1 \ldots d^4 x_n T_n(x_1,\ldots x_n)g(x_1)\ldots g(x_n)$$
$$ \=d \bf 1 \it + \it T.\rm \quad \hbox{\bf[Perturbative Ansatz (IV)] }\eqno(C.5)$$
The $T_{n}$ are operator-valued n-point distributions.

$\bullet$ The {\bf specific coupling} of the theory $T_{n=1}$ (\bf V \rm) is given.
\vskip 0,1cm
Examples:
\vskip 0,1cm
U(1) - gauge theory :
$$T_{1} = ie A_\nu :\overline \Psi \gamma ^\nu\Psi:\eqno(C.6)$$
$SU(N)$-gauge theory  :
$$T_1 = \frac{i}{2}g f_{abc}: (A_\nu^a A_\nu^b F_c^{\nu\mu}+A_\mu^a v_b \partial^\mu \tilde v_c):\eqno(C.7)$$
Note that all fields in $T_{n=1}$ are unterstood to be free fields!
\vskip0,3cm
Epstein and Glaser show that the whole perturbative S-matrix in the sense of a formal power series (IV) is already determined by the conditions of causality (I), translational invariance (II) and the specific coupling of the theory (V) except for a number of finite (!) free constants which have to be fixed by further physical conditions. Epstein and Glaser are able to work out the whole of perturbation theory by an explicit inductive construction in a mathematically rigorous way. The condition of Lorentz invariance (III) is optional because causal perturbative theory can be worked out without requiring (III).
The main steps are the following:

$\bullet$ Analogously to (C.5), Epstein and Glaser express the inverse S-matrix also by a formal power series:
$$S(g)^{-1}=1+\sum_{n=1}^\infty \frac {1}{n!} \int d^4x_{1}\ldots d^4 x_n \tilde{T}_n(x_1,\ldots x_n)g(x_1)\ldots g(x_n)$$
$$ =(\bf 1 \rm + \it T)^{-1}=\bf 1\rm + \sum_{n=1}^\infty (-\it T)^r.\eqno(C.8)$$ 
Since by definition $\tilde{T}(x_1, \ldots , x_n)$ and also $T_n (x_1, \ldots , x_n)$ are symmetric in $x_1, \ldots , x_n$, it is convenient to use a set-theoretical notation X = {$x_1, \ldots , x_n$}. The distributions $\tilde{T}$ can be computed by formal inversion of (C.5):
$$\tilde{T}_n(X)=\sum_{r=1}^n(-)^r\sum_{P_r}T_{n_1}(X_1)\ldots T_{n_r}(X_r),\eqno(C.9)$$ 
where the second sum runs over all partitions $P_r$ of X into r disjoint subsets
$$X=X_1\cup\ldots\cup X_r,\quad X_j\not=\emptyset,\quad \mid X_j \mid =n_j.\eqno(C.10)$$
Besides (C.9), one can deduce further relations between the n-point distrubutions:
$$\sum_{P_2^0} T_{n_1}(x)\tilde T_{n-n_1}(Z \backslash X) = 0 \eqno(C.10a)$$
for all Z with $\mid Z \mid = n \ge 1, \quad \mid X \mid = n_1$;
$$\sum_{P_2^0} T_{n-n_2} (Z \backslash Y) \tilde T_{n_2}(Y) = 0 \eqno(C.10b)$$
for all Z with $\mid Z \mid = n \ge 1, \quad \mid Y \mid = n_2$.
We stress the fact that all products of distributions are well-defined because the arguments are disjoint sets of points so that the products are tensor products of distributions.

$\bullet$ Epstein and Glaser translate the conditions imposed on S(g) into conditions on the n-point distributions $T_n (x_1,\ldots ,x_n)\quad$ and $\quad \tilde{T_n}(x_1,\ldots ,x_n) \quad n\epsilon \bf N \it$, according to the Bogoliubov's approach: 

{\bf(I)} Causality
$$T_n(x_1,\ldots ,x_n)=T_m(x_1,\ldots x_m)\cdot T_{n-m}(x_{m+1},\ldots x_n),\eqno(C.11)$$
$$\mbox{if} \qquad \{x_{m+1},\ldots x_n\}<\{x_1,\ldots x_m\} $$
$$\tilde{T}_n(x_1,\ldots ,x_n)=\tilde{T}_m(x_1,\ldots x_m)\tilde{T}_{n-m}(x_{m+1},\ldots x_n),$$
$$\mbox{if} \qquad \{x_1,\ldots x_m\}<\{x_{m+1},\ldots x_n\}$$

{\bf(II)} Translational Invariance
$$U(a,\bf 1\it) T_n(x_1,\ldots ,x_n)U(a,\bf 1\it )^{-1}=T_n(x_1+a,\ldots ,x_n+a) \quad  \forall a \epsilon \bf R^{\it 4} \eqno(C.12)$$
{\bf(III)} Lorentz Invariance
$$U(0,\Lambda)T_n(x_1,\ldots ,x_n)U(0,\Lambda)^{-1}=T_n(\Lambda x_1,\ldots ,\Lambda x_n) \quad \forall \Lambda \epsilon \rm L_+^\uparrow \eqno(C.13)$$
$\bullet$ Now Epstein and Glaser introduce the retarded and the advanced n-point distributions:
$$R_n(x_1,\ldots ,x_n)=T_n(x_1,\ldots ,x_n)+R'_n \qquad \mbox {where}\quad R'_n=\sum_{P_2}T_{n-n_1}(Y,x_n)\tilde{T}_{n_1}(X) \eqno(C.14)$$
$$A_n(x_1,\ldots ,x_n)=T_n(x_1,\ldots ,x_n)+A'_n \qquad \mbox{where} \quad A'_n=\sum_{P_2}\tilde{T}_{n_1}(X)T_{n-n_1}(Y,x_n). \eqno(C.15)$$
The sum runs over all partitions $P_2:\{x_1,\ldots x_{n-1} \}=X \cup Y, \quad X \not= \emptyset$ into disjoint subsets with $\mid X \mid =n_1 \ge 1, \mid Y \mid \le n-2.$

Both sums, $R'_n$ and $A'_n$, contain $T_j$'s with $j \le n-1$ only and are therefore known quantities in the inductive step from $n-1$ to $n \quad$ - in contrast to $T_n$.

Note that the last argument $x_n$ is marked as the reference point for the support of $R_n$ and $A_n$.

The following proposition is a consequence of the causality condition (I):
\vskip 0.2cm
{\bf Proposition C.1}
$$\mbox{supp} R_m(x_1,\ldots ,x_m) \subseteq \Gamma_{m-1}^+(x_m), 
\quad m < n \eqno(C.16)$$
where $\Gamma_{m-1}^+$ is in the (m-1)-dimensional closed forward cone
$$\Gamma_{m-1}^+(x_m)=\{(x_1,\ldots ,x_{m-1}) \mid (x_j - x_m)^2 \ge 0, x_j^0 \ge x_n^0, \forall j \}.$$
\vskip0.2cm
In the difference
$$D_n(x_1, \ldots ,x_n) \=d R_n-A_n = R'_n-A'_n\eqno(C.17)$$
the unknown n-point distribution $T_n$ cancels. Hence this quantity is also known in the inductive step. It should be added that $D_n$ has a causal support.
\vskip0.2cm
{\bf Proposition C.2}
$$\mbox{supp} D_n \subseteq \Gamma_{n-1}^+(x_n) \cup \Gamma_{n-1}^-(x_n) \eqno(C.18)$$
This crucial support property is preserved in the inductive step. It directly results from causality.  

$\bullet$ Given the aforegoing facts, the following inductive construction of the n-point distribution $T_n$ becomes possible:

Starting off with the known $T_m(x_1, \ldots , x_n), m \le n-1,$ one computes $A'_n, R'_n$ and $D_n = R'_n - A'_n.$ With regard to the supports, one can decompose $D_n$ in the following way:
$$D_n(x_1, \ldots , x_n) = R_n (x_1, \ldots , x_n) - A_n (x_1, \ldots , x_n)\eqno(C.19)$$
$$\mbox{supp} R_n \subseteq \Gamma_{n-1}^+(x_n), \quad \mbox{supp} A_n \subseteq \Gamma_{n-1}^-(x_n)$$
Then $T'_n$ is given by
$$T'_n = R_n - R'_n = A_n - A'_n \eqno(C.20)$$
One can verify that the $T'_n$ satisfy the conditions (C.11),(C.12) and (C.13) [13].\\
Because of the marked $x_n$-variable, we finally symmetrize:
$$T_n(x_1,\ldots x_n)=\sum_{\pi} \frac{1}{n!} T'_n(x_{\pi 1}, \ldots x_{\pi n}) \eqno(C.21)$$
The only nontrivial step in the construction is the splitting of the operator-valued distribution $D_n$ with support in $\Gamma^+ \cup \Gamma^-$ into a distribution $R_n$ with support in $\Gamma^+$ and a distribution $A_n$ with support in $\Gamma^-$. In causal perturbation theory the usual renormalization program is reduced to this conceptually simple and mathematically well-defined problem.

In fact this problem of distribution splitting was already solved in a general framework by the mathematician Malgrange in 1960 [40].
Epstein and Glaser used his general result for the special case of quantum field theory [13].
We follow a new formulation  of the splitting problem given by Scharf [42].

$\bullet \bullet $ {\bf The Theory of Distribution Splitting:} Let there be an operator-valued tempered \hyphenation{dis-tri-bu-tion} distribution with causal support:
$$D\epsilon \cal S' (\bf R^{\it 4n}),\quad \it supp D \subset \Gamma^+ (x_n) \cup \Gamma^- (x_n) \eqno (C.22)$$
The question is whether it is possible to find a pair (R, A) of tempered distributions on $\bf R^{\it 4n}$ with the following characteristics:
$$\bullet\quad R, A \epsilon \cal S \it' (\bf R^{\it 4n})\qquad \mbox{\bf (A)}$$
$$\bullet\quad R \subset \Gamma^+(x_n), \quad  A \subset \Gamma^-(x_n)\qquad \mbox{\bf (B)} \eqno (C.23)$$
$$\bullet\quad R - A = D\qquad \mbox{\bf (C)}$$
After normal ordering of the causal operator-valued distribution
$$D=\sum_k : \hat{\Theta}_k : d_k \eqno (C.24)$$
the numerical distributions $d_k$ remain to be split into a pair $(r_k, a_k)$ respectively:
$$d^k = r^k - a^k, \quad supp r_k \subseteq \Gamma^+_{n-1} (x_n), \quad supp a_k \subseteq \Gamma ^-_{n-1} (x_n) \eqno (C.25)$$
Without normal ordering this procedure is not well-defined.

From the translational invariance of $D_n$
$$U (a,\bf 1\it) D_n(x) U^{-1}(a,\bf 1 \it) = D (x + a) \eqno (C.26)$$
it follows
$$ d_k (x) = d_k (x + a)\quad \forall k \eqno (C.27)$$
We may set $x_n = 0$ and look at the splitting problem for distributions
$$d(x) = d_k (x_1, ..., x_{n-1}, 0) \epsilon \cal S' \it (\bf R^{\it 4n-1}) \eqno (C.28)$$
Because of $\Gamma^+_{n-1} (0) \cap \Gamma^-_{n-1} (0) = \{0\}$, the behaviour of the distribution at $x$ = 0 is crucial for the splitting problem. We therefore classify the singularities of distributions in this region. This can be a carried out with the help of the singular order of distributions which is a rigorous definition of the usual power-counting degree :

We assume $d(x)$ to be a tempered distribution $\epsilon \cal S'\it (\bf R^{\it m}),\it m = 4(n-1)$.
\vskip0,3cm
{\bf Definition C.1} The distribution $d(x) \epsilon \cal S' \it(\bf R^{\it m})$ has quasi-asymptotics $d_0(x)$ at $x = 0$, with regard to a positive continuous function $\rho(\delta), \delta > 0$ if the limit
$$\lim_{\delta\rightarrow 0} \rho(\delta)\delta^m d(\delta x) = d_0(x) \neq 0 \eqno (C.29)$$
exists in $S'(R^m)$.
\vskip0,3cm
We present the equivalent definition in momentum space:
\vskip0,3cm
{\bf Definition C.2} The distribution $d(p) \in \cal S' \it (\bf R^{\it m})$ has quasi-asymptotics $\hat{d}_0(p)$ at $p=\infty$, with regard to a positive continuous function $\rho(\delta), \delta > 0$ if the limit
$$\lim_{\delta\rightarrow 0} \rho(\delta) \Big{\langle} \hat{d} \Big{(} \frac{p}{\delta} \Big{)}, \check{\varphi}(p) \Big{\rangle} = \langle \hat{d}_0, \check{\varphi} \rangle \neq 0  \eqno (C.30)$$
exists for all $\check{\varphi} \in \cal S \rm (\bf R^{\it m}).$

By scaling transformation it follows that
$$\lim_{\delta\rightarrow 0} \frac{\rho(a\delta)}{\rho(\delta)}=a^\omega \eqno (C.31)$$
with some real $\omega.$ $\rho$ is called power-counting function.
\vskip0,3cm
{\bf Definition C.3} The distribution $\hat{d}(p)$ (resp. $d(x))\in \cal S' \rm (\bf R^{\it m})$ is called singular of order $\omega$ at $p = \infty$ (resp. $x = 0)$, if it has a quasi-asymptotics $\hat{d}(p)\quad (d_0(x))\quad at \quad p = \infty \quad (x = 0)$ with power-counting function $\rho(\delta)$ satisfying (C.31).
\vskip0,3cm
Note that this definition differs from the one introduced by Epstein and Glaser
 [13]. The latter definition is hampered by the fact that the corresponding definitions in the $x$-space and $p$-space are not completely equivalent. Our definition does not have this defect.
\vskip0,2cm
We specifiy the splitting problem by requiring
$$\bullet\qquad \omega(r) \le \omega(d)\quad \wedge \quad \omega(a) \le \omega(d).\qquad \mbox{{\bf (D)}} \eqno (C.32)$$
$\bullet$ Now we are in a position to construct explicit splitting solutions. One has to distinguish two cases:

{\bf 1. Case:} 
$$\omega(d) <0, \quad d \in \cal S' \it (\bf R^{\it m}) $$
 In this case we can construct a well-defined solution of the splitting problem by multiplying $d$ by
a discontinuous $\Theta$-step-function:
%$$ \big{\langle} r(x), \varphi \big{\rangle} \=d \big{\langle} \Theta(v x) 
%d(x), \varphi \big{\rangle}, \quad a(x)= r(x)-d(x), \quad \varphi \in \cal S, \eqno(C.33) $$
Let $v = (v_1, \cdots, v_{n-1}) \in \Gamma^+$, i.e. all four-vectors $v_j$ are time-like, $v_j \in \vee^+$. All products $v_j \cdot x_j$ are either $\ge 0$ for $x \in \Gamma^+$ or $\le 0$ for $x \in \Gamma^-.$  Using a continuous $C^\infty-$ function $\chi_0$ over $\bf R^{\it 1}$ with 
$$ \chi_0(t) = \left\{ \begin{array} {r@ {\quad\mbox{for}\quad}l} 0 & t \le 0 \\
<1 & 0<t<1 \\ 1 & t \ge 1 \end{array} \right. $$ 
one demonstrates [42] that the following limit exists (This limit defines (!) the multiplication of $d(x)$ by a $\Theta$-step-function.):
$$\lim_{\delta\rightarrow0}\chi_0\Big{(}\frac{vx}{\delta}\Big{)} d(x) \=d \Theta(vx)d(x) \=d r(x), \quad a(x)=r(x)-d(x) \eqno (C.33)$$
\vskip0.3cm
{\bf 2. Case}
$$\omega(d) \geq 0, \quad d \in \cal S'$$
In this case the trivial splitting as in case 1 is possible only if one  
replace the general testfunction $\varphi$ by 
$$(W\varphi)(x) \=d \varphi(x)-w(x)\sum_{\mid a \mid=0}^\omega \frac{x^a}{a!}(D^a\varphi)(0) \eqno(C.34)$$
(where $w(x) \in \cal S \it (\bf R^{\it m})$ with $w(0) = 1, \quad D^a w(0) = 0, \quad 1 \le \mid a \mid \le \omega$) (see [42]
On this way we arrive at a well-defined splitting solution $r$ :
$$\langle r(x),\varphi\rangle\=d\langle d(x),\Theta(v\cdot x) W\varphi\rangle,\quad a(x)=r(x)-d(x), \quad \varphi \in \cal S \eqno (C.35)$$
$r(x)$ defines a tempered distribution with $supp r \subseteq \Gamma^+(0).$ It should be stressed that $r(x) \in \cal S'$ is only a well-defined tempered distribution, provided that one sums up the additional terms in (C.34). In the case of $\omega(d) \ge 0$, formal trivial splitting leads to the well-known ultraviolet divergences in field theory.

$\bullet$ We still have to ask whether the splitting solution is unique:

Let $r_1 \in \cal S'$ and $r_2 \in \cal S'$ be two splitting solutions of a given distribution $d \in \cal S'$. After construction $r_1$ and $r_2$ have their support in $\Gamma^+$ and  agree with d on $\Gamma^+\setminus \{0\}$, from which follows that $(r_1 - r_2)$ is a tempered distribution with point support and with singular order $\omega \le \omega(d):$
$$\mbox{supp}(r_1 - r_2) \subset \{0\},\quad \omega(r_1-r_2) = \omega(d), \quad (r_1-r_2) \in \cal S' \eqno (C.36)$$
According to a well-known theorem in the theory of distributions, we have
$$r_1 - r_2 = \sum_{\mid a \mid=0}^{\omega_0} C_a D^a\delta(x). \eqno (C.37)$$
In case 1, $\omega(d)<0$, the splitting solution is thus unique, in particular
it is independent of the vector $v$ in (C.33).
In case 2, $\omega(d)\ge 0$, the splitting solution is only determined up to a local distribution with a fixed maximal singular degree $\omega_0\le \omega(d)$. The demands of causality (C.2) and translational invariance (C.3) leave the constants $C_a$ in (C.37) undetermined. They have to be fixed by additional physical normalization conditions such as Lorentz covariance or gauge invariance.

$\bullet$ In the causal approach the question of the normalizability of a quantum field theory does not involve the proof of its finiteness. Infinities do not appear at all in our formulation.
The problem of normalizability, however, consists above all in the proof of the statement that the number of the (of course finite) constants $C_a$ to be fixed by physical conditions stays the same in all orders $n$ of perturbation theory. This means that finitely many normalization conditions are sufficient to determine the S-matrix completely (see Chapter 4).

$\bullet \bullet $ {\bf Explicit Splitting Solutions in  Massless Theories:} We present some splitting formulae in momentum space :

Let $\hat r_0(p)\in \cal S' \it (\bf R^{\it m})$ be a splitting solution of $\hat{d}(p) \in \cal S' \it (\bf R^{\it m})$. We arrive at the general solution in momentum space by adding a polynomial in $p$ of degree $\omega = \omega(d)$ with undetermined coefficients $C_a$ (see (C.37)).
$$\hat r(p) = \hat r_0 (p) + \sum_{\mid a \mid=0}^{\omega(d)} C_a p^a \eqno (C.38)$$
Let the derivatives $D^b \hat r (q)$ exist in the usual sense of functions for all $\mid b \mid \le \omega.$ 
To put it more precisely, we require the following conditions:

({\bf 1}) $\hat {d}(p)$ ist $\omega$-times continuously differentiable in a neighbourhood of $p = q$.

({\bf 2}) The derivatives $(D^a \hat{d}) (p)$ are H\"older - continuous at $p = q \quad \forall \mid a \mid = \omega.$\\
\\
Under these conditions there exists a integral representation for $\hat r^{q}$ 
$$\hat r^{q}(p):=\hat r (p)- \sum_{\mid b \mid=0}^\omega \frac{(p)^b}{b!} D^b \hat{r}(0) \eqno (C.39)$$
This is the splitting solution of $\hat{d}(p)$ which is uniquely determined by the normalization conditions
$$ D^b \hat r^{q}(q)=0 \qquad \mid b \mid \le \omega \eqno (C.40)$$
For $p-q \in \tilde{\Gamma}_+^{n-1} (0)$ one deduces the following explicit representation of $\hat{r}^{q}$ as a dispersion integral. 
$$\hat r^{q} (p) = \frac{i}{2\pi} \int\limits_{-\infty}^\infty \frac{dt}{(t-i0)^{\omega+1}} \quad\frac{\hat{d}(tp+(1-t)q)}{(1-t+i0)} \quad \forall p-q \in \tilde{\Gamma}_+^{n-1}(0) \quad \omega(\hat{d}) \ge 0 \eqno (C.41)$$
A formula for other $p's$ can be derived by analytical continuation.

The unique splitting solutions in case (1)  $\quad \omega(d) < 0$ can be presented analogously as a dispersion integral
$$\hat r(p)=\frac{i}{2\pi} \int\limits_{-\infty}^\infty dt \frac{\hat{d}(tp)}{(1-t+i0)} \quad \forall p \in \tilde{\Gamma}_+^{n-1}(0) \quad w(\hat{d})<0 \eqno (C.42)$$
In both formulae (C.41,C.42) the t-integral is understood in the sense of distributions, i.e. one first has to smear out the integrand by a test function $\varphi \in \cal S$ and then the t-integration is to be carried out in order to arrive at $<r(p), \hat{\varphi}(p)>$.

$\bullet$ The splitting solution $r^{q=0}$ with the special choice $q=0$ is called the central or symmetrical splitting solution  which contains all relevant symmetrical characteristics. This solution is in particular $L_+^\uparrow$-covariant and also gauge invariant in case of gauge theories [25]. In theories with only massive fields the central splitting solution  exists.
But already for an abelian gauge theory with a massless gauge boson and massive matter fields there still is no complete proof of the existence of the central solution. For the proof of gauge invariance in this theory with the help of explicit splitting solutions one therefore has to pursue some additional consideration [25]. In [26] we have given a new proof of gauge invariance for the abelian theory which does not draw on explicit splitting solutions, is also valid for the case of massless matter fields, and thus yields important methodological suggestions for the nonabelian gauge theory.

$\bullet$ The splitting solution $\hat{r}^q$ (C.41) with normalization point $q$, however, $q \in \bf R^{\it 4l} \backslash \{0\}$ , also exists in the massless case under the above mentioned conditions. But considering the following equation
 
$$r^{\Lambda q}(\Lambda x) = D(\Lambda)r^q(x),\quad \Lambda \in L_+^\uparrow, \eqno (C.43)$$
it is obvious that $r^q$ is in general not covariant because the subtraction point $q$ is transformed with $\Lambda$. In order to obtain a covariant splitting solution, one has to perform a finite renormalization. Therefore, in applying (C.41), it is not necessary to compute precisely the $q$-depending terms which make the normalization (C.40): 
We define for an arbitrary retarded part $r(x)$ of $d(x)$:
$$r_q(x) \=d r(x)e^{iqx}. \eqno (C.44)$$
It is easy to see that $r_q$ is a splitting solution of $d_q$ defined by $d_q \=d e^{iqx} d(x)$. We assume $q$ to be totally space-like. Then there is the central solution $r_q^{q=0}$ of $d_q$ and
$$P_q(p) \=d \hat{r}_q(p) - \hat{r}_q^{q=0}(p) \eqno (C.45)$$
is a $q$-depending polynomial of degree $\omega$ in $p$. $\hat{r}(p)$ can be obtained from $\hat{r}_q(p)$ by taking the limit
$$\hat{r}(p) = \lim_{q\to0}[\hat{r}_q^{q=0}(p) + P_q(p)]. \eqno (C.46)$$
We must add a $q$-depending polynomial $P_q$ in such a way that the limit exists. Thus we obtain a splitting solution $\hat{r}$ of $d$.

In Chapter 3 (a) this splitting method is used for the analysis of the gauge boson self-energy.

$\bullet \bullet $ {\bf Lorentz Covariant Splitting and Cohomology:} In order to prove the nonabelian gauge invariance, we need the statement of the existence of a $L^{\uparrow}_+$-covariant splitting solution also in the case $m = 0$.
%
%First we consider the splitting of a Lorentz invariant causal distribution $d(x), \quad x = (x_1, \ldots, x_n)$, $x_j \in \bf R^{\it 4}$, satisfying
%$$d(x') = d(x), \eqno (5.1)$$
%where
%$$x' = \Lambda x \=d (\Lambda x_1, \ldots, \Lambda x_n) \eqno (5.2)$$
%for arbitrary proper Lorentz transformations $\Lambda \in L^{\uparrow}_+$. Let $r(x)$ be an arbitrary retarded part of the r.h.s. of (5.1). Then $r(\Lambda x)$ is a retarded part of the l.h.s. of (5.1). The two may differ by local terms
%$$r(\Lambda x) = r(x) + \sum_u c_u(\Lambda)D^u \delta (x). \eqno (5.3)$$
%The sum runs over all multi-indices $u$ with $\mid u \mid \le \omega$, where $\omega$ is the singular order of $d(x)$ at the origin. It is our aim to show that the splitting solution can be redefined (or normalized)
%$$\hat{r}(x) = r(x) + \sum_u b_uD^u \delta (x) \eqno (5.4)$$
%so that it becomes $L^{\uparrow}_+$ invariant.
%$$\hat{r}(\Lambda x) = \hat{r}(x). \eqno (5.5)$$
Epstein and Glaser give a sketch proof of this statement by integrating over a maximal compact subgroup of the complex Lorentz group [13,6.3]. In [27] it
is explicitly shown that the existence of a $ L_+^\uparrow$-covariant splitting solution is a direct consequence of a trivial cohomology of the Lorentz group.

$\bullet \bullet$ {\bf The Causal Construction in Case of Fermionic Vertices:} Our formulation of gauge invariance (see Chapter 2) also requires the inductive construction in case of fermionic vertices $T_{n=1}$. Therefore, we have to generalize the bosonic case treated by Epstein and Glaser (For a detailed analysis see [37]):

We start off with the theory given by
$$S_1(\underline{g})=\int d^4x \{T_0^1(x)g_0(x)+T_1^1(x)g_1(x)\} \qquad \mbox{where} \qquad \underline{g}:=(g_0,g_1). \eqno (C.47)$$
Let $T_0^1$ be a bosonic coupling and $T_1^1$ a fermionic one. Thus, $T_1^1$ anti-commute for space-like separated points:
$$\{T_1^1(x), T_1^1(y)\} = 0\quad \mbox{for} \quad (x-y)^2<0 \eqno (C.48)$$
However, the causality of the S-matrix (C.2) implies
$$[S_1(\underline{g}), S_1(\tilde{g})]=0 \eqno (C.49)$$
for space-like separation of the supports of $\underline{g}=(g_0,g_1)$ and $\tilde{g} = (\tilde{g}_0, \tilde{g}_1)$.

Inserting (C.47) and (C.48) in (C.49), we see that $g_1(x)$ and $\tilde{g}_1(y)$ must be anti-commuting Grassmann variables. 

For the S-matrix in n-th order, we make the ansatz
$$S_n(\underline{g})= \frac{1}{n!} \sum_{i_1,\ldots ,i_n=0,1} \int dx_1 \ldots dx_n \quad T_{i_1 \ldots i_n}^{(n)}(x_1, \ldots ,x_n) \quad g_{i_1}(x_1) \ldots g_{i_n}(x_n)\eqno (C.50)$$
where $i_e=0$ refers to a bosonic vertex and $i_e=1$ to a fermionic vertex.
Since the test function $g_{i_1}(x_1) \ldots g_{i_n}(x_n)$ in (C.50) has a mixed symmetry, we may assume $T_{i_1 \ldots i_n}^{(n)}(x_1, \ldots ,x_1)$ to have the same symmetry
$$T_{i_{\pi_1} \ldots i_{\pi_n}}(x_{\pi_1},\ldots x_{\pi_n})=(-1)^{Q(\pi)} \quad T_{{i_1} \ldots i_n}^{(n)}(x_1, \ldots ,x_n).\eqno (C.51)$$
Here, $Q(\pi)$ is the number of transpositions of fermionic vertices which are contained in $\pi$. Note that $Q(\pi)$ is not uniquely determined but $(-1)^{Q(\pi)}$ is. Because of (C.51), the set $X=\{(x_1,i_1),(x_2,i_2), \ldots ,(x_n,i_n)\}$ of pairs $(x_e,i_e)$ must be ordered in the fermionic vertices $i_e=1$. 

Following the inductive construction of the $T^{(n)}$'s in the pure bosonic case, it is now evident which modifications must be carried out for our mixed fermionic-bosonic case:
Every term must be multiplied with $(-1)^{Q(\pi)}$, where $\pi$ is the permutation, which puts $(x_1,i_1),\ldots ,(x_n,i_n)$ in that order in which they are in the considered term. 
Note that the central step in the inductive construction, the distribution splitting, is not affected by these additional factors $(-1)^{Q(\pi)}$. 
The final symmetrization of $T^{(n)}$'s turns partially into an antisymmetrization according to (C.51).

\vskip2cm

{\Large\bf Acknowledgements}
\vskip0.3cm
It is a pleasure to thank G\"unter Scharf for his continuous support. I profited a lot from his constructive criticism.  I am grateful to Michael D\"utsch for uncountable discussions and common work throughout the last years. I thank my sister Elisabeth and Michael D\"utsch for careful and critical reading of the manuscript. Financial support by {\it Cusanuswerk} and {\it Studienstiftung des deutschen Volkes} is gratefully acknowledged.

\vskip0.5cm
\newpage
{\Large\bf References} 

\vspace{0.5cm}

\begin{tabbing}

1. \quad\quad\quad\= N.N. Bogoliubov, A.A. Logunov, A.I. Oksak, I.T. Todorov,\\
 \>  General Principles of Quantum Field Theory,\\
 \> Kluwer Academic Publishers, Dordrecht 1990\\
2. \> V. Rivasseau,\\
 \>  From Perturbative to Constructive Renormalizations,\\
 \>  Princeton University Press, Princeton 1991\\
3. \> T. Balaban,\\
 \>  Communications in Mathematical Physics 122 (1989) 175, 355\\
4. \> T. Draper, S. Gottlieb, A. Soni , D. Toussant (eds.),\\
  \>  Lattice 93,\\
  \> Nuclear Physics (Proceedings Supplements) B34 (1994)\\
5. \> C.N. Yang, R.L. Mills, \\
 \> Progress of Theoretical Physics Supplement 37-38 (1966) 507 \\
6. \> G. 't Hooft, M. Veltman, \\
 \> Nuclear Physics B50 (1972) 318\\
7. \>  C. Becchi, A. Rouet, R. Stora,\\ 
 \>  Annals of Physics 98 (1976) 287\\
8. \> O.Piguet, A.Rouet,\\
 \>  Physics Reports 76 (1981) 1\\
9.  \> L. Baulieu,\\
 \>  Physics Reports 129 (1985) 1\\
10. \> P. Breitenlohner, D. Maison,\\
 \>  Communications in Mathematical Physics 52 (1976) 11, 39, 55\\
11. \> H. Balasin,\\
 \>  Das renormierte Wirkungsprinzip in der Quantenfeldtheorie,\\
 \>  Dissertation, Technical University Vienna 1990\\ 
12. \> G. Scharf, \\
 \> Finite Quantum Electrodynamics,\\
 \>  Texts and Monographs in Physics, Springer-Verlag, Berlin 1989\\
13. \> H. Epstein, V. Glaser, \\
 \> Annales de l'Institut Poincare 29 (1973) 211\\
 \> H. Epstein, V. Glaser,\\
 \>  in C. de Witt, R. Stora (eds.):\\
 \>  Statistical Mechanics and Quantum Field Theory,\\
 \>  Gordon and Breach, New York 1971, 501\\ 
14. \> H. Epstein, V. Glaser,\\
 \>  in G. Velo, A.S. Wightman (eds.):\\
 \>  Renormalization Theory,\\
 \>  D. Reidel Publishing Company, Dordrecht 1976, 193\\
15. \> R. Seneor,\\
 \>  in G. Velo, A.S. Wightman (eds.):\\
 \>  Renormalization Theory,\\
 \>  D. Reidel Publishing Company, Dordrecht 1976, 255\\
16. \> M. D\"utsch, F. Krahe, G. Scharf,\\
 \>  Journal of Physics G19 (1993) 503\\
17. \> J. Polchinski,\\
 \>  Nuclear Physics B231 (1984) 269\\
18. \> G. Gallavotti, F. Nicolo,\\
 \>  Communications In Mathematical Physics 100 (1985) 545; 101 (1986) 247\\
19. \> G. Keller, C. Kopper,\\
 \>  Physics Letters B273 (1991) 323\\
20. \> J.S. Feldman, T.R. Hurd, L. Rosen, J.D. Wright,\\
 \>  QED: A Proof of Renormalizability,\\
 \>  Lectures Notes in Physics 312, Springer-Verlag, Berlin 1988\\
21. \> M. Bonini, M. D'Attanasio, G. Marchesini,\\
 \>  Nuclear Physics B418 (1994) 81\\
22. \> M. Bonini, M. D'Attanasio, G. Marchesini,\\  
 \>  Nuclear Physics B421 (1994) 429\\  
23. \> O. Steinmann,\\
 \>  Nuclear Physics B350 (1991) 355, B361 (1991) 173\\
24. \> T. Hurth,\\
 \>  in preparation\\
25. \> M. D\"utsch, F. Krahe, G. Scharf, \\
 \> Nuovo Cimento 103A (1990) 903\\
26. \> M. D\"utsch, T. Hurth, G. Scharf,\\
 \> Physics Letters B327 (1994) 166\\
27. \> M. D\"utsch, T. Hurth, F. Krahe, G. Scharf,\\
 \>  Nuovo Cimento 106A (1993) 1029, 107A (1994) 375 \\
28. \> L. Schwartz, \\
\> Theorie des Distributions,\\
 \> Hermann, Paris 1966\\ 
29. \>  V.S. Vladimirov, \\
 \> Methods of the Theory of Functions of Many Complex Variables, \\ 
 \>  M.I.T. Press, Cambridge 1966\\ 
30. \> I. Schorn,\\
 \> Cutkosky-Regel und Zeitspiegelung in der kausalen St\"orungstheorie,\\
 \> Diploma Thesis, University Z\"urich 1991\\
31. \> N.V. Smolyakov,\\
 \> Theoretical and Mathematical Physics 50 (1982) 225\\
32. \> M. D\"utsch, T. Hurth, G. Scharf, \\
 \>  Nuovo Cimento A, to appear, ZU-TH-35/94\\
33. \> T. Kugo, I. Ojima, \\
 \> Progress of Theoretical Physics Supplement 66 (1979) 1\\
34. \> P. Cvitanovic, \\
 \> Physical Review D14 (1976) 1536\\
35. \> P. Dittner,\\
 \>  Communications in Mathematical Physics 22 (1971) 238\\
36. \> We thank G. Scharf for carrying out the calculation\\
 \>  of the determinants.\\
37. \> M. D\"utsch, T. Hurth, G. Scharf, \\
 \>  Nuovo Cimento A, to appear, ZU-TH-29/94\\
38. \> M. D\"utsch, F. Krahe, G. Scharf,\\
 \>  Nuovo Cimento 106A (1993) 277\\
39. \> O. Steinmann,\\
 \>  Perturbative Expansions in Axiomatic Field Theory,\\
 \>  Lectures Notes in Physics 11, Springer-Verlag, Berlin 1971\\
40. \> B. Malgrange, \\
 \> Seminaire Schwartz 21 (1960)\\
41. \> M. D\"utsch,\\
 \>   in preparation\\
42. \> G. Scharf,\\
 \>  Finite Quantum Electrodynamics (Second Edition),\\
 \>  Texts and Monographs in Physics, Springer-Verlag, Berlin 1995 \\
 
\\
\\
\end{tabbing}

\vspace{0.5cm}

\newpage

\newpage

\end{document}